\documentclass[manuscript]{acmart}
\AtBeginDocument{%
  \providecommand\BibTeX{{%
    \normalfont B\kern-0.5em{\scshape i\kern-0.25em b}\kern-0.8em\TeX}}}

\setcopyright{acmcopyright}
\copyrightyear{2025}
\acmYear{2025}
\acmDOI{XXXXXXX.XXXXXXX}

\acmJournal{JACM}
\acmVolume{37}
\acmNumber{4}
\acmArticle{111}
\acmMonth{8}

\acmPrice{15.00}
\acmISBN{978-1-4503-XXXX-X/18/06}

\usepackage{subcaption}

\usepackage{array, multirow} %%Table design
\usepackage{soul}
\usepackage{colortbl}

\usepackage{multicol}
\usepackage{lipsum}
\usepackage{mwe}
\usepackage{array}
\usepackage{booktabs} % for better spacing and formatting
\usepackage{multirow}
\usepackage{makecell}
\usepackage{threeparttable}
\usepackage{titlesec}
\usepackage{titlesec}
\titleformat{\subsubsection}[runin]{\normalfont\itshape}{\thesubsubsection.}{1em}{}[\newline]
\titlespacing*{\subsubsection}{0pt}{\baselineskip}{0pt}
\usepackage{hyperref} % fixing arxiv formatting issues
\usepackage{breakurl}

\usepackage{marginnote}

\makeatletter
\long\def\@mn@@@marginnote[#1]#2[#3]{%
  \begingroup
    \ifmmode\mn@strut\let\@tempa\mn@vadjust\else
      \if@inlabel\leavevmode\fi
      \ifhmode\mn@strut\let\@tempa\mn@vadjust\else\let\@tempa\mn@vlap\fi
    \fi
    \@tempa{%
      \vbox to\z@{%
        \vss
        \@mn@margintest
        \if@reversemargin\if@tempswa
            \@tempswafalse
          \else
            \@tempswatrue
        \fi\fi
          \rlap{%
            \if@mn@verbose
              \PackageInfo{marginnote}{xpos seems to be \@mn@currxpos}%
            \fi
            \begingroup
              \ifx\@mn@currxpos\relax\else\ifx\@mn@currxpos\@empty\else
                  \kern-\dimexpr\@mn@currxpos\relax
              \fi\fi
              \ifx\@mn@currpage\relax
                \let\@mn@currpage\@ne
              \fi
              \if@twoside\ifodd\@mn@currpage\relax
                  \kern\oddsidemargin
                \else
                  \kern\evensidemargin
                \fi
              \else
                \kern\oddsidemargin
              \fi
              \kern 1in
            \endgroup
            \kern\marginnotetextwidth\kern\marginparsep
            \vbox to\z@{\kern\marginnotevadjust\kern #3
              \vbox to\z@{%
                \hsize\marginparwidth
                \linewidth\hsize
                \kern-\parskip
                \marginfont\raggedrightmarginnote\strut\hspace{\z@}%
                \ignorespaces#2\endgraf
                \vss}%
              \vss}%
          }%
      }%
    }%
  \endgroup
}
\makeatother

\newcommand{\review}[1]{{#1}}
\renewcommand{\marginnote}[1]{{}}

\begin{document}

\newcommand{\subsubsubsection}[1]{\textbf{#1.}}

\title[Incorporating Procedural Fairness in Flag Submissions]{Incorporating Procedural Fairness in Flag Submissions on Social Media Platforms}

\author{Yunhee Shim}
\email{yunhee.shim@rutgers.edu}
\affiliation{%
  \institution{Rutgers University}
  \city{New Brunswick, NJ}
  \country{USA}
  }

\author{Shagun Jhaver}
\email{shagun.jhaver@rutgers.edu}
\affiliation{%
  \institution{Rutgers University}
  \city{New Brunswick, NJ}
  \country{USA}
  }

\renewcommand{\shortauthors}{Shim and Jhaver}

\begin{abstract}
Flagging mechanisms on social media platforms allow users to report inappropriate posts/accounts for review by content moderators.
These reports are pivotal to platforms' efforts toward regulating norm violations. This paper examines how platforms' design choices in implementing flagging mechanisms influence flaggers' perceptions of content moderation. We conducted a survey experiment asking US respondents (N=2,936) to flag inappropriate posts using one of 54 randomly assigned flagging implementations. After flagging, participants rated their fairness perceptions of the flag submission process along the dimensions of consistency, transparency, and voice (agency). We found that participants perceived greater transparency when flagging interfaces included community guidelines and greater voice when they incorporated a text box for open-ended feedback. Our qualitative analysis highlights user needs for improved accessibility, educational support for reporting, and protections against false flags. We offer design recommendations for building fairer flagging systems without exacerbating the cognitive burden of submitting flags.
\end{abstract}




\maketitle
% Paper structure

\section{Introduction}
%def: flagging
Flagging constitutes the most visible means through which social media users can directly influence platforms' content moderation decisions.
% It serves as an ex post moderation measure~\cite{grimmelmann2015virtues} that lets users report any post or account they deem inappropriate.
% On most platforms, flagging also requires submitters to specify how the reported item violates the site's content submission guidelines.
% 
% Flagging on social media platforms constitutes a crucial form of user engagement 
% with content moderation systems. As an ex-post moderation measure, it facilitates identifying and regulating inappropriate user-submitted content after it has been made visible to the public. The act of flagging lets users report their objections to a post to the platform, often allowing them to cite violations of the posting guidelines that the platform has in place \cite{crawford2016flag,zhang2023cleaning}. Such guidelines could address a variety of content-based harms, such as hate speech ~\cite{difranzo2018upstanding, ozanne2022shall,wilson2020hate}, self-harm content ~\cite{crawford2016flag,feuston2020conformity}, and misinformation ~\cite{seering2020reconsidering,westermann2022potential}. 
% 
Platforms maintain complex sociotechnical systems composed of automated mechanisms and human reviewers~\cite{gorwa2020algorithmic,roberts2016commercial,jhaver2019human} that continually review flagged posts and, when warranted, impose sanctions, such as removing, down-ranking, or shadow-banning the flagged posts ~\cite{goldman2021content,grimmelmann2015virtues}.
% By allowing users to participate directly in the content moderation process through flagging, platforms encourage them to share the responsibility of hosting appropriate content on their site \cite{ozanne2022shall}. 
Given the 
% large volume of posts submitted to popular platforms~\cite{huang2024opportunities} and the 
expenses associated with content review~\cite{li2022monetary,roberts2016commercial}, platforms rely on flags to identify and regulate norm violations and maintain the usability of their sites.

%current flagging mechanism in social media
Flags are now ubiquitously available as a feature across all social media platforms, but their description, placement, and implementation vary widely.
%placement
For example, Reddit offers a `report' button below each post, which triggers a one-step selection of the post's rule violation among 14 categories, including ``Hate,'' ``Sharing personal information,'' and ``Non-consensual intimate media.''
In contrast, each post on X has a `Report post' option embedded in a drop-down menu at the post's top-right corner; selecting this option prompts a multi-step information-gathering process about what is inappropriate in that post.
These design choices constrain and influence how flaggers can articulate their objections about inappropriate posts. Yet, little prior empirical research has examined how differences in flag implementations affect users' experiences of flag submissions. 
\marginnote{\review{2AC.3}}
We address this gap by focusing on social media users in their capacity as \textit{flag submitters (flaggers)}.
Prior research has shown that the regular contributions of flaggers are indispensable for most content moderation systems~\cite{crawford2016flag,gillespie2018custodians} --- their distributed labor collectively helps moderators address the daunting task of regulating 
vast and evolving content~\cite{grimmelmann2015virtues}.
Thus, it is vital that platforms design flags in ways that satisfy flaggers and encourage reporting of inappropriate posts.

We begin to fill this crucial research gap by examining how the design of four commonly deployed flagging components influences end-users.
These components include \textit{classification schemes}, which aid users in specifying their reason for the report through selecting a rule violation category; \textit{community guidelines}, which direct users and moderators alike in assessing a post's compliance with platform standards; a \textit{text box}, which gives users space to detail specifics of their objections; and \textit{moderator type}, which offers users some insight into who is administering the flag review process.
Current implementations of flagging mechanisms usually adopt one or more of these components ~\cite{crawford2016flag} to enhance the accuracy and perceived validity of reporting.
Therefore, we examine how various combinations of these design elements affect how users experience the flag submission process.

% Our study focuses on the design of the 'implementation of the flagging process.'
% This paper focuses on users' experience of the flagging mechanism. 
% %(e.g., rule violation categorization and information about the flag reviewer) that explicitly or implicitly offer flaggers information about the flag review process. 
% We evaluate how different flagging design elements shape users' experience reporting inappropriate posts.

%flagging elements + procedural fairness: research gap
Specifically, this paper explores users' perception of \textit{procedural fairness} in their interactions with flagging mechanisms. The procedural fairness perspective argues that the legitimacy of a decision-making system derives from public confidence in the fairness of the \textit{processes} through which decision-makers exercise their authority
% the procedures within a system operating in an unbiased and just manner, ensuring equitable treatment for all individuals involved 
~\cite{sunshine2003role, tyler2006psychological, vaccaro2020end}.
It has been widely argued that procedural fairness is a pivotal principle that should inform the design of social media governance systems \cite{schoenebeck2021drawing,suzor2019lawless,Katsaros2024justice,hartmann2020framework,jhaver2019did}. 
% Enhanced perceptions of procedural fairness foster users' support for moderation decisions ~\cite{valcke2020procedural,katsaros2022procedural,jhaver2019did} and correlate with norm-following behavior \cite{tyler2021social}, thereby strengthening trust between users and the platform ~\cite{jhaver2019did,jhaver2019does} and bolstering the legitimacy of governance mechanisms ~\cite{fan2020digital,gonccalves2021common}.
Prior research on enacting procedural fairness in content moderation systems has examined how moderation outcomes are delivered to and experienced by end-users~\cite{schoenebeck2021drawing,ma2023transparency,suzor2019we,jhaver2019did}. Building upon this rich literature, we concentrate on incorporating procedural fairness in flagging mechanisms from the perspective of flaggers.

%Research aims to fill the gap
We aim to (a) investigate how the commonly deployed design components of flagging mechanisms impact users' perceptions of procedural fairness and (b) derive empirically informed design recommendations for improving procedural fairness in flag submissions.
We examine three aspects of procedural fairness in this work --- \textbf{consistency}, \textbf{transparency}, and \textbf{voice} \cite{vaccaro2020end}.
Below, we discuss the treatment of each aspect in prior content moderation research and explain its relevance to flagging mechanisms to motivate our research inquiries.
\vspace{7px}

% The following defines each term and explores its context in flagging mechanisms before encapsulating it in a research question the paper investigates.

%Consistency
%\sandy{Does the first paragraph below address how moderators implement the decisions? If so, please clarify so readers can more easily differentiate first (moderator) and second (user) paragraphs below?}
\noindent
\textbf{Consistency} refers to the level of uniformity of the content moderation process, i.e., how content moderation policies are enacted to make moderation decisions regardless of the variable posting contexts \cite{vaccaro2020end,katsaros2022procedural}. 
Prior research shows that Black, Indigenous, and people of color (BIPOC), LBGTQIA+, disabled, plus-sized individuals, and sex workers experience more frequent and more severe moderation consequences than other users ~\cite{haimson2021disproportionate,ma2022m,are2023flagging,lyu2024blindtokers} despite being subjected to the same policies. Such inequitable moderation experiences have contributed to user perceptions that moderation systems discriminate against certain user groups based on their identity characteristics~\cite{thach2022visible}, further fueling the criticism that moderation has become a form of censorship \cite{ma2022m}.

Our study also addresses disproportionate moderation experiences, but we focus on a different stakeholder --- \textit{users who flag content}. 
% This will provide valuable insights into the underexplored perception of flaggers, who are key stakeholders in the use of flagging mechanisms. 
We posit that users' \textit{belief in} the platform's ability to enact consistent flag reviews, regardless of the flagger's identity, is crucial. Thus, prioritizing inter-flagger ``consistency'' would mitigate disproportionality in the review process and promote a perception of moderation fairness among users. Importantly, platforms should not just enact inter-flagger consistency in their flag review processes but also satisfy end-users that such consistency exists.
We examine factors that shape consistency perceptions by exploring the following research question:
%We focus here on one crucial posting context, \textit{inter-flagger consistency,} by which we mean that the flag review process remains consistent regardless of who flagged it (in terms of the flagger's identity attributes).
%\sandy{At this point, the reader is expecting you to discuss the other two attributes instead of launching into a long exploration of disproportionate decision making. Perhaps that belongs in a }
%Social media platforms have been criticized for enacting disproportionate content moderation decisions based on users' identities. Such effects have been studied from users' perspectives on the receiving end of moderation decisions \cite{haimson2021disproportionate,gorwa2020algorithmic,duguay2020queer,ma2022m}.

\vspace{5px}
    \textbf{RQ 1}: How do design choices in platforms' flagging components affect user perceptions of inter-flagger consistency?
\vspace{5px}

Note that while we focus on inter-flagger consistency, many other aspects of flagging consistency warrant attention but are beyond the scope of this study. 
For example, consistency regarding flag review outcomes for flagged users with different identity characteristics is important, but we chose not to examine it here because we are not concerned with flagged users in this study.
\vspace{7px}

%transparency
\noindent \textbf{Transparency} enables individuals to see how decisions are made and ensures social accountability~\cite{chan1998procedural,breton2007economics}.
%importance of transparency: +practical effectiveness
% Enacting transparency in decision-making positively influences users' perception of the decision-making system, even when the decisions are unfavorable~\cite{suzor2019we,ma2023transparency,thach2022visible,brockner2013and}.
In the context of content moderation, transparency has been conceptualized as the communicative steps that platforms may take to better explain the deeper complexities of moderation processes, policies, and outcomes to the many stakeholders implicated in them~\cite{suzor2019we}.
Content moderation scholars have raised concerns about the lack of transparency in moderation systems, citing insufficient details provided to sanctioned users about what rules they have violated and whether their content was removed by an algorithm or a human moderator~\cite{diaz2021double}.
% Studying the design of personal moderation tools that platforms offer, ~\citet{jhaver2023personalizing} highlighted that these tools do not effectively communicate the meanings of their interface elements and offer inadequate information about the inner logic and inputs to algorithms that drive them.
% and procedure information such as how the flag review process is conducted and by whom~\cite{zhang2023cleaning}.
%Researchers: Enhance transparency
Researchers have proposed various ways to enhance transparency in content moderation systems, including providing more details about decision-making processes~\cite{katsaros2022procedural,jhaver2019did,jhaver2019does}, describing 
how rule violations are detected ~\cite{suzor2019we} and clarifying how sanctions are determined for rule-violating posts \cite{chandrasekharan2019crossmod}.

Given that flagging is a key mechanism that drives content moderation systems, ensuring transparency about it is crucial for maintaining overall system fairness, especially since it can serve as a basis for explaining flagging outcomes.
Our study focuses on enacting transparency during the flag submission process, specifically the level of detail available or implied \textit{throughout the flagging process} about flag reviews. We conceptualize ``transparency'' as platforms using flagging components to inform flaggers how flag reviews are processed. We pursue the following research question:

\vspace{5px}
    \textbf{RQ 2}: How do design choices in platforms' flagging components affect user perceptions of transparency in the flagging process?
\vspace{7px}

\noindent
\textbf{Voice} refers to the extent to which users' input is accommodated in the decision-making process \cite{katsaros2022procedural,vaccaro2020end}. It has been identified as crucial to shaping individuals' fairness perceptions~\cite{drahos2017regulatory}. 
% When the decision-making process invites users' input, they are more likely to feel that they have been treated fairly---even when the outcome is unfavorable to them~\cite{thibaut1975procedural}---and they perceive greater satisfaction with the process outcome~\cite{van1998we}. 
In the context of content moderation research, ~\citet{ma2022m} found that YouTube creators associate the fairness of their moderation experiences with having their voice involved in the decision-making process.
Despite its crucial role, limited research examines the perception of voice/agency that users experience when interacting with flagging mechanisms. We conceptualize ``voice'' in the context of this paper as the users' perceived degree of involvement in the flagging process, especially regarding how well they can express their objections to an inappropriate post. We thus pursue the following research question:

\vspace{5px}
    \textbf{RQ 3}: How do design choices in platforms' flagging components affect user perceptions of voice in the flagging process?
\vspace{7px}

\noindent
Table \ref{table:attributes} presents these three attributes of procedural fairness, describes how previous literature has addressed each in the context of content moderation, and shows how our study conceptualizes them within the flagging mechanism.

 \par\null\par
\begin{table}[h] 
    \centering 
    \caption{Procedural Fairness Attributes. For each attribute, we present its definition from prior content moderation research and our conceptualization of that attribute in the context of flagging design.} 
    \label{table:attributes}
    {\fontsize{8}{10}\selectfont 
    \renewcommand{\arraystretch}{1.2}
    \setlength{\tabcolsep}{4pt}
    \begin{tabular}{m{1.6cm} m{5.5cm} m{6.5cm}}
\Xhline{1.5\arrayrulewidth}
       \multicolumn{1}{c}{\textbf{Attributes}} & \multicolumn{1}{c}{\textbf{Definition from Previous Research}} &  \multicolumn{1}{c}{\textbf{Our Conceptualization for Flagging}} \\
       \hline
      Consistency
        & Platform enforces content moderation policies uniformly, regardless of the specific post context~\cite{katsaros2022procedural,haimson2021disproportionate,ma2022m,are2023flagging,lyu2024blindtokers}.& Flagging mechanism applies the same rules and standards uniformly, in accordance with the platform’s values and norms, regardless of flaggers' identity characteristics.\\
       \hline
       Transparency 
        & Platform provides users with information about moderation process and reasoning behind decision-making outcomes~\cite{jhaver2019did,ma2023transparency,thach2022visible,suzor2019we,katsaros2022procedural}.
        & Flagging mechanism presents relevant information about the flagging process and flag reviews.\\
       \hline
       Voice 
        & Platform moderation process adopts measures to integrate users' opinion into the decision-making process ~\cite{vaccaro2020end,katsaros2022procedural,ma2022m}.& Flagging mechanism allows users to thoroughly express their objections to the flagged posts.\\
\Xhline{1.5\arrayrulewidth}
    \end{tabular}
    }
\end{table}

Besides the four flagging components indicated above, we aimed to more deeply understand the changes that end-users seek in the broader design and policy choices associated with flag implementations. Therefore, we ask:

\vspace{5px}
    \textbf{RQ 4}: How should flagging processes change to enhance users' fairness perceptions?
\vspace{7px}

To answer the preceding research questions, we conducted a controlled online experiment to evaluate the impact of different flagging implementations on users' fairness perceptions. We recruited 2,936 participants using Lucid Theorem\footnote{\url{https://lucidtheorem.com}} and randomly assigned each to experience one of 54 flagging scenarios simulated on Qualtrics. These scenarios were constructed by combining different implementations (or absence) of the four flagging components --- \textit{rule violation classification scheme, guidelines, a text box, and information about the moderator} (detailed in Sec. \ref{sec:components}, where we motivate our specific research hypotheses). 
After reporting a post in such simulated flagging scenarios, participants assessed the procedural fairness of their flagging experience along the dimensions of consistency, transparency, and voice. 
% Next, they answered open-ended questions that solicited their suggestions on how flagging systems could improve procedural fairness.

% Following data collection, we conducted a mixed-methods analysis to assess the relationships between the various design choices of flagging components and users' perceptions of fairness.
% Additionally, we evaluated how these different components influence flaggers' cognitive burden of submitting flags and their propensity to flag again.

Our statistical analyses (RQ 1, 2, 3) show that displaying community guidelines during the flag submission process raises users' perceptions of transparency. Additionally, we found that offering a text box where users can input their specific objections about the flagged post enhances their perception of having a voice in that  process. 
Our qualitative analysis of respondents' open-ended suggestions (RQ 4) to enhance flags' procedural fairness highlights users' interest in receiving information regarding how the flag review process works and whether the flag reviewers are politically neutral.
% , and how flaggers can offer platforms feedback about their flagging experience.  
Participants wanted flagging mechanisms to support greater expressivity, provide timely notifications of flag outcomes, and prevent the abuse of flags by bad actors wishing to take down otherwise appropriate content.

% e.g., by allowing flaggers to highlight sections of flagged posts that violate platform guidelines or offering alternative channels to submit flags, such as chat, emails, or phone calls. Finally, participants wanted to receive timely notifications of their flag outcomes, and they expected platforms to prevent the abuse of flags by bad actors wishing to take down otherwise appropriate content.

Our findings underscore the need for flagging systems to accommodate users' nuanced objections about why flagged posts deserve regulation. To achieve this, we recommend that flag implementations (1) integrate a text box for users to input their detailed perspectives to the platform, (2) enhance flags' vocabulary of complaints~\cite{crawford2016flag} by letting users highlight norm-violating portions of flagged posts and rate the violation severity, and (3) incorporate information and visualization systems that let flaggers track the review status of submitted flags. Additionally, platforms should (4) address flaggers' concerns regarding biased decision-making and false flagging by offering comprehensive information about posting guidelines, reviewers, and the review process. 
% Our empirical insights suggest that platforms could (5) reduce the burden of flagging and flag reviews by engaging in educational efforts to reform the authors of flagged posts, strengthening ex ante moderation measures~\cite{grimmelmann2015virtues}, and positively reinforcing constructive contributions.
% Further, providing mechanisms accessibility by diversifying reporting channels --- such as phone calls, emails, and chat --- and incorporating diverse users' perspectives through forums will enhance perceived fairness on the platform.

% Finally, we demonstrate that users evaluate the fairness of flags not in isolation but by contextualizing flagging within a broader set of platform-offered features that address harm, such as muting and blocking. Thus, platforms must clarify the distinctions between these tools regarding their affordances and required user inputs.
\section{Related Work}
\subsection{Content Moderation}
%Definition of content moderation
\textit{Content moderation} refers to the regulatory systems digital platforms deploy to promote effective communication among users while deterring exploitation of community attention ~\cite{grimmelmann2015virtues}. It ensures norm compliance by promoting exemplary posts, restricting the visibility of inappropriate posts, and educating users about appropriate conduct ~\cite{das2021jol,gillespie2018custodians,jhaver2019does}. Content moderation involves a range of \textit{ex ante} and \textit{ex post} measures---referring to actions taken before or after the content is published, respectively~\cite{gorwa2020algorithmic,grimmelmann2015virtues}---to address content-based harms~\cite{jhaver2023personalizing,scheuerman2021framework} and improve the quality of available posts online. 
Over the past few years, all large-scale social media platforms have developed ad hoc content moderation infrastructures to enact these measures. Such investments have often been made in response to widespread critiques of moderation deficiencies from various stakeholders --- lawmakers, researchers, news media, and the public at large \cite{gillespie2018custodians}.

% What addresses content moderation? (Meaning)
Our research addresses ex post moderation, which involves actions such as content removal, downranking, demonetizing, attaching warning labels to posts, suspending accounts, and restricting the visibility of posts or accounts to specific users ~\cite{goldman2021content,jiang2023trade}. Platforms implement such actions to reduce the negative influence of content such as hate speech ~\cite{difranzo2018upstanding,ozanne2022shall,wilson2020hate}, bullying ~\cite{das2021jol,gilbert2020run}, self-harm ~\cite{feuston2020conformity,crawford2016flag}, 
violence~\cite{tait2008pornographies},
and misinformation ~\cite{westermann2022potential,seering2020reconsidering} and to encourage healthy communication among users \cite{langvardt2017regulating}. 
We situate flagging as a crucial mechanism within the ex post moderation stage because it identifies posts that may require moderation intervention.

%How flagging is located in the overall content moderation environment
While moderation decisions are made on the platform side, flagging serves as a mechanism for users to participate in the governance process \cite{chipidza2022effectiveness}. 
\textit{Appeal mechanisms} ~\cite{vaccaro2020end, vaccaro2021contestability}, which let users express their dissatisfaction with content moderation decisions and request that they be reversed, also allow users to communicate with platform operators.
However, this mechanism only concerns moderated users and their sanctioned posts, leaving those choosing to flag posts with few or no options to appeal platforms' decisions.

Another bottom-up moderation mechanism that many platforms offer is \textit{personal moderation}~\cite{jhaver2023personalizing}, which lets users limit the visibility of undesirable posts on their feeds. 
Personal moderation includes actions like muting or blocking an account and customizing the sensitivity threshold for content on one's feed according to one's preferences \cite{jhaver2023personalizing,jhaver2023users}. It exclusively affects one person's feed without influencing others’ content consumption \cite{goldman2021content}. 
% For instance, users prioritizing free speech can opt for lenient moderation strategies by increasing the visibility levels of potentially offensive posts in their feeds \cite{jhaver2023users}. By offering the flexibility to adjust their own moderation threshold, personal moderation can balance platforms' efforts to uphold users' free speech values while safeguarding them from harm \cite{chipidza2022effectiveness}.
% 
Prior research examined the design choices involved in building personal moderation tools and recommended improvements in defining interface elements, incorporating cultural context, offering greater granularity, and leveraging example content \cite{jhaver2023personalizing}.
We add to this research by evaluating the design space of the flagging mechanism, another crucial bottom-up moderation measure that platforms offer.

\subsection{Flagging as a Content Moderation Tool}
%Flagging is an efficient tool for content moderation that current social media platforms frequently adopt.
As platforms grow, they face the challenges of scale when enacting content review of all submitted posts~\cite{gillespie2018custodians}.
Currently, flagging is employed by many social media platforms to achieve greater efficiency in their content review processes ~\cite{crawford2016flag,seering2020reconsidering,helberger2018governing,Thomas2022it's}. 
For instance, platforms can prioritize the review of posts flagged for containing inappropriate content such as misinformation or hate speech ~\cite{crawford2016flag,kou2021flag} and ensure immediate user safety through sanctions such as removing the flagged posts or limiting their visibility ~\cite{myers2018censored,zeng2022content}. 
Though we focus on social media platforms in this research, other digital platforms, such as online marketplaces (e.g., Amazon.com), financial apps (e.g., Venmo), and sharing economy services (e.g., Uber, Airbnb) also deploy flagging as a user-facing tool.

%Groups of users related to the flagging mechanism -- specifically, we focus on the flagger's flagging behavior and experiences.
When we consider the impact of flagging mechanisms on end-users, three categories of users stand out: 
(1) \textit{flagged users} whose posts are (either justifiably or unjustifiably) reported by another account for infringing on platform policies, 
(2) \textit{flaggers} who request content review of selected posts or accounts, and 
(3) \textit{silent bystanders} who witness policy violations but do not flag them. 
Our study focuses on flaggers, i.e., users who engage with the flagging mechanism to report a post and initiate its review.
While some flaggers might report content that violates platform guidelines (which is how platforms intend users to employ flags), others may use flags to express their social or political objections to the reported post~\cite{suzor2019lawless, kou2021flag,are2023flagging}, coordinate with others to collectively get a post sanctioned ~\cite{suzor2019we,seering2020reconsidering}, or pursue a personal vendetta against a poster ~\cite{myers2018censored}.
In this article, we investigate flaggers' experiences using the flagging mechanism to report posts that violate platform guidelines.\footnote{Note that our use of the term \textit{flaggers} refers to general end-users using the platform's flagging mechanism; we are not concerned with \textit{trusted flaggers}, who, as \citet{wilson2020hate} describe, are selected among end-users by platforms and contracted to perform moderation tasks because they possess the relevant linguistic facility or advanced knowledge of platform policies.}

\citet{zhang2023cleaning} identified three temporally distinct stages associated with the use of flagging mechanisms --- before, during, and after flagging --- and highlighted the various user needs and interface affordances available in each stage. Our research focuses on the ``during flagging'' stage, and examines the design opportunities within that stage to improve users' flagging experiences.

%How each flagging mechanism component (except for the moderator information) supports reporting a norm violating post 
Platforms would prefer that users flag only content that violates their existing guidelines to optimize the labor involved in content review.
As such, platforms seek to concentrate flaggers' attention on reporting authentic violations of platform guidelines.
They do this by designing flagging mechanisms in ways that emphasize their criteria for reviewing flagged content.  
For instance, flagging interfaces often prompt users to select from predefined categories of rule violations or present platform guidelines that may direct flaggers' attention to 
% platform norms and serve as an educational tool, making users aware of 
what platforms consider non-normative behaviors~\cite{chandrasekharan2018norms,matias2019preventing}.
On some platforms, flagging requires users to articulate their objections to the flagged post in a text box. This encourages users to reflect on the purpose of their flagging and elaborate on their perspectives regarding how the flagged post violates platform norms. 

\begin{figure*}[ht!]
    \centering  % Define a small rule width to ensure tight fitting
    \setlength{\fboxrule}{0.5pt} % Remove any additional space
    \setlength{\fboxsep}{0pt} 
        \framebox{\includegraphics[width=.3\textwidth]{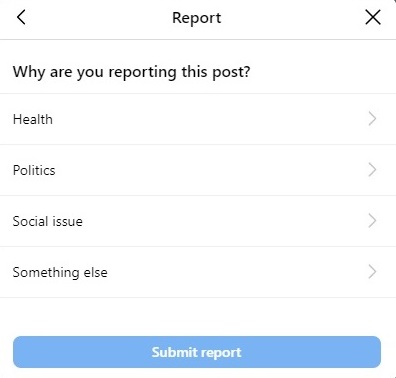}}
            \hspace{0.03\textwidth}
        \framebox{\includegraphics[width=.3\textwidth]{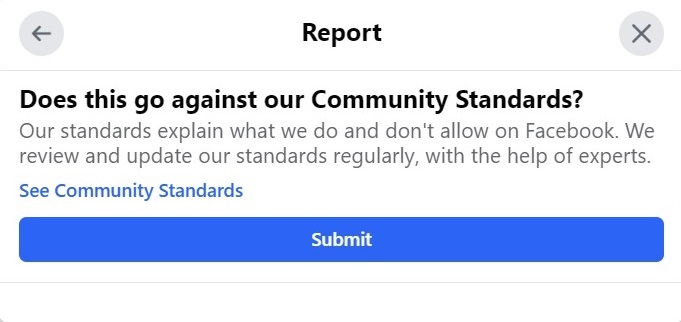}}
            \hspace{0.03\textwidth}
        \framebox{\includegraphics[width=.3\textwidth]{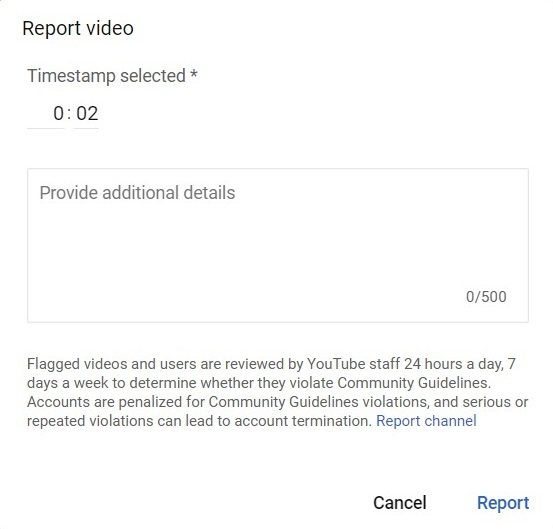}}
            \caption{When users initiate the flagging process on social media platforms, Instagram mandates that they specify the rule violation of the post (left), Facebook offers a link to its community guidelines (middle), and YouTube provides a text box to elaborate on the specific reasons for flagging content (right).}
                \label{fig:socialmedia}
        \end{figure*}

By incorporating these diverse elements, current social media sites demonstrate varied approaches to structuring their flagging mechanisms. For example, Instagram's flagging system lets users categorize rule violations and specify which rules the reported post has breached (Figure \ref{fig:socialmedia}).
Facebook offers a similar categorization interface and additionally prompts users to review the site's community standards when selecting a rule violation category (Figure \ref{fig:socialmedia}). YouTube, on the other hand, provides a text box, enabling users to articulate their reasons for flagging in their own words (Figure \ref{fig:socialmedia}). 

While these distinct components constitute the reporting mechanisms across different social media platforms, scant research attention has been paid to how each component influences flaggers' experiences. Our research integrates these diverse manifestations of commonly deployed flagging components into an experimental framework and investigates how different design choices for each component impact user perceptions.

\subsection{Procedural Fairness in Content Moderation}
%Def of fairness
Fairness is a multifaceted concept that includes procedural fairness \cite{vaccaro2020end,lee2019procedural}, outcome fairness \cite{ma2023transparency,vaccaro2020end}, and restorative fairness \cite{kou2021punishment,schoenebeck2021drawing}, emphasizing process, outcome, and feedback, respectively.
Scholars focused on incorporating \textit{fairness} in decision-making systems highlight enacting unbiased processes and outcomes, avoiding unjust situations \cite{shin2019role}, and treating every individual with respect, dignity \cite{katsaros2024online}, and equality \cite{ma2022m}. 
Content moderation scholars have argued that achieving fairness in moderation processes entails administering equitable treatment of all users~\cite{jhaver2019did} through clear and consistent criteria for content review that integrate the needs of marginalized groups, such as racial minorities and LGBTQ+ individuals~\cite{haimson2021disproportionate,ma2022m}. 

%Importance of fairness to the content moderation system:
Prior research has shown that enhancing fairness in social media platforms' decision-making systems can positively influence user attitudes \cite{jhaver2019did,ma2022m,ma2023litreview} and emotions, thereby improving the overall user experience. 
For instance, a higher perception of fairness regarding the moderation system enhances end-users' trust in it, which in turn lends greater legitimacy to both the decision-making process and its outcomes \cite{pan2022comparing}. Even in cases where moderation systems deliver unfavorable outcomes, users' perception of procedural fairness improves their satisfaction with the outcome \cite{vaccaro2020end}.
Further, moderation actions that induce fairness also encourage moderated users to remain active within the community~\cite{jhaver2019did}.

%Previous research on fairness in the content moderation system:
Acknowledging the preceding benefits of enacting fairness in content moderation systems, researchers have explored designing systems that emphasize fairness. Previous studies highlighted strategies such as adopting unbiased criteria for content reviews and incorporating transparency in platform protocols to enhance fairness \cite{schoenebeck2021drawing,bradford2019report}. 
For example, \citet{pan2022comparing} found that an unbiased moderation process could be achieved by involving an expert panel or a jury of users in the review process, which may increase users' perceptions of fairness. Several researchers also identified communication strategies to enhance moderation fairness, such as notifying users about content regulation via messages or emails \cite{ma2023transparency} and providing comprehensive information about content removal reasons and moderator type \cite{jhaver2019does}.

%A gap to spark our research: exploring the perception of end-users (flaggers) procedural fairness in the social media platform flagging system:
Though this literature provides valuable insights into what influences users' perceptions of fairness in platform-enacted moderation decisions, a gap remains in our understanding of what shapes fairness perceptions of user-driven content moderation mechanisms, especially flagging systems. Therefore, we explore how flaggers' perceptions of procedural fairness may vary based on the design components of the flagging mechanism.

%How to define procedural fairness in our study:
We focus on three key attributes to evaluate procedural fairness — \textit{consistency}, \textit{transparency}, and \textit{voice} — drawing from prior research by ~\citet{lee2019procedural} and \citet{vaccaro2020end}. Lee et al. introduced the fairness framework, including transparency and voice as key components, for assessing algorithmic fairness ~\cite{lee2019procedural}. 
Within this framework, transparency consists of standards clarity, standards validity, information representativeness, and outcome explanations, whereas voice comprises users having more control over the decision outcomes and the processes that lead to those outcomes. 
Vaccaro et al. developed a scale to measure fairness by drawing from both consequentialist and non-consequentialist approaches to justice, which differ in whether or not the rightness of an action should be judged based on its consequences alone ~\cite{vaccaro2020end}. They defined procedural justice as maintaining an equitable distribution of rewards or punishments and tied it to transparency, which ensures that the decision subjects understand the process, and voice, which allows individuals to express their opinions and arguments.
% process equity, which consists of transparency --- ensuring that the decision-making process is understood--- and voice, which allows individuals to express their opinions within the system. 

Inspired by these studies, we conceptualize procedural fairness in flagging systems as adopting a clear, equitable, and consistent standard for content review, enacting transparency in delivering procedural information, and ensuring robust user participation through input mechanisms while flagging. This non-consequentialist approach to justice reflects our interest in understanding how users perceive the fairness of flag submission procedures regardless of the final decision outcomes. As such, we refrain from including aspects related to outcome explanation and outcome control in our assessment of fairness perceptions. 

To summarize, our study investigates flaggers' perceptions of procedural fairness --- specifically, consistency, transparency, and voice — within flagging mechanisms designed with diverse components. We situate this work in conversation with other Human-Computer Interaction (HCI) studies ~\cite{kiesler2012regulating,jhaver2023personalizing,seering2017shaping,pan2022comparing,atreja2023remove,ma2023transparency} that experiment with interface and policy designs to improve platform governance outcomes.
\section{Components of Flagging Design}\label{sec:components}
Our study examines four components of flagging mechanisms commonly adopted by social media platforms. Drawing on the characteristics of each component and insights from previous literature, we formulate hypotheses about how these factors contribute to different fairness attributes in this section. We also describe how we operationalized each component through our survey questions.

\begin{figure*}[ht!]
    \centering
    \fbox{\includegraphics[width=.3\textwidth]{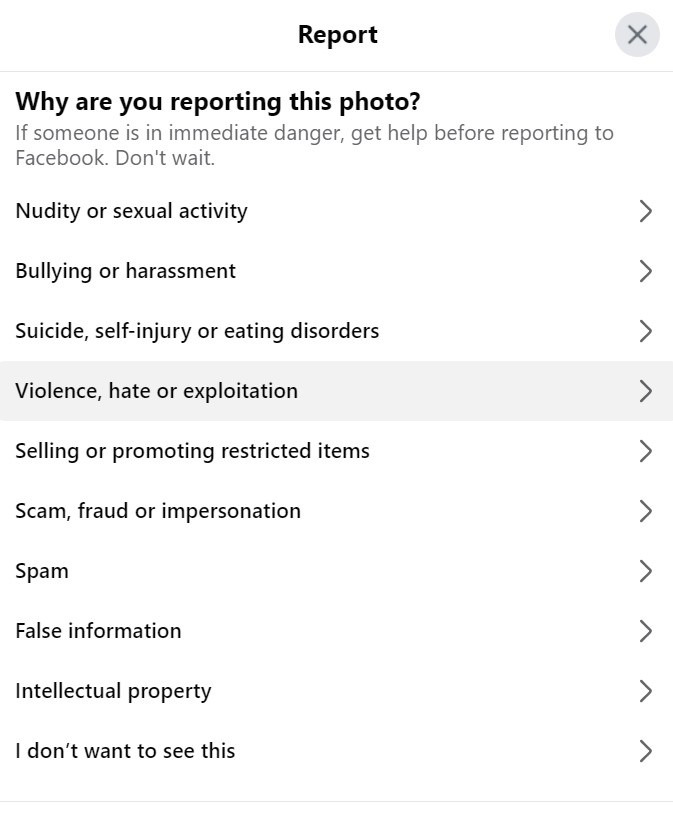}}
    \hspace{0.01\textwidth}
    \fbox{\includegraphics[width=.3\textwidth]{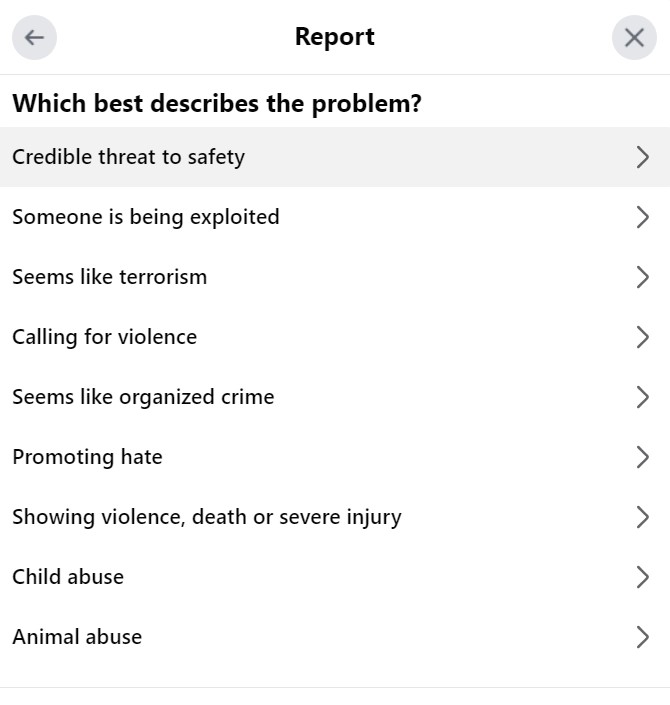}}
    \hspace{0.01\textwidth}
    \fbox{\includegraphics[width=.3\textwidth]{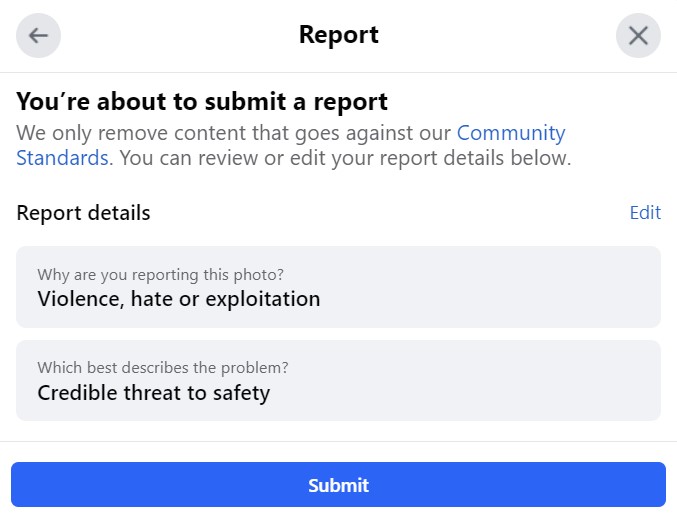}}
    \caption{This example illustrates Facebook's flag classification scheme after users initiate the flag submission process. In this example, the user categorizes her reason for flagging as `Violence, hate or exploitation' from the main menu; this selection displays a submenu from which the user selects the subcategory `Credible threat to safety.' Finally, the system displays these selections to the user, along with the submit button that prompts completing the report.}
    \label{fig:example_classification_scheme}
\end{figure*}

\subsection{Flag Classification Levels}
\label{sec:flag-classification-levels}
As users engage in the flagging process, they may encounter a classification scheme with primary menus and submenus (see Figure \ref{fig:example_classification_scheme}) that specify the range of rule violations the flagged post may have, such as hate speech and spam. Making selections from predefined rule violations lets users categorize their concerns with the flagged post~\cite{crawford2016flag}.
From platforms' perspective, imposing this precisely designed classification scheme on users is crucial to triage flagged posts, validate whether the selected rule violation occurred, and highlight that the content of flagged posts drives flag review decisions (instead of other factors, such as flaggers' identity).
However, it is not yet clear how the design of classification schemes affects flaggers' perceptions of fairness.

\textit{This study explores how adjusting the granularity of rule violation selection within the classification scheme impacts users' perceptions of fairness in the content moderation process.}
A more granular scheme would let users select not only one of the few broad rule categories (e.g., hate speech, misinformation) but also the precise subcategory within each category (e.g., race-based hate speech, health misinformation).

% We expect that requiring users to specify the violation within flagged content with higher granularity will prompt them to pay closer attention to the categories themselves and classify the rule violation categories cautiously. 
Since the rule violation scheme is derived from content moderation guidelines, 
%  established by the platform, 
% users may perceive that the moderation process relies on clear and precise rules when the classification scheme accommodates more specific rule violation selections. 
% Further, incorporating more comprehensive categories into the flagging mechanism may influence users to perceive that the moderation criteria are consistent rather than arbitrary or situational. Therefore, 
we expect that if the flagging mechanism offers users more detailed categories for flagging, they will perceive that the system prioritizes its formalized moderation criteria rather than other factors, such as flagger identity. Therefore, we hypothesize:

\begin{quote}
    \emph{H1-1: More granular designs of flag classification schemes in a flagging mechanism will increase users' perceptions of consistency in the flag review process.}\footnote{With this hypothesis, we also mean to include the assertion that providing even the least granular classification scheme will increase users' consistency perceptions when compared to not providing \textit{any} classification scheme. We have similarly merged other assertions throughout sec. \ref{sec:flag-classification-levels} and \ref{sec:posting-guidelines} to achieve conciseness.}    
\end{quote}

% Integrating the rule violation classification into the flagging mechanism serves a dual purpose: (a) it clarifies the reporting rationale from the user's perspective and (b) it reminds users of the platform's declared categories of rule violations. Offering users a more detailed classification of rule violations deepens their comprehension of the types of content subject to moderation. 
Transparency is linked to the amount of information about the process ~\cite{lee2019procedural,rader2018explanations}. Thus, furnishing a more detailed rule violation classification scheme may enhance the transparency of the flag review process. Therefore, we hypothesize:
\begin{quote}
    \emph{H1-2: More granular designs of flag classification schemes in a flagging mechanism will increase users' perceptions of transparency in the flag review process.}
\end{quote}

% Providing a more detailed rule violation classification scheme means that users can express their opinions more precisely on the platform. In other words, allowing users to select both a primary category and a subcategory sequentially ensures a more precise conveyance of their intentions during the review process compared to having no or only a single-level classification scheme option. 
Given that voice is closely tied to providing opportunities for users' perspectives to be expressed within the process \cite{katsaros2024online,katsaros2022procedural}, incorporating more detailed levels of rule violation classification scheme may lead users to perceive that the platform has a genuine interest in understanding their concerns, which may result in an elevated perception of having a voice in the process. We therefore hypothesize:
\begin{quote}
    \emph{H1-3: More granular designs of flag classification schemes in a flagging mechanism will increase users' perceptions of voice in the flag review process.}
\end{quote}

\begin{table}
\fontsize{8}{8}\selectfont
\vspace{-\baselineskip} % Reduce vertical space
\begin{center}
\renewcommand{\arraystretch}{1.2} % add vertical spacing
\caption{A taxonomy of rule violation categories developed by referencing flag interfaces across multiple platforms. We used this taxonomy to operationalize flag classification choices in our survey. Each category below also included an additional submenu option of `Something else.'}
\label{table:classification}
\begin{tabular}{l|l}
\Xhline{1.5\arrayrulewidth}
\multicolumn{1}{c|}{\textbf{Primary category}} & \multicolumn{1}{c}{\textbf{Subdivision category}} \\
\hline
    \multirow{4}{*}{Child safety} 
                    & Child exploitation \\
                    & Child neglect \\
                    & Child nudity \\
                    & Inappropriate interaction with children \\
\hline
    \multirow{3}{*}{False news or misinformation } 
                    & Health  \\
                    & Politics \\
                    & Social issue \\
\hline
    \multirow{2}{*}{Harassment or bullying} 
                    & Me \\
                    & Someone I know \\
\hline
    \multirow{7}{*}{Hate speech} 
                    & Race or ethnicity \\
                    & National origin \\
                    & Religious affiliation \\
                    & Social caste \\
                    & Sexual orientation \\
                    & Sex or gender identity \\
                    & Disability or disease \\
\hline
    \multirow{3}{*}{Impersonation} 
                    & High profile impersonation \\
                    & Private impersonation \\
                    & Unauthentic behavior \\
\hline
    \multirow{4}{*}{Unauthorized sale} 
                    & Drugs \\
                    & Weapons \\
                    & Endangered animals \\
                    & Other animals \\
\hline
    \multirow{2}{*}{Self-injury} 
                    & Suicide \\
                    & Eating disorder \\
\hline
    \multirow{3}{*}{Sexual activity}
                    & Nudity or pornography \\
                    & Sexual exploitation or solicitation \\
                    & Sharing private images\\
\hline
    \multirow{3}{*}{Violence or incitement} 
                    & Animal abuse \\
                    & Riot or terrorism \\
                    & Death or severe injury \\
\Xhline{1.5\arrayrulewidth}

\end{tabular}
% \caption*{Each category includes an option for `Something else.'\\}
\end{center}
\end{table}

\vspace{8pt}

\noindent \textbf{Operationalization} \\
In our survey, we organized rule violations into nine distinct primary categories, each with an additional submenu, as shown in Table \ref{table:classification}. This classification scheme was developed by referencing rule violation categories in flagging mechanisms across major social media platforms, including Facebook, Reddit, X, and YouTube.
By synthesizing rule violation categories across multiple platforms, we aimed to 
allow users to explore flagging mechanisms without focusing on the norms for any particular platform.

When designing rule violation classification schemes, we constructed three types based on the number of steps involved in rule violation category selection. Each participant interacted with one of these three scheme types:

\begin{enumerate}
    \item Users are not required to select a flag category. 
    \item Users are required to select only a primary flag category.
    \item Users are required to select a primary flag category and a corresponding subcategory.
\end{enumerate}

\subsection{Posting Guidelines}
\label{sec:posting-guidelines}
% Today, all major social media platforms publicly list their guidelines for appropriate conduct or terms of use on their sites.
Posting guidelines are crucial for designing accountable content moderation systems because they serve as the official criteria for the platform's moderation decisions \cite{gillespie2018custodians, zolides2021gender, maddox2020guidelines}.
Prior research also shows that drawing users' attention to posting guidelines contributes to pro-social community outcomes. 
For example, \citet{matias2019preventing} demonstrated through an online experiment on Reddit that announcing posting guidelines in public discussions increases the likelihood of compliance with them and encourages newcomer participation.
% \citet{bhandari2021you} show that users who look at the posting guidelines are more likely to flag harassment posts. 
Further, when guidelines are easily noticeable, users are more likely to accept the platform's moderation decisions ~\cite{bradford2019report, kiesler2012regulating} and perceive them as fair ~\cite{jhaver2019did}. 

While this literature has established the benefits of how users' attention to posting guidelines enhances their fairness perception of moderation outcomes, little attention has been given to integrating these guidelines into flagging mechanisms  and exploring its impact.
% Further, \citet{zhang2023cleaning} noted deficiencies in the clarity and details of how rule violations are defined in the current social media reporting systems. 
% Therefore, we posit that enhancing guideline visibility and details may improve user perceptions about the flagging process. 
We investigate how different granularity levels of posting guidelines influence users' perceptions, specifically regarding the fairness of the flagging process. 

% Guidelines function as the rubric for content moderation, lending greater legitimacy to the review process \cite{crawford2016flag}.
Offering clear posting guidelines in the flagging mechanism may lead users to perceive that the review process adheres to a predetermined rubric and is largely influenced by the specified guidelines and not other external factors. 
% Therefore, the visibility of guidelines in this process can augment users' perception of consistent review procedures grounded in established principles. 
Additionally, we posit that more detailed guidelines, presented with specific examples of rule violations, could strengthen the perception of a robust rule-based review process. 
Therefore, we hypothesize:
\begin{quote}
    \emph{H2-1: Integrating more granular designs of posting guidelines into a flagging mechanism will increase users' perceptions of consistency in the flag review process.}
\end{quote}

Integrating posting guidelines in the flagging mechanism is a way of disclosing content review criteria by illustrating possible flaggable posts, thereby enhancing users' understanding of the content review process \cite{ozanne2022shall}. 
% Thus, more detailed guidelines, including rule-violation example posts, will give users more information on the review criteria, thereby promoting greater transparency. 
We therefore hypothesize that:

\begin{quote}
    \emph{H2-2: Integrating more granular designs of posting guidelines into a flagging mechanism will increase users' perceptions of transparency in the flag review process.}
\end{quote}

Showing guidelines in the flagging procedure may impose a sense of the need for strict adherence to predefined rules.  
%in assessing the necessity of flagging the content compared to not providing the guidelines. 
Since users sometimes flag in ways that may not align with platforms' notion of flaggability~\cite{kou2021flag}, 
% e.g., to express their objections to others' opinions \cite{gillespie2018custodians}, 
requiring them to follow the guidelines may lead to perceptions that they cannot express their opinions freely.
% discourage users from expressing their opinions freely. 
% flagging objections that have not yet been formalized as rule violations in the guidelines. Therefore, when a flagging mechanism prompts users to assess flaggability by providing clear posting guidelines, they may perceive a lack of opportunities to express their opinions freely. 
We thus hypothesize:
%\sandy{It seems like the hypothesis below does not reflect the sentence above, which seems to suggest users would have less freedom/voice/agency. Please clarify the hypothesis statement to explain what you mean wrt users v moderators.}
\begin{quote}
    \emph{H2-3: Integrating more granular designs of posting guidelines into a flagging mechanism will decrease users' perception of voice in the flag review process.}
\end{quote}

\vspace{8pt}
\noindent \textbf{Operationalization} \\
We developed a set of platform guidelines that explicitly describe the types of content prohibited on the platform. Table \ref{table:guidelines} shows this list and their corresponding examples. 
%We use the term `posting guidelines' instead of `community guidelines' as we do not assume the flagging circumstances of a particular target community during the survey.
We developed these guidelines by synthesizing Facebook, Reddit, X, and YouTube guidelines. 
We retrieved the guidelines from webpages of these sites under titles such as `community standards,' `content policy,' `rules and policy,' or `community guidelines \& policies.' By identifying common guideline themes across platforms, we compiled a core list of community guidelines and incorporated specific language from different platforms to clarify them. 
% For example, all platforms address violence-related activities, categorizing them as \textit{violence and incitement} (Facebook), \textit{threats of violence} and \textit{incite violence} (Reddit), \textit{glorification of violence}, \textit{violent and hateful entities}, and \textit{violent speech} (X), and \textit{violent criminal organizations} and \textit{violent or graphic content} (YouTube). We consolidated these various forms of violent activities into one overarching category of `violence,' recognizing it as a fundamental aspect of platform guidelines.

To construct detailed guidelines, we gathered illustrative examples for each rule violation category through two means: (1) using examples officially offered in platforms' description of community guidelines and (2) conducting a content search on platforms with keywords borrowed from guidelines. 

% For many of these guidelines, we used examples officially offered by the platforms we considered (Facebook, Reddit, X, and YouTube) when such examples were available. 
% For instance, YouTube provides several examples of hate speech, from which we selected and adjusted two for our guidelines.
% We could not find official examples for some rule violation categories in our list.
% In such cases, we conducted a content search on the platforms with keywords borrowed from platform guidelines to find explicit examples illustrating prohibited activities. For instance, X addresses destructive behavior with keywords such as physical injuries, cutting, eating disorders, bulimia, and anorexia, which we adopted to search for explicit examples to include in our guidelines.

\begin{table}[h]
\fontsize{8}{8}\selectfont
\renewcommand\arraystretch{1.2}
\begin{center}
\caption{Posting guidelines developed by synthesizing guidelines across Facebook, Reddit, X, and YouTube and used in our survey interface. A prompt of `We do not allow content that' appeared below these guidelines were shown.}
\label{table:guidelines}
\begin{tabular}{m{5cm}|m{9cm}} 
\Xhline{1.5\arrayrulewidth} 
\multicolumn{1}{c|}{\textbf{Posting Guidelines}} & \multicolumn{1}{c}{\textbf{Examples of Posts Violating the Guideline}} \\
\hline
    \multirow{2}{*}{\parbox{5cm}{Depicts or encourages harm against children, including maltreatment and exploitation.}} 
                            & -“[Sadistic video toward a child] Being strict with your child at an early age will bring you some benefits.” \\
                            & -“Leave a child alone at home. They need to be strong by themselves.” \\
                            \hline
                            
    \multirow{3}{*}{\parbox{5cm}{Contains false news or inaccurate information.}} 
                            & -	“There is no climate emergency. It's another scam. Time to wake up.” \\
                            & -	“Covid-19 vaccines can cause injury and Death. Save people from being vaccinated.” \\
                            \hline
                            
    \multirow{2}{*}{\parbox{5cm}{Contains bullying or threats against anyone.}}
                            & -	“Look at this dog of a woman! She’s not even a human being — she must be some sort of mutant or animal!” \\
                            & -	“I hate her so much. I wish she’d just get hit by a truck and die.” \\
                            \hline
                            
    \multirow{2}{*}{\parbox{5cm}{Demeans, defames, or promotes discrimination against individuals or groups of people.}}
                            & -	“A shit Muslim bigot like you would recognize history if it crawled up you cunt.” \\
                            & -	“\#LGBT community is full of whores spreading AIDS link the Black Plague.” \\
                            \hline
                            
    \multirow{2}{*}{\parbox{5cm}{Solicits any transaction or gift of illegal/regulated goods.}} 
                            & -	“[A picture of a firearm] Order a custom gun today—DM for purchase.” \\
                            & -	“Having cigarettes, tobacco today \#Teens \#studentDiscount.” \\
                            \hline
                            
    \multirow{3}{*}{\parbox{5cm}{Celebrates or encourages destructive behavior.}}
                            & -	“All my problems will disappear if I become skinnier.” \\
                            & -	“Please participate in Momo challenges [self-harm challenges] for your BEAUTY.” \\
                            \hline
                            
    \multirow{3}{*}{\parbox{5cm}{Contains sexually explicit images/videos.}} 
                            & -	“[External page links]: Who wants sexual gratification?  Come and enjoy!” \\
                            & -	“Here are some [celebrity’s name] wardrobe accidents \& nude photo leaks. Check them out today.” \\
                            \hline
                            
    \multirow{3}{*}{\parbox{5cm}{Depicts or facilitates violence or aggression.}}
                            & -	“Here is useful information about how to hit a woman so no one knows.” \\
                            & -	“[Video showing a white nationalist punching a black BLM activist] There's no better feeling than eliminating the enemy.” \\
\Xhline{1.5\arrayrulewidth}
\end{tabular}
\end{center}
\end{table}

We designed three levels of posting guidelines in our survey interface, each differing in the depth of information provided:

\begin{enumerate}
    \item Users do not see any posting guidelines.
    \item Users see posting guidelines consisting of simple descriptions for each type of guideline violation.
    \item Users see posting guidelines consisting of simple descriptions and example posts for each type of guideline violation.
\end{enumerate}

\subsection{Text Box}
\label{sec:text-box}
From the perspective of end-users, most content moderation processes have rigid protocols that offer limited opportunities for expression or interaction with platform administrators~\cite{jhaver2023personalizing,shahid2023decolonizing}. However, some mechanisms offer flexibility e.g.,
% such as enabling users to express their opinions freely in their own words. One example is 
a text box in a flagging mechanism that invites users to express their opinions before a flagging decision is made. 
% Flaggers could clarify their rationale for reporting by providing relevant context and detailing the rule violations that they believe have occurred. Although \citet{wilson2020hate} found that the flagging mechanism can enhance users' trust in moderation decisions by offering opportunities to explain the reasons for reporting, such as the post's context, most social media platforms do not let users express their reasons for flagging a post in an open-ended manner. 
% 
We examine the utility of offering such a text box in the flagging mechanism, enabling users to explain their reasons for flagging, including how toxic posts impact them. We expect that this text box would empower flaggers to actively participate in content moderation by detailing their objections, and its unrestricted nature would let them express their opinions in diverse and personalized ways.

However, users may also perceive that evaluating flagged content by incorporating the content of a text box could significantly increase the moderators' degree of freedom in their assessment criteria, which could shift the review process away from standardized protocols. 
This flexibility may lead users to perceive greater uncertainty about the flag review process and how it would be shaped by the length and quality of their flag submissions. Therefore, we hypothesize:

\begin{quote}
    \emph{H3-1: Offering a text box in a flagging mechanism will decrease perceived consistency in the flag review process.}
\end{quote}

By detailing the context of any flagged post and why it is inappropriate in a text box, flaggers provide moderators with a basis for post review. 
The availability of this text box may signal to users that moderators incorporate this input into their decisions --- this offers additional insight into the review process.
% It may reinforce users' perceptions that their input strengthens the justification for the flag review and its outcome. 
%their flagging choices are taken seriously and that clear and detailed information through their inputs would strengthen the justification for flagging and the review process's rationale. 
% Thus, we expect that providing users with a text box would enhance their perception of transparency about the review of flagged content, and 
We thus hypothesize:

\begin{quote}
    \emph{H3-2: Offering a text box in a flagging mechanism will increase users' perceptions of process transparency.}
\end{quote}

Including a text box in the flagging process lets users articulate their reasons for flagging with higher precision.
This feature empowers users by granting them substantial autonomy to express their flagging intentions beyond the confines of predefined options provided by the platform's classification scheme. 
% By enabling users to elaborate on their opinions rather than relying solely on predetermined categories of inappropriate content, this approach addresses the limitations of the classification scheme. Consequently, providing a text box may lead users to perceive that the flagging mechanism values their opinions and intentions more than when they cannot express their views in their own words. Based on this, we expect that users will have a higher perception of voice when the flagging mechanism offers a text box, and 
We thus hypothesize:
\begin{quote}
    \emph{H3-3: Offering a text box in a flagging mechanism will increase users' perceptions of voice in moderation decisions.}
\end{quote}

\vspace{8pt}
\noindent \textbf{Operationalization} \\
We integrated a text box in our survey's flagging mechanism using the text entry feature provided by Qualtrics. This text box enabled participants to input their responses without any word limit, and it was furnished with the prompt, ``\textit{Please describe your problems with this post. Your response will let us know what's happening and help us review this post.}'' Figure \ref{fig:free-writing} displays how this box appeared during the survey. Each participant experienced one of the following two conditions:

\begin{figure}
    \centering
    \fbox{\includegraphics[width = 10cm]{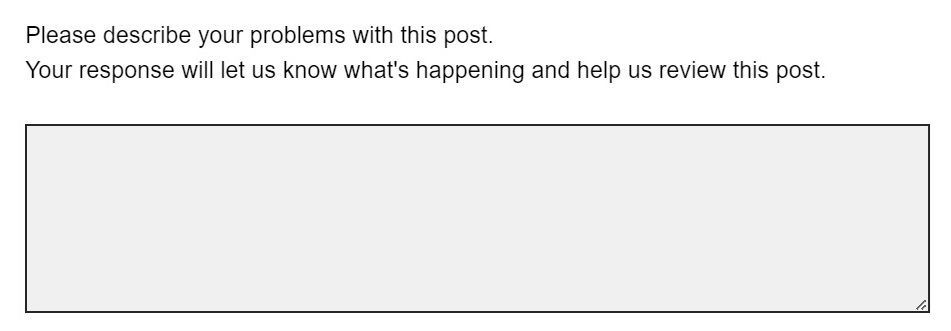}}
    \caption{Participants can express their opinions about the flagged post using the text entry feature provided by Qualtrics.}
    \label{fig:free-writing}
\end{figure}

\begin{enumerate}
    \item Users are not given a text box.
    \item Users are given a text box to express their opinions regarding the flagged post.
\end{enumerate}

\subsection{Moderator Type}
After users submit a flag for moderation, it undergoes assessment by either a human or automated moderator \cite{pan2022comparing, seering2020reconsidering} who evaluates whether the flagged content should be kept on the site or sanctioned. 
Several studies indicate a preference among users for human decision-makers over AI counterparts due to perceptions that humans can better recognize emotions~\cite{lyons2022s, wojcieszak2021can}. 
% Further, \citet{pan2022comparing} found that users perceive moderation decision-making from a human expert panel as more legitimate than decisions made by paid platform reviewers or algorithms. 
However, some research suggests that whether a moderator is human or AI does not significantly affect perceptions of fairness~\cite{jhaver2019did}, transparency \cite{ozanne2022shall, gonccalves2021common} and trust \cite{molina2022ai} in moderation decisions.

Though these prior studies explore how the choice of AI versus human decision-makers shapes user perceptions of moderation outcomes, we know relatively little about how different types of moderators impact user participatory processes like flagging. 
% In addition, prior research has mainly focused on users' perceptions when provided with moderation information alongside decision outcomes \cite{pan2022comparing, jhaver2019does, molina2022ai}. In contrast, our study examines the impact of revealing moderator information \textit{before} decision-making. 
We investigate how different scenarios in flagging mechanisms involving three types of moderator information (human, bot, not specified) affect users' perceptions of the fairness of flag review.

Human moderators are proficient in incorporating the context and their domain knowledge when regulating posts, but they risk introducing potential biases in decision-making \cite{pan2022comparing}. In contrast, automated moderators make decisions based on objective, pre-defined criteria \cite{ozanne2022shall}, even though they may be unable to account for some relevant context. Thus, we 
% posit that users will perceive the flag review process conducted by an automated bot as more consistent than a review conducted by a human or an unknown moderator and thus 
hypothesize:
\begin{quote}
    \emph{H4-1: Providing a moderator's identity as a bot compared to when it is unknown or human in a flagging mechanism will increase users' perceptions of consistency in the flag review process.}
\end{quote}

Transparency in content moderation fosters a meaningful understanding of the moderation processes and improves accountability~\cite{suzor2019we}. Lack of information about the moderator type may lead users to speculate about who reviews their flagged posts, thereby reducing transparency about the process \cite{myers2018censored,suzor2019lawless}. Therefore, we hypothesize:
% that providing information about whether a human or a bot would review the flag can improve the perceived transparency of the flagging system compared to withholding such information.
\begin{quote}
    \emph{H4-2: Providing a moderator's identity as a human or bot compared to when it is unknown in a flagging mechanism will increase users' perceptions of transparency in the flag review process.}
\end{quote}

Prior studies have shown that users perceive decisions made by human moderators, such as expert juries, to be more legitimate than those made by AI \cite{pan2022comparing}. Additionally, users prefer expressing their opinions to human moderators during the appeal process since they believe that humans are more likely to examine their opinions with empathy and compassion \cite{yurrita2023disentangling}. We thus hypothesize:
% that in the context of flagging interactions, users would feel they have a greater voice when a human moderator is involved in the review process instead of an automated moderator or an unknown source.
\begin{quote}
    \emph{H4-3: Providing a moderator's identity as a human compared to when it is unknown or a bot in a flagging mechanism will increase users' perceptions of voice in the flag review process.}
\end{quote}

\vspace{8pt}
\noindent \textbf{Operationalization} \\
Following the other flagging prompts, we included a thank you message to users for their flagging action in our survey, optionally accompanied by moderator information. Specifically, each participant saw one of the following messages:

% In the first condition, participants received only a message expressing gratitude for flagging. In the second condition, the gratitude message was exhibited alongside the information that a human moderator will be responsible for reviewing the flag. Finally, in the third condition, the gratitude message noted that an automated moderator will review the flagged post. 
% Specifically, we used the following message texts in each condition:

\begin{enumerate}
    \item No moderator information: ``Thank you for reporting.''
    \item Human moderator review: ``Thank you for reporting. Our team of human moderators will review your flagged content.''
    \item Automated moderator review: ``Thank you for reporting. Our automated moderator will review your flagged content. Automation enables us to manage the hundreds of millions of reports we receive annually more effectively.''
\end{enumerate}

\section{Methods}
We begin this section by describing how we implemented 54 distinct flagging scenarios using Qualtrics to survey our participants.
% and explain how we operationalized our variables of interest. 
Next, we offer information about our survey deployment and detail our ethical considerations.
Finally, we describe the methods we used to analyze the Likert scale and open-ended responses from our survey data.
We preregistered this study at OSF.\footnote{\url{https://osf.io/982wd/?view_only=91e6792f777b44b582fd6bb480ab142c}}

\subsection{Survey Design} %%%%PAST TENSE%%%%
Our mixed methods study deployed a triangulation design~\cite{ivankova2009mixed} in which we collected both quantitative and qualitative data using a questionnaire that contained close-ended and open-ended questions. 
Both methods were given equal weight~\cite{ivankova2009mixed} and the mixing of the two methods occurred during the interpretation of the study results (Section \ref{sec:Discussion}). 
This approach allowed us to compare and contrast our findings to produce well-validated conclusions and gain an in-depth understanding of how the overall design and procedural space of flag mechanisms shape flaggers' fairness attitudes.

% To test how the availability and granularity of different flagging components affect users' perceived fairness, 
To collect data for this study, we created an online survey using Qualtrics software and
%\sandy{Briefly state survey purpose and audience before going into further detail.}
distributed it via Lucid Theorem, an academically-oriented survey platform, to collect responses from U.S. adult internet users. 
The survey began with a page displaying seven examples of rule-violating posts, shown in Table \ref{table:postsselected}.
%\sandy{Warn readers about nature of the content in the table, which is certainly offensive.}
\textbf{\textit{Warning:}} The example posts in Table \ref{table:postsselected} may be disturbing to some readers.
\begin{table}[h]
    \centering
    \captionsetup{justification=centering}
    \caption{Rule-violating Posts that Survey Participants Were Asked to Choose From to Initiate the Flagging Mechanism.}
    \label{table:postsselected}
    \resizebox{0.9\textwidth}{!}{%
    \begin{tabular}{@{\hspace{10pt}} c p{15cm} @{\hspace{10pt}}}
        \toprule
   \multicolumn{1}{c}{\textbf{No.}} & \multicolumn{1}{c}{\textbf{Post}} \\
        \midrule
        1 & Mexicans come from an uncivilized, backward society. They are filthy criminals, molesting innocent American women and menacing entire neighborhoods. For the sake of our safety, they should all be beaten up and rot in jail forever. \\
        \midrule
        2 & @Sean11 I hate all you F***er and your mother all of you should go die in a hole! \\
        \midrule
        3 & Drinking bleach has been scientifically proven to prevent COVID-19. You should drink a 250ml glass of bleach with every meal. We will get through this together!\\
        \midrule
        4 & Today is a great day! I ate 723 calories today and have been exercising since 10 a.m. It's better to die than to never be skinny. Maintaining 40 pounds is a mandate for an adult woman. \#Anorexia \#annorexya \#anoreskya \\
        \midrule
        5 & Download the software for credit card number generator today - [website URL]\\
        \midrule
        6 & Omg just got tons of Bucks from here! - [xyz phishing site].com. \\
        \midrule
        7 & I have a masturbation video of @Janny12. **Cute Yellow Chink Asian! ** Selling for \$100 for just 5 minutes! If anyone wants 37 minutes of her video, send me a private message.\\
        \bottomrule
    \end{tabular}
    }
\end{table}

We borrowed and modified these rule-violating examples from previous research \cite{nycyk2016enforcing,gonccalves2021common,dineva2022managing,kraut2012building} as well as posts we encountered during our regular use of social media. These posts reflected the diversity of norm violations users might encounter in their everyday use of social media, including instances of hate speech, bullying, self-injury, and misinformation. 
We constructed three of these posts in ways that pose classification challenges due to either the absence of a clear rule violation category or multiple possible selections within the rule violation classification scheme. For instance, post \#7 in Table \ref{table:postsselected} is an example of race-based hate speech that is also sexually explicit. 
This approach aimed to encourage participants to consciously explore the flagging process and invest more effort into elaborating upon their intentions. 

The survey instructed participants to read all seven posts and choose one to report. 
At this point, we also described the concept of flagging on social media, explained that our survey questions would simulate the steps involved in flagging a post, and requested participants to follow these steps to submit a flag in ways similar to how they would flag their selected post on a social media platform.
After selecting a post, participants were randomly assigned to one of the 54 flagging scenarios simulated via Qualtrics questions (i.e., we did not direct participants to an external site) as described below. 
This selected post was displayed at the top of the screen to guide participants throughout the flagging procedure.

The flag submission choices shown to participants followed a between-subjects factorial design created by combining different implementation choices for the four flagging components. It comprised 3 (Rule violation classification schemes) $\times$ 3 (Granularity of posting guidelines) $\times$ 2 (Availability of a text box) $\times$ 3 (Types of moderators) different conditions. 
% We implemented each flagging choice within Qualtrics, simulating how users report posts on social media sites.
% 
% Each participant was randomly assigned to one of the 54 conditions described above.
Depending on their assigned scenario, participants experienced variations in the design of the rule violation classification scheme, the level of detail regarding platform guidelines, the presence or absence of an open-ended text box, and information about the moderator (human, bot, or left unspecified).

\subsubsection{Flagging Scenarios}
\begin{figure}[h!]
    \centering
    \includegraphics[width = 0.8\textwidth]{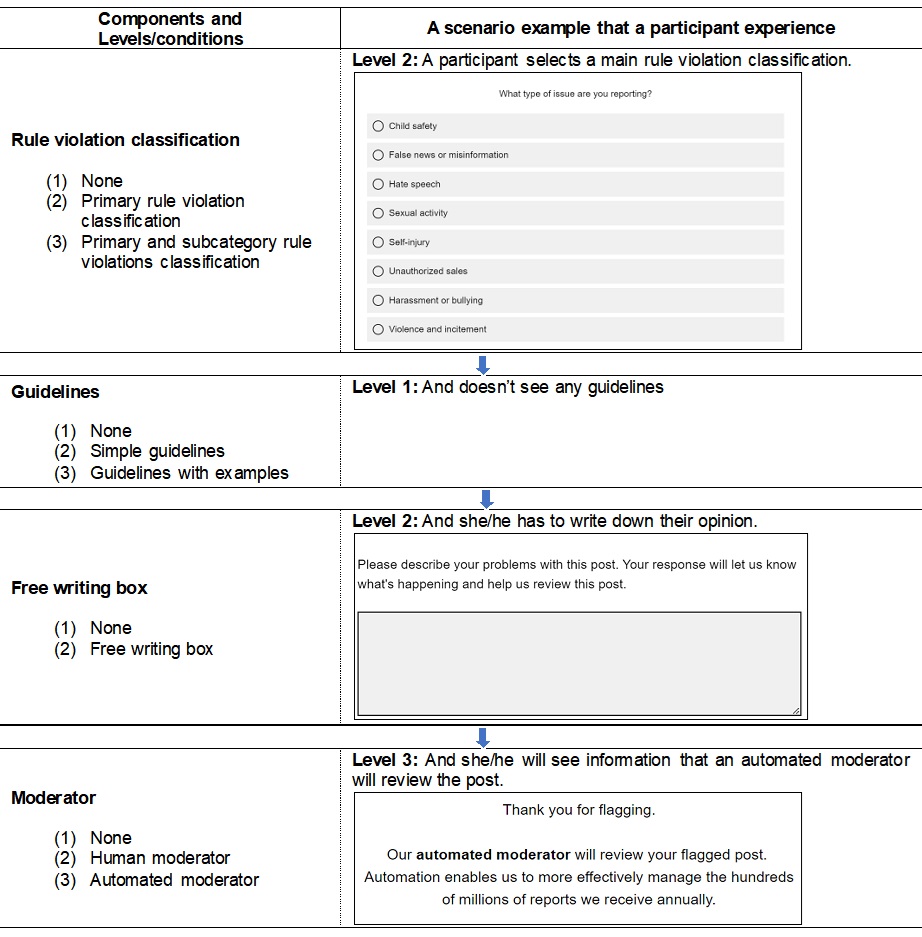}
    \caption{The four components, the different levels of each component, and an example scenario of the flagging mechanism experienced by participants. The left column outlines each element of the flagging mechanism along with its corresponding levels, while the right column exemplifies a scenario comprising a randomly selected combination of these levels.}
    \label{fig:scenario}
\end{figure}

Section \ref{sec:components} details how we operationalized the four components of flagging designs within the Qualtrics software.
In summary, Figure \ref{fig:scenario} illustrates the flagging components and an example of flagging design flow a participant could have encountered in the survey. The left column enumerates each component and its different levels (i.e., conditions).
The right column shows one example of the flagging scenario a participant might have encountered. Once a participant selected one of the seven examples of inappropriate posts listed above, the flagging process (scenario) started. In this particular scenario, a participant would have been required to choose one of the rule violations from the primary classification scheme for the post they selected. Then, the participant would have been offered a text box to describe the rule violation in detail. The final step would have been a message of gratitude for flagging, informing them that an auto-moderator will review the post they flagged.

\subsubsection{Dependent Variables: Procedural Fairness} \label{sec:methods_dv}
In each scenario, after participants completed the flagging, they were prompted to evaluate their perceptions of procedural fairness regarding the flag mechanism they encountered. The survey directed participants to respond to three questions assessing procedural fairness, as outlined below:

\begin{enumerate}
    \item``This flagging mechanism will review each flagged post consistently regardless of the flagger’s personal characteristics, including gender, age, and account history.'' 
    \item ``This flagging mechanism offers sufficient visibility into the important elements of the flag review process.''
    \item ``This flagging mechanism allows me to fully express my objections with the flagged post.''
\end{enumerate}

We used responses to these questions to operationalize participants' perceptions of consistency, transparency, and voice, respectively.
Participants indicated their responses to these questions on a 7-point Likert scale that ranged from \textit{Strongly Disagree} to \textit{Strongly Agree}. 

\vspace{7pt}
\noindent Additionally, to answer RQ 4, the survey featured three open-ended questions asking participants to provide suggestions on how the flagging process could be further enhanced. These questions were: 
\begin{enumerate}
    \item ``What changes would you suggest to this flagging mechanism process to increase your trust that the flag review is consistent regardless of who flagged the post?''
    \item ``What changes would you suggest to improve the transparency of this flagging mechanism process?''
    \item ``What changes would you suggest so that users can more fully express their objections about the concerned posts?''
\end{enumerate}

\subsubsection{Usability}
As detailed above, we designed 54 distinct flagging scenarios for our study, each comprising a combination of varying levels of components, resulting in differing workloads for each scenario. According to the Technology Acceptance Model (TAM), which examines user interaction with systems, the cognitive effort expended during system use significantly influences subsequent system adoption \cite{davis1989perceived}. In accordance with this model, we expect that the diverse scenarios in our survey could impose varying cognitive burdens on the users and influence their future use of flags. Thus, in each scenario, we included two questions to assess usability as part of the survey: one regarding the effort required by users to use the flagging mechanism and the other regarding participants' intentions to use this mechanism in the future. Participants were asked to rate the following two questions using a 7-point Likert scale:
\begin{enumerate}
    \item ``How demanding was it for you to use this flagging mechanism to report the post you selected?'' (rated on the scale of `Very demanding' to `Very undemanding')
    \item ``How likely are you to use this flagging mechanism to report any inappropriate post on a social media platform you use?'' (rated on the scale of `Very likely' to `Very unlikely'.)
\end{enumerate}

\subsection{Deployment of the Survey}
Our study was considered exempt from review by our institute's\footnote{We will reveal the institute name after the peer review is completed.} IRB on November 16th, 2023. 
Before launching the survey, we conducted a pilot survey from May 1 to May 8 to gather feedback on the survey's wording and organization. Based on this feedback, we revised the survey questionnaire by adjusting certain words and adding signposts. After this, we rolled out our main survey, which targeted U.S. adult internet users aged 18 and above. Participants were recruited through Lucid Theorem (\url{https://lucidtheorem.com}), a survey company providing access to nationally representative samples, from May 16 to May 18, 2024. Participants who completed the survey received \$1.50 in compensation via the Lucid system. In total, we collected responses from 3,650 participants.

We excluded respondents who opted out during the survey (N = 496),\footnote{Our review of collected data shows that 253 participants opted out on the first page where we requested participation consent, 7 dropped out at the next question requesting selection of a post that participants would like to flag, 148 dropped out in the flag submission stage, and 88 dropped out after answering the flag submission questions. We found no noticeable trends in the number of dropouts across the different flag implementation scenarios.} spent less than 1 minute or more than 50 minutes on the survey (N = 40), and exhibited straight-lining behavior by selecting the same response for all questions (N = 178).
Following these pre-processing steps, we retained 2,936 survey responses, which we used for our subsequent analyses.
Appendix \ref{sec:appendix_survey_sample} describes the demographic distribution of our survey sample.
On average, the participants took 393.2 seconds to complete the survey.
As mentioned above, the survey began with participants selecting a post to flag from the list of rule-violating posts presented in Table \ref{table:postsselected}. 
% Descriptive analysis showed the frequency distribution of choices made by survey participants to be as follows: Post 1 (21.2\%), Post 2 (9.3\%), Post 3 (23.6\%), Post 4 (5.6\%), Post 5 (6.1\%), Post 6 (4.6\%), and Post 7 (29.7\%).
Table \ref{table:rule_violation_example_means} (Appendix \ref{sec:appendix_posts}) shows the distribution of post choices made by survey participants and how average fairness perceptions vary across these choices --- we did not find any noticeable trends in fairness perceptions based on participants' post choices.

\subsubsection{Ethical Considerations}
Our survey contained several examples of social media posts that breach platform guidelines and are regarded as inappropriate. These cases were included to direct participants' attention to the rationale behind reporting and to provide a realistic context for navigating the flag submission procedures. When obtaining consent to participate in this study, we informed participants of the potential risk of encountering these inappropriate posts before they began the survey. Participants who preferred not to view these posts could choose not to participate and opt out of the survey. 

We included only textual content in our examples, refraining from using visually disturbing material to mitigate the risk of participants experiencing psychological harm.
To safeguard the mental health of our participants, we provided information about mental health resources, including the contact information of organizations such as the National Institute of Mental Health and Mental Health America, for participants to use if they needed help.

\subsection{Data Analysis}
\subsubsection{Quantitative Analysis}
We used SPSS Version 29 to analyze our quantitative data. 
Sec. \ref{sec:quant_findings} presents the results of our quantitative analyses.
To test hypotheses \textit{H1-3}, \textit{H2-1}, and \textit{H2-3}, we conducted one-way Analysis of Variance (ANOVA) tests. ANOVA examines whether statistically significant differences exist between three or more groups.
% relationships between multiple nominal independent variables and a continuous dependent variable. 
In cases where homogeneity of variance was violated (Levene's test resulted in a \textit{p}-value below .05), we employed Welch's ANOVA test (\textit{H1-1}, \textit{H1-2}, and \textit{H2-2}). 
For testing hypotheses \textit{H3-1}, \textit{H3-2}, \textit{H3-3}, \textit{H4-1}, \textit{H4-2}, and \textit{H4-3}, we conducted independent samples \textit{t}-tests. This method examines whether
statistically significant differences exist between two groups.

%additional analysis section
As an additional exploration, we built a General Linear Model (GLM) to examine the interaction effects among our independent variables on participants' perceived consistency, transparency, and voice. This analysis demonstrated how different combinations of component choices impact users' perceived fairness. 
Additionally, we evaluated how each flagging component impacted flaggers' cognitive burden. 
To test whether different choices of classification schemes, posting guidelines, and moderator type affect participants' cognitive load, we conducted three separate ANOVA tests, one for each of those components.
In the case of the text box, we conducted an independent samples \textit{t}-test to see how the availability of the text box affects users' perceived cognitive burden.

\subsubsection{Qualitative Analysis}
The survey included open-ended questions (Sec. \ref{sec:methods_dv}) to elicit suggestions for enhancing consistency, transparency, and voice in the flagging mechanism. After cleaning the data, we collected 1,741 valid responses for these three questions: Consistency (N = 657), Transparency (N = 691), and Voice (N = 393). Next, we performed an inductive analysis~\cite{corbin2015basics} on these responses using NVivo v.14. 

Although the survey provided separate boxes for suggestions related to each aspect of fairness (consistency, transparency, and voice), respondents frequently combined and expressed their responses in a single long answer, highlighting the interconnected nature of these aspects. 
Thus, we integrated responses to these separate sections and analyzed them together to understand the participants' intentions better. In addition, we excluded the responses that merely pointed out the importance of each flagging component without any additional elaboration, e.g., a response just stating `text box' was excluded. This step helped us surface more nuanced insights. 

Next, the two coauthors independently analyzed the initial 20\% of open-ended responses concerning consistency, transparency, and voice, and coded them. 
Some responses were so detailed and complex that we attached multiple codes to them. 
Subsequently, we engaged in iterative discussions to compare and refine our codes, often reflecting on emerging concepts and quoted responses, and achieving consensus on code definitions and applications through collaborative discussions.
After coding the entire dataset, we refocused our analysis at the broader level of themes rather than codes. Specifically, we iteratively sorted related codes into potential themes by using mind-maps and attending to the relationships between codes and between emerging themes~\cite{braun2006using}.
Through this process, we next reviewed and refined our candidate themes and arrived at five key themes we present as our findings. 
Sec. \ref{sec:qual_findings} presents the results of our qualitative analysis.
\section{Findings}
\subsection{Quantitative Findings}
\label{sec:quant_findings}
\begin{table}[h]
    \centering
    \captionsetup{justification=centering}
    \caption{Mean Values of Perceived Consistency, Transparency, and Voice Across Different Conditions of Flagging Components \\ 
    (Likert Scale: Strongly Disagree = 1 to Strongly Agree = 7).
    Standard deviation (SD) values are in parentheses. \textit{N} represents the number of participants placed in the condition.}
    \label{table:flagstats}
    \resizebox{0.9\textwidth}{!}{%
    \begin{tabular}{@{\hspace{10pt}}l l c c c c@{\hspace{10pt}}}
        \toprule
        \textbf{Condition Group} & \textbf{Condition} & \textbf{\textit{N}} & \textbf{Consistency (M, SD)} & \textbf{Transparency (M, SD)} & \textbf{Voice (M, SD)} \\
        \midrule
        \multirow{3}{*}{Classification Scheme} 
        & None & 968 & 5.53 (1.45) & 5.29 (1.43) & 5.23 (1.66) \\
        & Simple & 980 & 5.61 (1.39) & 5.35 (1.32) & 5.37 (1.45) \\
        & Detailed & 988 & 5.48 (1.51) & 5.27 (1.39) & 5.30 (1.52) \\
        \midrule
        \multirow{3}{*}{Guidelines Level} 
        & None & 993 & 5.51 (1.48) & 5.15 (1.45) & 5.24 (1.61) \\
        & Simple & 971 & 5.53 (1.45) & 5.33 (1.37) & 5.31 (1.54) \\
        & Detailed & 972 & 5.59 (1.43) & 5.44 (1.29) & 5.35 (1.49) \\
        \midrule
        \multirow{2}{*}{Text Box Availability} 
        & Absent & 1,492 & 5.54 (1.44) & 5.27 (1.38) & 4.94 (1.64) \\
        & Present & 1,444 & 5.55 (1.46) & 5.33 (1.38) & 5.67 (1.35) \\
        \midrule
        \multirow{3}{*}{Moderator Type} 
        & Not Available & 988 & 5.56(1.46) & 5.32 (1.38) & 5.33 (1.54)\\
        & Human & 972 & 5.52(1.45) & 5.27(1.41) & 5.26 (1.55) \\
        & Bot & 976 & 5.55 (1.45) & 5.33(1.34) & 5.32(1.55) \\
        \bottomrule
    \end{tabular}
    }
\end{table}

%\sandy{I suggest you add to the end of 5.1 a list of all hypotheses with those accepted or rejected marked with check/xs. Readers might appreciate this.} \yunhee{I feel like this is redundant; we already have summaries about testing Hypotheses in each subsection.} \shagun{No need to add this.}
Table \ref{table:flagstats} shows the mean values of perceived consistency, transparency, and voice reported by participant groups that encountered different conditions for the four flagging components.
Given that we ran three tests for each flagging component, we estimated the statistical significance of each result following Bonferroni correction ($\alpha$ < .05/3).
Table \ref{table:results_summary} summarizes the results for our hypothesis.

\begin{table}[h!]
    \centering
    \captionsetup{justification=centering}
    \caption{Summary of Research Hypotheses, Variables, and Outcomes}
    \label{table:results_summary}
    \resizebox{\textwidth}{!}{%
    \begin{tabular}{m{10cm} m{2.5cm} m{2.5cm} m{3cm}} 
        \toprule
        \multicolumn{1}{c}{\textbf{Hypothesis}} 
        & \multicolumn{1}{c}{\textbf{Independent Variable}} 
        & \multicolumn{1}{c}{\textbf{Dependent Variable}} 
        & \multicolumn{1}{l}{\textbf{Outcome}} \\
        \midrule
        \textbf{H1-1} More granular classification schemes will increase perceived consistency. 
        & \multirow{3}{*}{\parbox{2.5cm}{Granularity of classification scheme}} & Consistency 
        & Not supported \\
        \textbf{H1-2} More granular classification schemes will increase perceived transparency. 
        & 
        & Transparency 
        & Not supported \\
        \textbf{H1-3} More granular classification schemes will increase perceived voice. 
        & 
        & Voice 
        & Not supported \\
        \midrule
        \textbf{H2-1} More granular posting guidelines will increase perceived consistency. 
        & \multirow{3}{*}{\parbox{2.5cm}{Granularity of posting guidelines}}
        & Consistency 
        & Not supported \\
        \textbf{H2-2} More granular posting guidelines will increase perceived transparency. 
        & & Transparency 
        & \textbf{Partially supported} \\
        \textbf{H2-3} More granular posting guidelines will decrease perceived voice. 
        & 
        & Voice 
        & Not supported \\
        \midrule
        \textbf{H3-1} Offering a text box will decrease perceived consistency. 
        & \multirow{3}{*}{\parbox{1.5cm}{Text box availability}}
        & Consistency 
        & Not supported \\
        \textbf{H3-2} Offering a text box will increase perceived transparency. 
        & & Transparency 
        & Not supported \\
        \textbf{H3-3} Offering a text box will increase perceived voice. 
        & 
        & Voice 
        & \textbf{Supported} \\
        \midrule
        \textbf{H4-1} A bot moderator will have higher perceived consistency than an unknown or human moderator. 
        & \multirow{4}{*}{Moderator type} 
        & Consistency 
        & Not supported \\
        \textbf{H4-2} A bot or human moderator will have higher perceived transparency than an unknown moderator. 
        & 
        & Transparency 
        & Not supported \\
        \textbf{H4-3} A human moderator will have higher perceived voice than a bot or unknown moderator. 
        & 
        & Voice 
        & Not supported \\
        \bottomrule
    \end{tabular}
    }
\end{table}

\subsubsection{Rule Violation Classification}
\label{sec:Rule Violation Classification Scheme Levels}

We tested Hypotheses 1 by examining how different implementations of rule violation classification schemes affect user perceptions through ANOVA.
First, we rejected \textit{\textbf{H1-1}} given that contrary to our hypothesis, participants shown a simple classification scheme have a higher perceived consistency than those shown a detailed one. 
Further, no significant differences were observed between the absent and simple conditions, nor between the absent and detailed conditions.
We also rejected \textit{\textbf{H1-2}} since the different classification schemes do not significantly impact perceived transparency. 
Finally, we rejected \textit{\textbf{H1-3}}, given that differences in classification scheme do not significantly impact perceived voice.

\subsubsection{Posting Guidelines}
\label{sec:Granularity of the Platform's Posting Guidelines}

To test Hypotheses 2, we used ANOVA tests to examine how displaying posting guidelines with different granularity levels affect user perceptions. 
We found that these differences do not significantly impact participant perceptions of consistency and voice, thus rejecting \textit{\textbf{H2-1}} and \textit{\textbf{H2-3}}. However, we found partial support for \textit{\textbf{H2-2}} --- participants encountering simple (M = 5.33) or detailed guidelines (M = 5.44) have a significantly higher perception of transparency than those in the absent guidelines condition (M = 5.15, 
\textit{F}(2, 1952.81) = 11.36, \textit{p} < .001).

\subsubsection{Text Box}
\label{sec:Availability of a Free-Text Box}

We conducted t-tests to examine how the availability of a text box affects user perceptions to test Hypotheses 3. 
Our analysis shows that the availability of open-ended text boxes that let users describe their reasons for flagging does not significantly affect their perceptions of consistency and transparency, leading us to reject \textit{\textbf{H3-1}} and \textit{\textbf{H3-2}}. However, we found support for \textit{\textbf{H3-3}} --- participants in the text box condition (M = 5.67) report a significantly higher perceived voice compared to those without such a box (M = 4.94,
\textit{t}(2859.50) = -13.32, \textit{p} <.001).

\subsubsection{Moderator Type}
\label{sec:Moderator Type}

To test Hypotheses 4, we conducted t-tests to examine how different moderator types affect user perceptions. 
Our analyses found that different moderator identities do not significantly impact perceived consistency, transparency, and voice. Therefore, we reject the hypotheses \textit{\textbf{H4-1}}, \textit{\textbf{H4-2}}, and \textit{\textbf{H4-3}}.

\subsubsection{Additional Analyses: Interaction Effects and Usability}
\label{sec:Additional Analyses}

\textbf{Interaction Effects.} In addition to testing our proposed hypotheses, 
% we built a General Linear Model (GLM) to examine how different combinations of flagging components affect perceived fairness. Using this model, 
we explored the interaction effects among the four flagging components on perceived consistency, transparency, and voice. Table \ref{table:interaction_effect} in Appendix \ref{sec:appendix_additional_analyses} shows the results of this analysis.
We estimated the statistical significance of these results following Bonferroni correction ($\alpha$ < .05/11) and concluded that the interaction between
% Results show that the interaction between guidelines and the availability of a text box significantly affected perceived transparency (\textit{F}(2, 2936) = 4.22, \textit{p} = .02).
classification schemes and the availability of a text box significantly affected perceived voice (\textit{F}(2, 2936) = 8.29, \textit{p} < .001). 
Specifically, for participants in the condition \textit{without} a text box, providing either a simple (\textit{MD} = 0.34, \textit{SE} = 0.10, \textit{p} < .001) or detailed (\textit{MD} = 0.35, \textit{SE} = 0.10, \textit{p} < .001) classification scheme elicited significantly higher perceptions of voice than providing no classification scheme (Table \ref{table:Interaction_voice}, Appendix \ref{sec:appendix_additional_analyses}).
All other interactions did not significantly impact any of the three aspects of procedural fairness.

\textbf{Usability.}  
% Given that usability is a crucial factor in designing interactive processes like flagging, 
We examined how variations in each component of flagging mechanisms differentially impacted the cognitive burden on our participants, thereby potentially influencing flag usability.
Our analysis (Table \ref{table:cognitive_burden}, Appendix \ref{sec:appendix_additional_analyses}) shows that participants in the condition with a text box 
(\textit{M} = 3.86, \textit{SD} = 1.73, \textit{EM} = .05)
experienced a significantly higher cognitive burden compared to those without such a box (\textit{M} = 3.66, \textit{SD} = 1.80, \textit{EM} = .05).
No other effects approached statistical significance.
We also found that variations in how each flagging component is presented (or not presented) did not significantly affect participants' self-reported likelihood of using flagging mechanism (Table \ref{table:future_use}, Appendix \ref{sec:appendix_additional_analyses}).

\subsection{Qualitative Findings} 
\label{sec:qual_findings}
Our inductive analysis surfaced five themes through our combining and distilling of codes that together provide a nuanced understanding of user perspectives on enhancing fairness in flagging mechanisms, thus addressing RQ 4. Table \ref{table:Qual_results} summarizes these themes and their frequency.
 
% Qualitative analysis
\begin{table}[ht]
\centering
\resizebox{0.7\textwidth}{!}{%
\begin{threeparttable}
\caption{Analysis of Responses to Open-ended Questions: Suggestions for Enhancing Fairness.}
\label{table:Qual_results}
\begin{tabular}{m{8cm} m{2cm}}
\toprule
\textbf{Theme} & \textbf{Frequency} \\
\midrule
Desired attributes of flag reviewers & 138 (11\%) \\
Needing support for greater expressivity & 297 (24\%) \\
Demanding outcome notifications with timely review & 361 (29\%) \\
Expectations regarding review procedures and statistics & 298 (24\%) \\
Preventing flagging abuse and protecting flaggers & 144 (12\%) \\
\bottomrule
\end{tabular}
\end{threeparttable}}
\end{table}

\subsubsection{Desired Attributes of Flag Reviewers}
\label{sec:Desired Attributes of Flag Reviewers}

Participants frequently expressed their concerns about flag reviewers’ biases and detailed their preferences regarding whether they prefer a human or AI moderator and their reasons for such preferences. Some participants (N = 61) mentioned that they \textbf{prefer a human reviewer} who can understand the context and nuance of the flagged post. Further, a few (N = 9) expressed their interest in \textbf{communicating with moderators} about the flagging process and the results. This suggests that users need thorough explanations for flagging outcomes through detailed conversations, revealing a preference for human interaction. For instance, participant P312 suggested:

\begin{quote}
     \textit{Enable direct messaging with moderators for in-depth discussions.}
\end{quote}

However, other respondents (N = 24) expressed a \textbf{preference for AI-based reviewers}, such as a bot or an algorithmic moderator, to enhance fairness. These divergent views regarding the types of reviewers are consistent with our quantitative findings, which indicate no significant differences in perceived fairness based on the moderator type.

Many respondents offered suggestions to reduce biases in decision-making. 
For example, 17 participants suggested that platforms implement collaborative decision-making by involving multiple moderators for flagged posts. Some participants specified their suggestions for a certain combination or number of moderators (often three) or a mix of human and bot moderators. They believed that such cooperative decision-making in the flagging process would increase fairness and reduce biases. For example, P405 wrote:

\begin{quote}
    \textit{Form a diverse review team comprising individuals with varying backgrounds, perspectives, and experiences. This diversity helps mitigate biases and ensures a more balanced evaluation of flagged content.}
\end{quote}

Further, to reduce a biased moderation review process, 27 participants emphasized the desired qualifications and characteristics of individual moderators. This includes moderators’ expertise in the subject matter of reviewed content, their understanding of ethical standards regarding rule violations, and their political neutrality.

\subsubsection{Needing Support for Greater Expressivity}
\label{sec:Desiring Flagging Mechanisms to Support Greater Expressivity}

Respondents often noted that the currently available flagging affordances do not allow them to fully communicate their objections to the flagged posts, and they
% In addition to appreciating the different components they encountered within the reporting mechanism (e.g., posting guidelines, text box), many participants 
proposed many constructive suggestions as to how flagging systems could better support expressivity.
Among these, the most frequently suggested feature referenced the classification scheme, such as adopting a \textbf{broader classification scheme} (N = 45) with a wider range of rule violation categories for selections and the inclusion of \textbf{multiple choices for flagging reasons} (N = 90), which would enable flaggers to select several rule violation categories within the classification scheme. 
% Our survey was designed to include some examples of rule violations involving multiple infractions, making it challenging for participants to select only one rule violation classification category when the classification scheme was provided. This restriction created a dilemma for some respondents regarding which category they should select, often inducing them to consider how the ability to select multiple options for categorizing rule violations would shape the flagging experience. 
% P343 felt that this feature would help prioritize the flagging of severely inappropriate posts:
% 
% \begin{quote}
%     \textit{For example, I could only choose hate speech, but if it was also considered harassment and could fall under multiple categories, it could be removed or reviewed faster.}
% \end{quote}
% 
For example, Participant P158 advocated for allowing the selection of multiple categories, anticipating that it would induce changes at the flag review stage:
\begin{quote}
    \textit{Allow to select more than one answer. Sometimes flagging should be reviewed for more than one concern: this might increase the likelihood of a thorough review.}
\end{quote}

% Further, 45 respondents also mentioned preferences for a \textbf{broader classification scheme} with more options for rule violation selections. For instance, P485 wrote: 

% \begin{quote}
%      \textit{Implementing a wider range of specific categories for flagging posts.}
% \end{quote}

Such responses reflect participants' desire to classify and elaborate on the types of rule violations they encounter when making their reporting decisions. 
Other suggestions on how to achieve this included an ability to \textbf{rate the severity of posts} (N = 15) or \textbf{highlight specific sections of posts that violate rules} (N = 14). For instance, P708 noted:

\begin{quote}
   \textit{Even within flagged items, the severity level is not touched. Some flagged categories are more harmful than others.}
\end{quote}

In addition to emphasizing their need for clear and detailed reporting, some participants stressed the importance of maximizing user participation in the flagging process. For instance, they desired an ability to \textbf{use alternative channels} (N = 21) for flagging inappropriate posts, such as chat, email, or phone calls. Some felt that platforms should \textbf{encourage users to flag more} (N = 19). This could be achieved by providing clear instructions, step-by-step explanations, or incentivizing flag submissions for users. Another suggestion to enhance user participation and promote inclusivity was to \textbf{simplify the flagging process} (N = 73), ensuring that it is user-friendly and accessible to a wide range of users.

\begin{quote}
    \textit{Use easy words to describe so that everyone will understand easily. (P782)}
\end{quote}

Further, some participants advocated for adopting measures that enable flaggers to corroborate a post's toxicity by sharing their perspectives and referencing previously flagged posts.
This finding suggests that incorporating multiple perspectives, not only from flag reviewers but also among flaggers themselves, could contribute to designing a fairer flagging mechanism.
As specific methods, respondents expressed interest in \textbf{discussing toxic posts in a forum} (N = 21), where they could share their thoughts on the regulation of such posts. 
%This would facilitate a peer review process, allowing users to specify their reasons for flagging a post and understand others' perspectives.
%Participants believed that involving diverse perspectives prior to the flag submission could lead to more accurate and reliable decisions in the post-review stage about whether to remove the flagged content. For instance, P705 commented:
%\begin{quote}
%\textit{Possibly creating a forum where some users can discuss their objections about a concerning post.}
%\end{quote}
Another approach to ensuring accurate judgment of the post's toxicity involved \textbf{reviewing similar previously flagged posts and their post-review outcomes before submitting a flag. }
%They felt that this feature, along with the feedback from other users, could ensure that their flags are sufficiently justified for removal by allowing them to more clearly indicate the extent to which flagged content violates rules. 
Participants believed that implementing such features would enhance the credibility of their objections and provide the platform with clear insights into the urgency of removing flagged content.

Besides communicating their objections to individual posts, some participants desired an ability to \textbf{provide feedback on their flagging experience} (N = 6). That is, they wanted to voice their opinions not only about inappropriate posts on the platform but also about the flag submission process and its results. P249 elaborated on how the platform must listen to users' opinions on the flagging mechanism:

\begin{quote}
    \textit{The [flagging] mechanism should be reevaluated at least every three months in case any revisions are indicated over time.}
\end{quote}

This shows that participants want to contribute and express their perspectives on how the flagging mechanism should operate. In line with this, some participants recommended including a \textbf{rebuttal or appeal option in the flagging process} that would allow flaggers to contest the flag outcome decisions (N = 12). P708 specifically emphasized that an appeal process for flagging outcomes is crucial for both parties --- the flagger and the author of the flagged post:

\begin{quote}
    \textit{You should be able to dispute the flag on both sides of it. }
\end{quote}

\subsubsection{Demanding Outcome Notifications with Timely Review}
\label{sec:User Demands for Outcome Notification with Timely Review}

Many respondents (N = 284) expressed a strong preference for receiving \textbf{notifications of flagging outcomes and explanations} for those outcomes. Given that our survey questionnaire was designed to gauge perceptions of procedural fairness prior to decision-making for the flagged content, this finding highlights that participants closely associate the decisions regarding flagged posts with the overall fairness of the flagging mechanism. P312 emphasized the importance of transparency in decision-making, stating:

\begin{quote}
    \textit{The review committee should write back to the flagger why they agreed or disagreed with the flagger's decision to flag certain content. }
\end{quote}

% These results contribute to our understanding that flagging outcomes impact perceived fairness not only for the \textit{authors} whose posts were flagged for rule violations \cite{jhaver2019did} but also for the \textit{flaggers }who reported the violations. 
Some respondents expressed a strong desire to receive a timely reply to their flags, which relates to the `voice' aspect of procedural fairness.
This suggests that flagging mechanisms should not only allow users to express their concerns during flagging but also ensure that these concerns are promptly considered and addressed in a responsive manner. 
This desire was especially reflected in participant suggestions for an \textbf{expedited review process} to inform flaggers of flag outcomes (N = 31).
For instance, P174 noted the importance of timely reviews, recommending that:

\begin{quote}
    \textit{Ensure follow-up or decision is made within 24 hours.}
\end{quote}

Additionally, some respondents (N = 46) called for \textbf{immediate actions}, such as post removal or account suspension upon flagging. 
For instance, P304 suggested making flagged posts invisible until a decision is reached about their removal:

\begin{quote}
    \textit{Delete the post until you confirm it is something that violates the rules.}
\end{quote}

%We observed that these participants emphasize the procedural fairness aspect of having their requests promptly addressed by the platform, associating `fairness' with responsiveness to their concerns.
%In a similar vein, some participants  of their flagging promptly. 

To sum up, users perceived fairness in the flagging process as being related to the platform's immediate responsiveness and timely notification of the flagging results. We infer from these suggestions that users' perceptions of fairness in flagging may depend on how seriously the platform takes their flagging requests, how clearly they explain the decision outcome, and the time it takes to review the requests.

\subsubsection{Expectations Regarding Review Procedures and Statistics}
\label{sec:User Expectations Regarding Review Procedures and Statistics}
Respondents often desired to learn more about platforms' processing of flags at the system-level.
In addition to the flagging results, a significant portion of participants (N = 131) shared their need for more information on the flagging process, specifically about \textbf{how the review process works}. They argued that the flagger should be informed about the review criteria the platform adopts, who conducted the review, what steps are entailed in the review process, and how long they should expect to wait until the decision-making occurs. For instance, P500 proposed:
\begin{quote}
    \textit{Probably more specifics. "Your report will be reviewed with[in] X period of time."}
\end{quote}

In addition, 15 respondents mentioned that the flagging mechanism needs to provide \textbf{information about the possible outcomes of flag review}, such as the removal of flagged posts and the additional sanctions that flagged users may face.
They felt such information would help end-users decide whether and how to flag inappropriate posts they encounter.
Further, 137 respondents conveyed their specific needs to have regular updates on the submitted flag by \textbf{tracking changes in the flag review status}. They desired a flag tracking system that could offer information such as the confirmation of flag submission, the review steps already taken, the current review stage, and the remaining steps. They hoped that such information could be provided via emails, message notifications, or a dedicated flag tracking page on the platform. For instance, P540 mentioned:

\begin{quote}
   \textit{[Platforms should] offer an incident number that the person making the flag can refer to, then allow a process to follow the incident number through completion.} 
\end{quote}

% \begin{quote}
%    \textit{Updates on the review, and just showing that the company cares. (P194)}
% \end{quote}

Some participants suggested that platforms
% One of the noteworthy aspects users commented on regarding the flagging mechanism, particularly concerning outcomes, was the suggestion to 
provide \textbf{statistics on flag submissions and flag review outcomes} (N = 15). They felt that platforms could enhance transparency about flagging by disclosing statistics related to the number of flags submitted and deleted as well as the relative frequencies of rule violation categories of submitted flags in the form of monthly or annual reports. 
% This initiative ensures that presenting flagging results would contribute to users' perceived fairness. 

\subsubsection{Preventing Flagging Abuse and Protecting Flaggers} \label{sec:Concerns about Flagging Abuse and Suggestions to Address it}
In addition to the suggestions about various components of the flagging mechanism, respondents provided feedback on improving overall platform management, including the flagging mechanism. Some respondents (N = 20) emphasized the need to \textbf{prevent abuse of flagging} by malicious users. This concern was specifically raised about potential misuse, where flags might be submitted without any valid reasons for post removal. They noted that the platforms must verify each flag to ensure it originates from a human user, i.e., it is not automated, and that it is not driven by specific political agendas. P109 underscored this by stating:

\begin{quote}
    \textit{Make sure the flag is correctly verified because some people just flag others for no reason.}
\end{quote}

Additionally, respondents advocated for implementing \textbf{preventive measures} (N = 38) to curb the dissemination of toxic content. Suggestions included employing advanced filtering systems that block posts containing certain keywords and displaying some sanctioned posts as examples of what constitutes norm violations for educational purposes. Some respondents also suggested that flaggers should \textbf{directly communicate with the authors of rule-violating posts} independent of the flagging mechanism, which could reduce reliance on flagging as the primary means of enforcement. P816 suggested:

\begin{quote}
    \textit{ Instead of reporting, go directly to the person and talk to them. If that doesn't work, then report them.}
\end{quote}

% These insights align with findings from \citet{zhang2023cleaning}, highlighting that some users wish to have direct communication to effectively mitigate rule violations rather than submitting a flag, thereby achieving the educational purpose of rule enforcement. 

In contrast, some respondents emphasized the importance of \textbf{protecting flaggers} (N = 60) to ensure fairness. They stressed the need for secure flagging processes that shield the identity of flaggers and prevent potential repercussions by authors of flagged posts. Some participants expressed that this confidentiality is also crucial in minimizing biases during content review, i.e., moderators should review flagged posts without knowing who flagged them. Alongside user protection, considerations for safeguarding \textbf{free speech} (N = 23) were also prominent. While acknowledging the role of flagging in community moderation, users cautioned against overly restrictive flagging practices that could stifle online expression. P555 wrote:

\begin{quote}
    \textit{People need to understand that everyone has a right to their own opinions.}
\end{quote}

%Further, some participants expressed an interest in using personal moderation tools following flag submissions, reflecting concerns about fairness and its intersection with free speech.
\section{Discussion} \label{sec:Discussion}
%Conceptual implications?

We began this study with the goal of examining how flagging mechanisms should be designed to enhance fairness perceptions among flaggers.
Our statistical analysis of responses from a large-scale survey experiment shows that including posting guidelines and a text box for feedback within flag implementations helps enhance users' fairness perceptions, whereas offering classification schemes or providing information about whether the flag reviewer is a human or a bot does not significantly influence users attitudes. 
Our qualitative analysis of open-ended responses shows that users feel concerned about reviewers' biases, desire flagging systems to support greater expressivity, demand timely notifications and explanations of flag outcomes, wish to see information about flag processing at the system-level, and expect platforms to prevent flag abuse.

These results contribute empirically informed guidance on how social media platforms should design different components of their flagging interfaces and how these design choices could impact users' attitudes toward flagging. 
% Critically, we show that users recognize flagging as a critical tool to express their requests for removing inappropriate content online. 
We document the key information and security needs of flag submitters and offer insights for how platforms could address them.
We show that users' engagement with flags triggers a range of sociopolitical concerns regarding platforms' responsibilities, freedom of speech, algorithmic evaluation, safety against online harms, and privacy.
We also provide a methodological framework that others can adopt to evaluate the effectiveness of new components and affordances that seek to address users' reporting needs.

% We began this study with the goal of examining how users engage with flagging mechanisms and how these mechanisms could be designed to enhance fairness perceptions among flaggers.
% %Fairness evaluation situated in a broader context.
% Our mixed-methods analysis of responses from a large-scale survey suggests that many users recognize flagging as a critical tool to express their requests for removing inappropriate content online. 
% Perhaps surprisingly, users' engagement with flimsy flag icons, now ubiquitously available across most platforms~\cite{crawford2016flag,zhang2023cleaning}, although often deeply embedded in action menus, triggers a range of sociopolitical concerns regarding platforms' responsibilities, freedom of speech, algorithmic evaluation, safety against online harms, and privacy.

Prior empirical research on enacting fairness in content moderation~\cite{myers2018censored,jhaver2019did,jhaver2019does,ma2023litreview,vaccaro2020end} has largely focused on how moderation decisions are enacted and communicated to moderated users. Our work contributes to this research by examining procedural fairness during flag submissions.
% 
% While flags offer a means for regular users to participate in the complex and opaque platform governance processes~\cite{suzor2019lawless,ma2023transparency}, users are keenly aware that flagging outcomes are largely determined by governance policies, mechanisms, and moderators instituted by platforms.
% This reflects the power and information asymmetry between users and platforms observed in prior research~\cite{are2023flagging} and emphasizes how platforms conceal, obscure, and strategically disclose details about their governance practices~\cite{cotter2023shadowbanning}. 
% % The complex and opaque nature of moderation infrastructures interferes with the goal of enhancing users' fairness.
We show that
from the flaggers' perspective, each aspect of flagging mechanisms --- the procedural elements available during flagging, flag review criteria, flagging outcomes, and how they are communicated --- require greater consistency, transparency, and support for user expression. Below, we discuss our specific design insights about how to achieve these goals, elaborate upon our conceptual contributions regarding flagging as a content moderation mechanism, and suggest promising avenues for further research.

\subsection{Enhancing Users' Perceptions of Being Heard and Allowing Detailed Expression}
%Implementing a Writing Box: being heard
\subsubsection{Incorporating a Text Box in Flagging Mechanisms}
Our analysis reveals that flaggers' fairness perceptions, especially regarding having a voice in the content moderation process, improve when they have an opportunity to express in detail their objections to the post during the flagging process. 
Specifically, our quantitative analysis demonstrates that the availability of a text box, which lets users articulate their thoughts in their own words within the flagging mechanism, significantly enhances their sense of being heard (Sec.~\ref{sec:Availability of a Free-Text Box}).
Contrary to our hypothesis \textit{H3-1} (Sec.~\ref{sec:text-box}), adding this text box to flag designs does not come at the cost of reducing users' perceived consistency (Sec.~\ref{sec:Availability of a Free-Text Box}).
Besides procedural fairness, a text box could also enhance outcome fairness by providing flag reviewers additional context to fairly evaluate flagged posts. 
Flag outcomes that are accompanied by explanations regarding decision-making~\cite{jhaver2019did,jhaver2019does} and how flaggers' voice was considered could improve flaggers' and flagged users' perceptions of both procedural and outcome fairness.

We found that a text box increases the cognitive load on users when flagging a post (Sec.~\ref{sec:Additional Analyses}), which may reduce the mechanism's overall usability.
Although we did not find a direct association between this feature and users' intention to flag again, prior research indicates that mental fatigue during the flagging process can lead to negative user experiences \cite{gillespie2018regulation}. 
Therefore, platforms should carefully consider adopting this feature since it may limit user participation.
One way for flag mechanisms to reduce unwarranted cognitive load could be to make user input in this text box optional and clarify that it is to be used only if flaggers want to provide additional necessary context.

Further, we recognize that rigorous implementation of this feature requires platform moderators to carefully consider user-submitted text when making their moderation decisions, which presents challenges of scale~\cite{gorwa2020algorithmic,gillespie2020content}.
The results of our interaction effects suggest a likely compromise: 
we found that when flagging mechanisms do not offer a text box, the ability to specify rule violation through a classification scheme becomes significantly more crucial to satisfying users' voice needs (Sec. ~\ref{sec:Additional Analyses}).
Therefore, we suggest that if platforms are concerned about the negative impact of a text box on usability or cannot afford to implement one due to labor constraints, they should at least implement a robust classification scheme for rule violations (or another analogous feature that lets users adequately specify their post objections) in their flagging mechanisms.

\subsubsection{Establishing Mechanisms to Track Flag Status}
Our qualitative analysis underscores users' desire to not only voice their objections to the flagged post but also to ensure that their flag request has been successfully submitted, that they can monitor the flag processing status (Sec. ~\ref{sec:User Expectations Regarding Review Procedures and Statistics}), and that their flags receive timely decision-making by flag reviewers (Sec. ~\ref{sec:User Demands for Outcome Notification with Timely Review}). This emphasizes users' need for platforms to take their input in the moderation process seriously and to meet their expectations that platforms address their concerns without undue delays. 

We echo participants' suggestion that platforms implement a streamlined tracking system to monitor the review progress of submitted flags.
Previous HCI research efforts to design for contestability in moderation procedures~\cite{vaccaro2021contestability,vaccaro2020end} could serve as a blueprint for taking a user-centered approach to building such systems.
Some participants expressed a desire to provide feedback on their flagging experiences (Sec. ~\ref{sec:Desiring Flagging Mechanisms to Support Greater Expressivity}). This suggests that offering a chance to be heard on the entire flagging process (i.e., beyond their objections to individual flagged posts) may influence users' fairness perceptions.
Therefore, platforms could deploy a feedback system to collect recommendations for improvements and regularly evaluate the feedback received to update their current practices.

\subsubsection{Widening Flags' Vocabulary of Complaint}
Prior theoretical analysis of flagging mechanisms emphasized that flags offer ``a narrow vocabulary of complaint'' and do not account for the many complex reasons that people might choose to flag or let users specify their degree of concern with the flagged posts~\cite{crawford2016flag}. 
Our quantitative analysis did not find a relationship between the granularity of classification schemes and fairness perceptions (Sec. \ref{sec:Rule Violation Classification Scheme Levels}).
However, we found that offering a classification scheme (either simple or detailed) enhances voice perceptions when a text box is not provided (Table \ref{table:Interaction_voice}) and providing this scheme does not impact usability (Sec. \ref{sec:Additional Analyses}).
% 
% One might expect that offering a selection of more granular rule violation categories would empower users to better express their objections.
% In line with this, we found that a simple classification scheme enhances the perceived sense of voice more than having no classification scheme (Sec. \ref{sec:Rule Violation Classification Scheme Levels}). Further, 
% However, we found that voice perceptions do not significantly differ between simple and detailed schemes.
% Our findings also reveal that raising the granularity of rule violations does not significantly increase users' perceived consistency and transparency (Sec. \ref{sec:Rule Violation Classification Scheme Levels}). 
Overall, these results suggest that 
% although deploying simple classification schemes helps flaggers voice their objections to platform operators as compared to having no such schemes, 
asking users to categorize their objections using a classification scheme remains fundamentally a constricted way to submit reports from users' perspectives\textemdash and adding submenus to these schemes does not aid in addressing users' expression and fairness needs (Sec. \ref{sec:Rule Violation Classification Scheme Levels}).

Our qualitative results offer some clues on how platforms could expand flags' narrow vocabulary of complaint. Our participants suggested that flagging mechanisms deploy a classification scheme that allows the selection of multiple rule violations, an ability to highlight portions of the flagged post that violate platform rules, and a way to rate how severely inappropriate the post is (Sec. ~\ref{sec:Desiring Flagging Mechanisms to Support Greater Expressivity}).
Participants' demand for such features indicates that users feel a need to exert a greater level of control over the moderation procedures, echoing previous research findings on users interacting with personal moderation~\cite{jhaver2023personalizing} and community management tools~\cite{jhaver2019human}.

While users desire expressive flagging options, it is crucial that platforms provide such options only when they can effectively incorporate that expression. For instance, allowing flaggers to provide free-text responses may enhance their sense of having a voice (perceived fairness) during flag submission, but it does not improve actual procedural fairness if the flag reviewers lack the ability to process that input when making its moderation decisions.

\vspace{5px}
\noindent In sum, enacting flagging solutions that let users clarify the post's context and offer greater flexibility than having to shoehorn complex feelings into a single category selection~\cite{crawford2016flag}
would enhance flaggers' fairness perceptions.

\subsection{Incorporating Transparency in Review Criteria, Reviewers, and Review Outcomes.}
\subsubsection{Integrating Posting Guidelines in Flagging Mechanisms} 
Our statistical analyses show that incorporating posting guidelines into the flagging process enhances participants' perceptions of fairness (Sec. ~\ref{sec:Granularity of the Platform's Posting Guidelines}).
Specifically, providing these guidelines improves users' perceptions of moderation transparency, and providing additional information—such as examples of rule-violating posts in these guidelines—further contributes to users recognizing the flagging process as more transparent. 
This aligns with prior research on the design of personal moderation interfaces~\cite{jhaver2023personalizing}, where including examples of rule-violating posts enhanced users' perceptions of control over the moderation process.
We also found that contrary to our hypothesis \textit{H2-3} (Sec. \ref{sec:posting-guidelines}), integrating posting guidelines does not compromise users' voice perceptions (Sec. \ref{sec:Granularity of the Platform's Posting Guidelines}), and neither does it raise users' cognitive burdens (Sec. \ref{sec:Additional Analyses})---these results further incentivize showing posting guidelines to flaggers.
% Further, our interaction effects results (Sec. ~\ref{sec:Additional Analyses}) show that when flag mechanisms do not include a text box, offering these guidelines (either with or without examples) becomes especially critical to supporting users' transparency perceptions. 

However, most mainstream platforms currently \textit{do not} include posting guidelines in their flagging designs; we found that, currently, only Facebook and Instagram link users to these guidelines. 
We suggest that platforms list or link posting guidelines in the flagging mechanism to enhance their users' understanding of the flag review process and improve their transparency perceptions.

\subsubsection{Offering Information About Flag Reviewers} 
Our quantitative analysis shows that fairness perceptions of flagging procedures are not influenced by information about the moderator type (a human, a bot, or no information) (Sec. ~\ref{sec:Moderator Type}).
This result aligns with prior research findings 
%that fairness perceptions of moderation procedures or outcomes are not influenced by the choice of whether human or AI reviewers are involved
~\cite{ozanne2022shall,gonccalves2021common,molina2022ai}.
However, our qualitative findings suggest that offering other types of information about moderators who review flags may enhance users' perceptions of fairness.
Participants suggested that disclosing moderators' personal characteristics, 
%e.g., their political affiliations or 
e.g., information about their experiences and skills, such as the training moderators receive or their expertise in specific subjects, can foster greater trust in the flag review procedures (Sec. ~\ref{sec:Desired Attributes of Flag Reviewers}). 
Thus, we suggest that flagging mechanisms offer some information about flag reviewers while still respecting their privacy.
Such information could also enhance moderation fairness from the perspective of flagged users.

Our findings highlight the utility of involving multiple reviewers in content moderation, reflecting users' preferences for incorporating diverse perspectives (Sec. ~\ref{sec:Desired Attributes of Flag Reviewers}). Participants suggested that using more than one moderator or combining human and bot reviewers could enhance fairness in the flag review process. This aligns with Fan and Zhang's finding that group-based moderation improves perceptions of fairness \cite{fan2020digital} and Katsaros et al.'s observation that users prefer systems that combine humans with algorithmic decisions \cite{katsaros2024online}, supporting the need for hybrid review models.
Our qualitative findings, as well as prior research \cite{duguay2020queer,pan2022comparing}, suggest that users' preference for involving multiple moderators stems from their concerns about potential biases in decisions made by a single moderator, particularly human moderators who might be influenced by their personal values when making moderation decisions (Sec. ~\ref{sec:Desired Attributes of Flag Reviewers}).
Therefore, flagging mechanisms can enhance their fairness by instituting procedures that incorporate the perspectives of multiple flag reviewers and enacting substantive measures to prevent biases in moderators' decision-making.

\subsubsection{Reforming Post-flag Submission Steps} 
%Flag review process info+Anonymous (secures)users
Participants emphasized the need for greater transparency in the flag review process and the disclosure of information about post-flag submission steps. This involves revealing the review criteria, the specifics of each step in the flag review, the expected timeline, and the information visible to reviewers about flaggers (Sec. ~\ref{sec:User Expectations Regarding Review Procedures and Statistics}). 
%Specifically, flaggers are concerned about how much information reviewers can access and seek assurance that their identities remain anonymous throughout the process. Addressing these concerns by incorporating detailed information on anonymity would further enhance users' perception of transparency.

%Enhancing the accuracy of the system
Additionally, our qualitative insights show users' concerns that the flag review process may not adequately address the problem of malicious~\cite{griffin2022sanitised} or organized~\cite{crawford2016flag} flagging, i.e., flagging of content that does not violate platform guidelines (Sec. ~\ref{sec:Concerns about Flagging Abuse and Suggestions to Address it}).
Since a flag may not accurately indicate the post's actual inappropriateness~\cite{kou2021flag}, e.g., some people may exploit flags as a form of `digilantism' or politically motivated extrajudicial practice ~\cite{jane2017dude}, users feel concerned about unjust sanctions against norm-complying content.
We suggest that platforms assuage such concerns by informing users about the measures they take to prevent malicious users from abusing flagging. For example, they could specify their procedures for sanctioning users who repeatedly engage in false flagging.

% Participants also considered it important that reviewers rigorously assess whether the flagged post indeed violates the inappropriate content criteria specified in the flag (Sec. ~\ref{sec:Concerns about Flagging Abuse and Suggestions to Address it}). 
% Thus, we recommend that platforms develop a publicly accessible protocol for handling situations in which the flagged post does not actually fall into the category (e.g., hate speech, misinformation) that flaggers choose.
% For instance, if the flag review determines that the flagged post is inappropriate but for a different reason than that  included in the flag, flag reviewers may still sanction that post and additionally use this opportunity to inform flaggers about how to make accurate selections from the rule violation classification scheme; such educational measures may reduce the flag review burden for subsequent flags.
% Alternatively, flag reviewers may allow flaggers to re-classify the flag since this could increase the certainty of the review's legitimacy.

\subsection{Supporting Flaggers with Different Technological Competencies and Diverse Perspectives.}
%Educating Users on Flagging: encourage users+higher accessibility
\subsubsection{Improving Flagging Visibility and Accessibility}
Participants' open-ended responses indicate a desire for flagging systems to have greater accessibility and improved usability to empower more users to report inappropriate posts (Sec. ~\ref{sec:Desiring Flagging Mechanisms to Support Greater Expressivity}).
One approach to extend awareness is to educate users on how to flag objectionable content effectively, why such flagging is important, and how their feedback is processed.
Platforms themselves can play an important role in such educational initiatives.
As noted by \citet{naab2018flagging}, platforms often fail to encourage user engagement by not providing clear, accessible information on flagging uncivil posts, highlighting the need for more visible guidance on flagging.
Therefore, it is vital to promote user participation through clear descriptions of flagging steps, offering incentives, and ensuring the simplicity and convenience of the flagging process.  
Additionally, establishing diverse channels for reporting inappropriate content, such as phone calls, emails, and chats, could help users with different technological competencies and preferences to flag in ways they find intuitive and accessible.

\subsubsection{Hosting Discussion Forums for Flagging}
%Forum
As participants suggested, platforms could also create special forums for discussions centered around flagging (Sec. ~\ref{sec:Desiring Flagging Mechanisms to Support Greater Expressivity}).
For example, such forums could host conversations about whether certain controversial posts should be flagged or how platforms' moderation policies and flagging classification schemes do not accommodate certain norm violations.
They could also offer users a converging space to discuss their flagging experiences, e.g., forum members could share their flagging history regarding the outcomes of posts they previously flagged.
Such spaces could allow users to appreciate diverse perspectives regarding content moderation and develop a shared understanding of how platforms respond to flagging efforts.

One challenge with hosting such forums is that a narrow set of influential individuals with strong viewpoints or even bad actors may unduly influence discussions about flagging norms, e.g., conversations about whether certain posts should be flagged \cite{gaozhao2021flagging}.
Therefore, platforms should carefully design such gatherings in collaboration with a diverse set of stakeholders, state their purpose clearly, and keep them well moderated.

\subsection{Addressing Online Harms Holistically}
\label{subsection:Outcome_fairness}
%Our study aimed to focus on procedural fairness
\subsubsection{Enacting Outcome Fairness in Flagging Mechanisms}
Our study investigated how different components of the flagging system influence users' perceptions of procedural fairness. 
Procedural fairness, as studied by several researchers \cite{sunshine2003role,tyler2006psychological, vaccaro2020end}, is characterized by its non-consequential nature and concerns users' experience of the procedural steps~\cite{ma2022m}.
%But.. the results suggest that users' perception of fairness is tied to the outcome.
However, our qualitative analysis revealed that many participants link fairness in flagging with its decision outcomes (Sec. ~\ref{sec:User Demands for Outcome Notification with Timely Review}). 

Specifically, participants observed that flagging mechanisms can enhance their fairness by providing clear information about outcomes accompanied by detailed reasoning for outcome decisions.
Some participants also indicated a need for a rebuttal system that lets flaggers challenge flagging outcomes (Sec. ~\ref{sec:Desiring Flagging Mechanisms to Support Greater Expressivity}).
Others were curious to see comprehensive statistics about the regulation of flagged content, including the proportion of submitted posts that are flagged and the ratio of flagged posts sanctioned within specific timeframes (Sec. ~\ref{sec:User Expectations Regarding Review Procedures and Statistics}).
These suggestions indicate that users conceptualize fairness of flagging mechanisms in a holistic manner, especially attending to flagging outcomes and how they are communicated as well as platform-wide measures associated with flagging.

%Educational measures to violators, thus preventing the re-occurrence of rule violation
\subsubsection{Encouraging Norm Compliance Among Flagged Users}
Beyond concerns about whether the flagged post was sanctioned~\citep{crawford2016flag}, flaggers may be invested in how effective their flagging efforts are at preventing further norm violations by the flagged users.
Our findings show that some users prefer educating rule-violating users rather than merely taking punitive measures against them, such as removing their content after flag review (Sec. ~\ref{sec:Concerns about Flagging Abuse and Suggestions to Address it}).

Therefore, platforms should consider investing in educational measures, such as helping the authors of flagged posts better understand the community rules and how to adhere to them.
As prior research shows, moderator messages that explain to the sanctioned users \textit{why} their posts are removed helps improve their attitudes and future behaviors~\cite{jhaver2019did,jhaver2019does}.
Highlighting, rewarding, and incentivizing desirable behaviors also reinforce constructive contributions~\cite{lambert2024positive,choi2024creator}.
Further, platforms may strengthen ex ante moderation measures \cite{grimmelmann2015virtues}, such as surfacing posting guidelines while a user is writing a post or using AI-based tools to warn users when their post draft is likely to be sanctioned~\cite{jerasa2024writing,katsaros2022reconsidering}.
In line with findings from \citet{zhang2023cleaning}, some of our respondents preferred directly communicating with rule-violators over reporting them to persuade norm compliance.
Such measures may reduce the burden of flagging on regular users and flag review on moderators.

% Additionally, platforms might consider adopting soft moderation measures (instead of content removal) in some cases, such as putting information labels to contextualize the information in flagged posts~\cite{morrow2022emerging} or enacting a strike system for flagged accounts. Such measures would contribute to educating flagged users and encouraging constructive posts in the future \cite{singhal2023sok}. 
% Platforms that adopt these approaches as outcomes of the flagging process would demonstrate their commitment to preventing future violations by utilizing the flagging mechanism.

%Uphold free speech
\subsubsection{Supporting Free Speech Values}
Our qualitative analysis also highlighted that upholding free speech values in moderation mechanisms is critical for many end-users (Sec. ~\ref{sec:Concerns about Flagging Abuse and Suggestions to Address it}). 
Some participants stressed that using flags as a tool to induce post removals may violate free speech principles.
As prior research points out, free speech proponents support the use of personal moderation tools like muting and blocking, especially when compared to platform-enacted moderation, because these tools affect only the configuring user's newsfeed ~\cite{jhaver2023users,jhaver2023personalizing}. 
Thus, platforms might consider informing users about these alternative options to address content-based harms~\cite{jhaver2022filterbuddy} and clarifying their distinct affordances when they attempt to use the flagging tools.

\subsection{Limitations and Future Work}
%Limitations
% The demographic profile of our study participants was skewed toward White individuals, which may not represent the overall user population.
During data collection, we did not ground our questions in a specific social media site to increase the generalizability of our findings. 
Each platform implements its flagging design differently and has established different levels of trust with its end-users.
Therefore, future work could adopt our methods to study flagging interactions in specific platforms and surface additional nuances.

We captured our participants' demographic details directly from Lucid's pre-collected data, which does not allow reporting of non-binary gender identities. Future studies using Lucid Theorem can capture this information using a separate question in the survey. 
While we undertook pre-processing steps to discard inappropriate participant responses, implementing attention checks in our survey questionnaire would have further improved the validity of this study.

Our experiments evaluated participants' flagging of a single inappropriate post. However, users' behaviors may vary in their daily social media use as they encounter multiple instances of norm violations.
Additionally, while the use of highly offensive stimuli in our survey creates a strong setting for evaluating flag mechanisms, less extreme content might shape flagging needs and actions differently.
Therefore, longitudinal or in situ analyses that examine how users interact with flagging tools in their day-to-day settings would be valuable to further inform the tradeoffs between incorporating fairness and reducing cognitive burden.

\section{Conclusion}
This paper examines how flag components that provide different types of information about the flag review process shape users' attitudes toward flagging.
Our analysis shows that including posting guidelines in flag designs enhances users' transparency perceptions and offering a text box improves their voice perceptions.
We found that 
users desire flag mechanisms to support greater expressivity, timely notifications of flag review updates, increased visibility into flag review procedures and reviewers, and stronger protections against flag abuse. 
We discuss how innovations in flagging systems, such as improving their accessibility and offering support for highlighting the severity of rule violations, could better support end-users' fairness demands.
This investigation demonstrates how the design and policy choices made in the implementation of flagging infrastructures deeply shape users' daily experiences of social media use and address (or fail to address) their vital needs to combat content-based harms. We call for future studies to deploy similar user-centered approaches and social justice orientations to improve the current practices of platform governance.

\begin{acks}
Awaiting paper acceptance.
\end{acks}

\bibliographystyle{ACM-Reference-Format}
\bibliography{references}


\begin{thebibliography}{93}

%%% the Web address.

\ifx \showCODEN    \undefined \def \showCODEN     #1{\unskip}     \fi
\ifx \showDOI      \undefined \def \showDOI       #1{#1}\fi
\ifx \showISBNx    \undefined \def \showISBNx     #1{\unskip}     \fi
\ifx \showISBNxiii \undefined \def \showISBNxiii  #1{\unskip}     \fi
\ifx \showISSN     \undefined \def \showISSN      #1{\unskip}     \fi
\ifx \showLCCN     \undefined \def \showLCCN      #1{\unskip}     \fi
\ifx \shownote     \undefined \def \shownote      #1{#1}          \fi
\ifx \showarticletitle \undefined \def \showarticletitle #1{#1}   \fi
\ifx \showURL      \undefined \def \showURL       {\relax}        \fi
% invisible to TeX
\providecommand\bibfield[2]{#2}
\providecommand\bibinfo[2]{#2}
\providecommand\natexlab[1]{#1}
\providecommand\showeprint[2][]{arXiv:#2}

\bibitem[Are(2023)]%
        {are2023flagging}
\bibfield{author}{\bibinfo{person}{Carolina Are}.} \bibinfo{year}{2023}\natexlab{}.
\newblock \showarticletitle{Flagging as a silencing tool: Exploring the relationship between de-platforming of sex and online abuse on Instagram and TikTok}.
\newblock \bibinfo{journal}{\emph{New Media \& Society}} (\bibinfo{year}{2023}), \bibinfo{pages}{14614448241228544}.
\newblock


\bibitem[Atreja et~al\mbox{.}(2023)]%
        {atreja2023remove}
\bibfield{author}{\bibinfo{person}{Shubham Atreja}, \bibinfo{person}{Libby Hemphill}, {and} \bibinfo{person}{Paul Resnick}.} \bibinfo{year}{2023}\natexlab{}.
\newblock \showarticletitle{Remove, reduce, inform: what actions do people want Social Media platforms to take on potentially misleading content?}
\newblock \bibinfo{journal}{\emph{Proceedings of the ACM on Human-Computer Interaction}} \bibinfo{volume}{7}, \bibinfo{number}{CSCW2} (\bibinfo{year}{2023}), \bibinfo{pages}{1--33}.
\newblock


\bibitem[Bradford et~al\mbox{.}(2019)]%
        {bradford2019report}
\bibfield{author}{\bibinfo{person}{Ben Bradford}, \bibinfo{person}{Florian Grisel}, \bibinfo{person}{Tracey~L Meares}, \bibinfo{person}{Emily Owens}, \bibinfo{person}{Baron~L Pineda}, \bibinfo{person}{Jacob Shapiro}, \bibinfo{person}{Tom~R Tyler}, {and} \bibinfo{person}{Danieli~Evans Peterman}.} \bibinfo{year}{2019}\natexlab{}.
\newblock \showarticletitle{Report of the Facebook data transparency advisory group}.
\newblock \bibinfo{journal}{\emph{Yale Justice Collaboratory}} (\bibinfo{year}{2019}).
\newblock


\bibitem[Braun and Clarke(2006)]%
        {braun2006using}
\bibfield{author}{\bibinfo{person}{Virginia Braun} {and} \bibinfo{person}{Victoria Clarke}.} \bibinfo{year}{2006}\natexlab{}.
\newblock \showarticletitle{Using thematic analysis in psychology}.
\newblock \bibinfo{journal}{\emph{Qualitative research in psychology}} \bibinfo{volume}{3}, \bibinfo{number}{2} (\bibinfo{year}{2006}), \bibinfo{pages}{77--101}.
\newblock


\bibitem[Breton et~al\mbox{.}(2007)]%
        {breton2007economics}
\bibfield{author}{\bibinfo{person}{Albert Breton}, \bibinfo{person}{Gianluigi Galeotti}, \bibinfo{person}{Pierre Salmon}, \bibinfo{person}{Ronald Wintrobe}, {et~al\mbox{.}}} \bibinfo{year}{2007}\natexlab{}.
\newblock \bibinfo{booktitle}{\emph{The economics of transparency in politics}}.
\newblock \bibinfo{publisher}{Ashgate Aldershot}.
\newblock


\bibitem[Chan~Kim and Mauborgne(1998)]%
        {chan1998procedural}
\bibfield{author}{\bibinfo{person}{W Chan~Kim} {and} \bibinfo{person}{Ren{\'e}e Mauborgne}.} \bibinfo{year}{1998}\natexlab{}.
\newblock \showarticletitle{Procedural justice, strategic decision making, and the knowledge economy}.
\newblock \bibinfo{journal}{\emph{Strategic management journal}} \bibinfo{volume}{19}, \bibinfo{number}{4} (\bibinfo{year}{1998}), \bibinfo{pages}{323--338}.
\newblock


\bibitem[Chandrasekharan et~al\mbox{.}(2019)]%
        {chandrasekharan2019crossmod}
\bibfield{author}{\bibinfo{person}{Eshwar Chandrasekharan}, \bibinfo{person}{Chaitrali Gandhi}, \bibinfo{person}{Matthew~Wortley Mustelier}, {and} \bibinfo{person}{Eric Gilbert}.} \bibinfo{year}{2019}\natexlab{}.
\newblock \showarticletitle{Crossmod: A cross-community learning-based system to assist reddit moderators}.
\newblock \bibinfo{journal}{\emph{Proceedings of the ACM on human-computer interaction}} \bibinfo{volume}{3}, \bibinfo{number}{CSCW} (\bibinfo{year}{2019}), \bibinfo{pages}{1--30}.
\newblock


\bibitem[Chandrasekharan et~al\mbox{.}(2018)]%
        {chandrasekharan2018norms}
\bibfield{author}{\bibinfo{person}{Eshwar Chandrasekharan}, \bibinfo{person}{Mattia Samory}, \bibinfo{person}{Shagun Jhaver}, \bibinfo{person}{Hunter Charvat}, \bibinfo{person}{Amy Bruckman}, \bibinfo{person}{Cliff Lampe}, \bibinfo{person}{Jacob Eisenstein}, {and} \bibinfo{person}{Eric Gilbert}.} \bibinfo{year}{2018}\natexlab{}.
\newblock \showarticletitle{The Internet's Hidden Rules: An Empirical Study of Reddit Norm Violations at Micro, Meso, and Macro Scales}.
\newblock \bibinfo{journal}{\emph{Proc. ACM Hum.-Comput. Interact.}} \bibinfo{volume}{2}, \bibinfo{number}{CSCW}, Article \bibinfo{articleno}{32} (\bibinfo{date}{nov} \bibinfo{year}{2018}), \bibinfo{numpages}{25}~pages.
\newblock
\urldef\tempurl%
\url{https://doi.org/10.1145/3274301}
\showDOI{\tempurl}


\bibitem[Chipidza and Yan(2022)]%
        {chipidza2022effectiveness}
\bibfield{author}{\bibinfo{person}{Wallace Chipidza} {and} \bibinfo{person}{Jie Yan}.} \bibinfo{year}{2022}\natexlab{}.
\newblock \showarticletitle{The effectiveness of flagging content belonging to prominent individuals: The case of Donald Trump on Twitter}.
\newblock \bibinfo{journal}{\emph{Journal of the Association for Information Science and Technology}} \bibinfo{volume}{73}, \bibinfo{number}{11} (\bibinfo{year}{2022}), \bibinfo{pages}{1641--1658}.
\newblock


\bibitem[Choi et~al\mbox{.}(2024)]%
        {choi2024creator}
\bibfield{author}{\bibinfo{person}{Frederick Choi}, \bibinfo{person}{Charlotte Lambert}, \bibinfo{person}{Vinay Koshy}, \bibinfo{person}{Sowmya Pratipati}, \bibinfo{person}{Tue Do}, {and} \bibinfo{person}{Eshwar Chandrasekharan}.} \bibinfo{year}{2024}\natexlab{}.
\newblock \showarticletitle{Creator Hearts: Investigating the Impact Positive Signals from YouTube Creators in Shaping Comment Section Behavior}.
\newblock \bibinfo{journal}{\emph{arXiv preprint arXiv:2404.03612}} (\bibinfo{year}{2024}).
\newblock


\bibitem[Corbin and Strauss(2015)]%
        {corbin2015basics}
\bibfield{author}{\bibinfo{person}{Juliet Corbin} {and} \bibinfo{person}{Anselm Strauss}.} \bibinfo{year}{2015}\natexlab{}.
\newblock \bibinfo{booktitle}{\emph{Basics of qualitative research}}. Vol.~\bibinfo{volume}{14}.
\newblock \bibinfo{publisher}{sage}.
\newblock


\bibitem[Crawford and Gillespie(2016)]%
        {crawford2016flag}
\bibfield{author}{\bibinfo{person}{Kate Crawford} {and} \bibinfo{person}{Tarleton Gillespie}.} \bibinfo{year}{2016}\natexlab{}.
\newblock \showarticletitle{What is a flag for? Social media reporting tools and the vocabulary of complaint}.
\newblock \bibinfo{journal}{\emph{New Media \& Society}} \bibinfo{volume}{18}, \bibinfo{number}{3} (\bibinfo{year}{2016}), \bibinfo{pages}{410--428}.
\newblock


\bibitem[Das et~al\mbox{.}(2021)]%
        {das2021jol}
\bibfield{author}{\bibinfo{person}{Dipto Das}, \bibinfo{person}{Carsten {\O}sterlund}, {and} \bibinfo{person}{Bryan Semaan}.} \bibinfo{year}{2021}\natexlab{}.
\newblock \showarticletitle{" Jol" or" Pani"?: How Does Governance Shape a Platform's Identity?}
\newblock \bibinfo{journal}{\emph{Proceedings of the ACM on Human-Computer Interaction}} \bibinfo{volume}{5}, \bibinfo{number}{CSCW2} (\bibinfo{year}{2021}), \bibinfo{pages}{1--25}.
\newblock


\bibitem[Davis(1989)]%
        {davis1989perceived}
\bibfield{author}{\bibinfo{person}{Fred~D Davis}.} \bibinfo{year}{1989}\natexlab{}.
\newblock \showarticletitle{Perceived usefulness, perceived ease of use, and user acceptance of information technology}.
\newblock \bibinfo{journal}{\emph{MIS quarterly}} (\bibinfo{year}{1989}), \bibinfo{pages}{319--340}.
\newblock


\bibitem[D{\'\i}az and Hecht-Felella(2021)]%
        {diaz2021double}
\bibfield{author}{\bibinfo{person}{{\'A}ngel D{\'\i}az} {and} \bibinfo{person}{Laura Hecht-Felella}.} \bibinfo{year}{2021}\natexlab{}.
\newblock \showarticletitle{Double standards in social media content moderation}.
\newblock \bibinfo{journal}{\emph{Brennan Center for Justice at New York University School of Law. https://www. brennancenter. org/our-work/research-reports/double-standards-socialmedia-content-moderation}} (\bibinfo{year}{2021}).
\newblock


\bibitem[DiFranzo et~al\mbox{.}(2018)]%
        {difranzo2018upstanding}
\bibfield{author}{\bibinfo{person}{Dominic DiFranzo}, \bibinfo{person}{Samuel~Hardman Taylor}, \bibinfo{person}{Franccesca Kazerooni}, \bibinfo{person}{Olivia~D Wherry}, {and} \bibinfo{person}{Natalya~N Bazarova}.} \bibinfo{year}{2018}\natexlab{}.
\newblock \showarticletitle{Upstanding by design: Bystander intervention in cyberbullying}. In \bibinfo{booktitle}{\emph{Proceedings of the 2018 CHI conference on human factors in computing systems}}. \bibinfo{pages}{1--12}.
\newblock


\bibitem[Dineva and Breitsohl(2022)]%
        {dineva2022managing}
\bibfield{author}{\bibinfo{person}{Denitsa Dineva} {and} \bibinfo{person}{Jan Breitsohl}.} \bibinfo{year}{2022}\natexlab{}.
\newblock \showarticletitle{Managing trolling in online communities: an organizational perspective}.
\newblock \bibinfo{journal}{\emph{Internet Research}} \bibinfo{volume}{32}, \bibinfo{number}{1} (\bibinfo{year}{2022}), \bibinfo{pages}{292--311}.
\newblock


\bibitem[Drahos(2017)]%
        {drahos2017regulatory}
\bibfield{author}{\bibinfo{person}{Peter Drahos}.} \bibinfo{year}{2017}\natexlab{}.
\newblock \bibinfo{booktitle}{\emph{Regulatory theory: Foundations and applications}}.
\newblock \bibinfo{publisher}{ANU Press}.
\newblock


\bibitem[Duguay et~al\mbox{.}(2020)]%
        {duguay2020queer}
\bibfield{author}{\bibinfo{person}{Stefanie Duguay}, \bibinfo{person}{Jean Burgess}, {and} \bibinfo{person}{Nicolas Suzor}.} \bibinfo{year}{2020}\natexlab{}.
\newblock \showarticletitle{Queer women’s experiences of patchwork platform governance on Tinder, Instagram, and Vine}.
\newblock \bibinfo{journal}{\emph{Convergence}} \bibinfo{volume}{26}, \bibinfo{number}{2} (\bibinfo{year}{2020}), \bibinfo{pages}{237--252}.
\newblock


\bibitem[Fan and Zhang(2020)]%
        {fan2020digital}
\bibfield{author}{\bibinfo{person}{Jenny Fan} {and} \bibinfo{person}{Amy~X Zhang}.} \bibinfo{year}{2020}\natexlab{}.
\newblock \showarticletitle{Digital juries: A civics-oriented approach to platform governance}. In \bibinfo{booktitle}{\emph{Proceedings of the 2020 CHI conference on human factors in computing systems}}. \bibinfo{pages}{1--14}.
\newblock


\bibitem[Feuston et~al\mbox{.}(2020)]%
        {feuston2020conformity}
\bibfield{author}{\bibinfo{person}{Jessica~L Feuston}, \bibinfo{person}{Alex~S Taylor}, {and} \bibinfo{person}{Anne~Marie Piper}.} \bibinfo{year}{2020}\natexlab{}.
\newblock \showarticletitle{Conformity of eating disorders through content moderation}.
\newblock \bibinfo{journal}{\emph{Proceedings of the ACM on Human-Computer Interaction}} \bibinfo{volume}{4}, \bibinfo{number}{CSCW1} (\bibinfo{year}{2020}), \bibinfo{pages}{1--28}.
\newblock


\bibitem[Gaozhao(2021)]%
        {gaozhao2021flagging}
\bibfield{author}{\bibinfo{person}{Dongfang Gaozhao}.} \bibinfo{year}{2021}\natexlab{}.
\newblock \showarticletitle{Flagging fake news on social media: An experimental study of media consumers' identification of fake news}.
\newblock \bibinfo{journal}{\emph{Government Information Quarterly}} \bibinfo{volume}{38}, \bibinfo{number}{3} (\bibinfo{year}{2021}), \bibinfo{pages}{101591}.
\newblock


\bibitem[Gilbert(2020)]%
        {gilbert2020run}
\bibfield{author}{\bibinfo{person}{Sarah~A Gilbert}.} \bibinfo{year}{2020}\natexlab{}.
\newblock \showarticletitle{" I run the world's largest historical outreach project and it's on a cesspool of a website." Moderating a Public Scholarship Site on Reddit: A Case Study of r/AskHistorians}.
\newblock \bibinfo{journal}{\emph{Proceedings of the ACM on Human-Computer Interaction}} \bibinfo{volume}{4}, \bibinfo{number}{CSCW1} (\bibinfo{year}{2020}), \bibinfo{pages}{1--27}.
\newblock


\bibitem[Gillespie(2018a)]%
        {gillespie2018custodians}
\bibfield{author}{\bibinfo{person}{Tarleton Gillespie}.} \bibinfo{year}{2018}\natexlab{a}.
\newblock \bibinfo{booktitle}{\emph{Custodians of the Internet: Platforms, content moderation, and the hidden decisions that shape social media}}.
\newblock \bibinfo{publisher}{Yale University Press}.
\newblock


\bibitem[Gillespie(2018b)]%
        {gillespie2018regulation}
\bibfield{author}{\bibinfo{person}{Tarleton Gillespie}.} \bibinfo{year}{2018}\natexlab{b}.
\newblock \showarticletitle{Regulation of and by platforms}.
\newblock \bibinfo{journal}{\emph{The SAGE handbook of social media}} (\bibinfo{year}{2018}), \bibinfo{pages}{254--278}.
\newblock


\bibitem[Gillespie(2020)]%
        {gillespie2020content}
\bibfield{author}{\bibinfo{person}{Tarleton Gillespie}.} \bibinfo{year}{2020}\natexlab{}.
\newblock \showarticletitle{Content moderation, AI, and the question of scale}.
\newblock \bibinfo{journal}{\emph{Big Data \& Society}} \bibinfo{volume}{7}, \bibinfo{number}{2} (\bibinfo{year}{2020}), \bibinfo{pages}{2053951720943234}.
\newblock


\bibitem[Goldman(2021)]%
        {goldman2021content}
\bibfield{author}{\bibinfo{person}{Eric Goldman}.} \bibinfo{year}{2021}\natexlab{}.
\newblock \showarticletitle{Content moderation remedies}.
\newblock \bibinfo{journal}{\emph{Mich. Tech. L. Rev.}}  \bibinfo{volume}{28} (\bibinfo{year}{2021}), \bibinfo{pages}{1}.
\newblock


\bibitem[Gon{\c{c}}alves et~al\mbox{.}(2021)]%
        {gonccalves2021common}
\bibfield{author}{\bibinfo{person}{Jo{\~a}o Gon{\c{c}}alves}, \bibinfo{person}{Ina Weber}, \bibinfo{person}{Gina~M Masullo}, \bibinfo{person}{Marisa Torres~da Silva}, {and} \bibinfo{person}{Joep Hofhuis}.} \bibinfo{year}{2021}\natexlab{}.
\newblock \showarticletitle{Common sense or censorship: How algorithmic moderators and message type influence perceptions of online content deletion}.
\newblock \bibinfo{journal}{\emph{new media \& society}} (\bibinfo{year}{2021}), \bibinfo{pages}{14614448211032310}.
\newblock


\bibitem[Gorwa et~al\mbox{.}(2020)]%
        {gorwa2020algorithmic}
\bibfield{author}{\bibinfo{person}{Robert Gorwa}, \bibinfo{person}{Reuben Binns}, {and} \bibinfo{person}{Christian Katzenbach}.} \bibinfo{year}{2020}\natexlab{}.
\newblock \showarticletitle{Algorithmic content moderation: Technical and political challenges in the automation of platform governance}.
\newblock \bibinfo{journal}{\emph{Big Data \& Society}} \bibinfo{volume}{7}, \bibinfo{number}{1} (\bibinfo{year}{2020}), \bibinfo{pages}{2053951719897945}.
\newblock


\bibitem[Griffin(2022)]%
        {griffin2022sanitised}
\bibfield{author}{\bibinfo{person}{Rachel Griffin}.} \bibinfo{year}{2022}\natexlab{}.
\newblock \showarticletitle{The sanitised platform}.
\newblock \bibinfo{journal}{\emph{J. Intell. Prop. Info. Tech. \& Elec. Com. L.}}  \bibinfo{volume}{13} (\bibinfo{year}{2022}), \bibinfo{pages}{36}.
\newblock


\bibitem[Grimmelmann(2015)]%
        {grimmelmann2015virtues}
\bibfield{author}{\bibinfo{person}{James Grimmelmann}.} \bibinfo{year}{2015}\natexlab{}.
\newblock \showarticletitle{The virtues of moderation}.
\newblock \bibinfo{journal}{\emph{Yale JL \& Tech.}}  \bibinfo{volume}{17} (\bibinfo{year}{2015}), \bibinfo{pages}{42}.
\newblock


\bibitem[Haimson et~al\mbox{.}(2021)]%
        {haimson2021disproportionate}
\bibfield{author}{\bibinfo{person}{Oliver~L Haimson}, \bibinfo{person}{Daniel Delmonaco}, \bibinfo{person}{Peipei Nie}, {and} \bibinfo{person}{Andrea Wegner}.} \bibinfo{year}{2021}\natexlab{}.
\newblock \showarticletitle{Disproportionate removals and differing content moderation experiences for conservative, transgender, and black social media users: Marginalization and moderation gray areas}.
\newblock \bibinfo{journal}{\emph{Proceedings of the ACM on Human-Computer Interaction}} \bibinfo{volume}{5}, \bibinfo{number}{CSCW2} (\bibinfo{year}{2021}), \bibinfo{pages}{1--35}.
\newblock


\bibitem[Hartmann(2020)]%
        {hartmann2020framework}
\bibfield{author}{\bibinfo{person}{Ivar~A. Hartmann}.} \bibinfo{year}{2020}\natexlab{}.
\newblock \showarticletitle{A new framework for online content moderation}.
\newblock \bibinfo{journal}{\emph{Computer Law \& Security Review}}  \bibinfo{volume}{36} (\bibinfo{year}{2020}), \bibinfo{pages}{105376}.
\newblock
\showISSN{0267-3649}
\urldef\tempurl%
\url{https://doi.org/10.1016/j.clsr.2019.105376}
\showDOI{\tempurl}


\bibitem[Helberger et~al\mbox{.}(2018)]%
        {helberger2018governing}
\bibfield{author}{\bibinfo{person}{Natali Helberger}, \bibinfo{person}{Jo Pierson}, {and} \bibinfo{person}{Thomas Poell}.} \bibinfo{year}{2018}\natexlab{}.
\newblock \showarticletitle{Governing online platforms: From contested to cooperative responsibility}.
\newblock \bibinfo{journal}{\emph{The information society}} \bibinfo{volume}{34}, \bibinfo{number}{1} (\bibinfo{year}{2018}), \bibinfo{pages}{1--14}.
\newblock


\bibitem[Ivankova and Creswell(2009)]%
        {ivankova2009mixed}
\bibfield{author}{\bibinfo{person}{Nataliya~V Ivankova} {and} \bibinfo{person}{John~W Creswell}.} \bibinfo{year}{2009}\natexlab{}.
\newblock \showarticletitle{Mixed methods}.
\newblock \bibinfo{journal}{\emph{Qualitative research in applied linguistics: A practical introduction}}  \bibinfo{volume}{23} (\bibinfo{year}{2009}), \bibinfo{pages}{135--161}.
\newblock


\bibitem[Jane(2017)]%
        {jane2017dude}
\bibfield{author}{\bibinfo{person}{Emma~A Jane}.} \bibinfo{year}{2017}\natexlab{}.
\newblock \showarticletitle{‘Dude… stop the spread’: antagonism, agonism, and\# manspreading on social media}.
\newblock \bibinfo{journal}{\emph{International Journal of Cultural Studies}} \bibinfo{volume}{20}, \bibinfo{number}{5} (\bibinfo{year}{2017}), \bibinfo{pages}{459--475}.
\newblock


\bibitem[Jerasa and Burriss(2024)]%
        {jerasa2024writing}
\bibfield{author}{\bibinfo{person}{Sarah Jerasa} {and} \bibinfo{person}{Sarah~K Burriss}.} \bibinfo{year}{2024}\natexlab{}.
\newblock \showarticletitle{Writing with, for, and against the algorithm: TikTokers’ relationships with AI as audience, co-author, and censor}.
\newblock \bibinfo{journal}{\emph{English Teaching: Practice \& Critique}} \bibinfo{volume}{23}, \bibinfo{number}{1} (\bibinfo{year}{2024}), \bibinfo{pages}{118--134}.
\newblock


\bibitem[Jhaver et~al\mbox{.}(2019a)]%
        {jhaver2019did}
\bibfield{author}{\bibinfo{person}{Shagun Jhaver}, \bibinfo{person}{Darren~Scott Appling}, \bibinfo{person}{Eric Gilbert}, {and} \bibinfo{person}{Amy Bruckman}.} \bibinfo{year}{2019}\natexlab{a}.
\newblock \showarticletitle{"Did you suspect the post would be removed?" Understanding user reactions to content removals on Reddit}.
\newblock \bibinfo{journal}{\emph{Proceedings of the ACM on human-computer interaction}} \bibinfo{volume}{3}, \bibinfo{number}{CSCW} (\bibinfo{year}{2019}), \bibinfo{pages}{1--33}.
\newblock


\bibitem[Jhaver et~al\mbox{.}(2019b)]%
        {jhaver2019human}
\bibfield{author}{\bibinfo{person}{Shagun Jhaver}, \bibinfo{person}{Iris Birman}, \bibinfo{person}{Eric Gilbert}, {and} \bibinfo{person}{Amy Bruckman}.} \bibinfo{year}{2019}\natexlab{b}.
\newblock \showarticletitle{Human-machine collaboration for content regulation: The case of reddit automoderator}.
\newblock \bibinfo{journal}{\emph{ACM Transactions on Computer-Human Interaction (TOCHI)}} \bibinfo{volume}{26}, \bibinfo{number}{5} (\bibinfo{year}{2019}), \bibinfo{pages}{1--35}.
\newblock


\bibitem[Jhaver et~al\mbox{.}(2019c)]%
        {jhaver2019does}
\bibfield{author}{\bibinfo{person}{Shagun Jhaver}, \bibinfo{person}{Amy Bruckman}, {and} \bibinfo{person}{Eric Gilbert}.} \bibinfo{year}{2019}\natexlab{c}.
\newblock \showarticletitle{Does transparency in moderation really matter? User behavior after content removal explanations on reddit}.
\newblock \bibinfo{journal}{\emph{Proceedings of the ACM on Human-Computer Interaction}} \bibinfo{volume}{3}, \bibinfo{number}{CSCW} (\bibinfo{year}{2019}), \bibinfo{pages}{1--27}.
\newblock


\bibitem[Jhaver et~al\mbox{.}(2022)]%
        {jhaver2022filterbuddy}
\bibfield{author}{\bibinfo{person}{Shagun Jhaver}, \bibinfo{person}{Quan~Ze Chen}, \bibinfo{person}{Detlef Knauss}, {and} \bibinfo{person}{Amy~X. Zhang}.} \bibinfo{year}{2022}\natexlab{}.
\newblock \showarticletitle{Designing Word Filter Tools for Creator-led Comment Moderation}. In \bibinfo{booktitle}{\emph{Proceedings of the 2022 CHI Conference on Human Factors in Computing Systems}} (New Orleans, LA, USA) \emph{(\bibinfo{series}{CHI '22})}. \bibinfo{publisher}{Association for Computing Machinery}, \bibinfo{address}{New York, NY, USA}, Article \bibinfo{articleno}{205}, \bibinfo{numpages}{21}~pages.
\newblock
\showISBNx{9781450391573}
\urldef\tempurl%
\url{https://doi.org/10.1145/3491102.3517505}
\showDOI{\tempurl}


\bibitem[Jhaver et~al\mbox{.}(2023)]%
        {jhaver2023personalizing}
\bibfield{author}{\bibinfo{person}{Shagun Jhaver}, \bibinfo{person}{Alice~Qian Zhang}, \bibinfo{person}{Quan~Ze Chen}, \bibinfo{person}{Nikhila Natarajan}, \bibinfo{person}{Ruotong Wang}, {and} \bibinfo{person}{Amy~X Zhang}.} \bibinfo{year}{2023}\natexlab{}.
\newblock \showarticletitle{Personalizing content moderation on social media: User perspectives on moderation choices, interface design, and labor}.
\newblock \bibinfo{journal}{\emph{Proceedings of the ACM on Human-Computer Interaction}} \bibinfo{volume}{7}, \bibinfo{number}{CSCW2} (\bibinfo{year}{2023}), \bibinfo{pages}{1--33}.
\newblock


\bibitem[Jhaver and Zhang(2023)]%
        {jhaver2023users}
\bibfield{author}{\bibinfo{person}{Shagun Jhaver} {and} \bibinfo{person}{Amy~X Zhang}.} \bibinfo{year}{2023}\natexlab{}.
\newblock \showarticletitle{Do users want platform moderation or individual control? Examining the role of third-person effects and free speech support in shaping moderation preferences}.
\newblock \bibinfo{journal}{\emph{New Media \& Society}} (\bibinfo{year}{2023}), \bibinfo{pages}{14614448231217993}.
\newblock


\bibitem[Jiang et~al\mbox{.}(2023)]%
        {jiang2023trade}
\bibfield{author}{\bibinfo{person}{Jialun~Aaron Jiang}, \bibinfo{person}{Peipei Nie}, \bibinfo{person}{Jed~R Brubaker}, {and} \bibinfo{person}{Casey Fiesler}.} \bibinfo{year}{2023}\natexlab{}.
\newblock \showarticletitle{A trade-off-centered framework of content moderation}.
\newblock \bibinfo{journal}{\emph{ACM Transactions on Computer-Human Interaction}} \bibinfo{volume}{30}, \bibinfo{number}{1} (\bibinfo{year}{2023}), \bibinfo{pages}{1--34}.
\newblock


\bibitem[Katsaros et~al\mbox{.}(2024)]%
        {katsaros2024online}
\bibfield{author}{\bibinfo{person}{Matthew Katsaros}, \bibinfo{person}{Jisu Kim}, {and} \bibinfo{person}{Tom Tyler}.} \bibinfo{year}{2024}\natexlab{}.
\newblock \showarticletitle{Online content moderation: does justice need a human face?}
\newblock \bibinfo{journal}{\emph{International Journal of Human--Computer Interaction}} \bibinfo{volume}{40}, \bibinfo{number}{1} (\bibinfo{year}{2024}), \bibinfo{pages}{66--77}.
\newblock


\bibitem[Katsaros et~al\mbox{.}(2022a)]%
        {katsaros2022procedural}
\bibfield{author}{\bibinfo{person}{Matthew Katsaros}, \bibinfo{person}{Tom Tyler}, \bibinfo{person}{Jisu Kim}, {and} \bibinfo{person}{Tracey Meares}.} \bibinfo{year}{2022}\natexlab{a}.
\newblock \showarticletitle{Procedural justice and self governance on Twitter: Unpacking the experience of rule breaking on Twitter}.
\newblock \bibinfo{journal}{\emph{Journal of Online Trust and Safety}} \bibinfo{volume}{1}, \bibinfo{number}{3} (\bibinfo{year}{2022}).
\newblock


\bibitem[Katsaros et~al\mbox{.}(2022b)]%
        {katsaros2022reconsidering}
\bibfield{author}{\bibinfo{person}{Matthew Katsaros}, \bibinfo{person}{Kathy Yang}, {and} \bibinfo{person}{Lauren Fratamico}.} \bibinfo{year}{2022}\natexlab{b}.
\newblock \showarticletitle{Reconsidering tweets: Intervening during tweet creation decreases offensive content}. In \bibinfo{booktitle}{\emph{Proceedings of the International AAAI Conference on Web and Social Media}}, Vol.~\bibinfo{volume}{16}. \bibinfo{pages}{477--487}.
\newblock


\bibitem[Kiesler et~al\mbox{.}(2012)]%
        {kiesler2012regulating}
\bibfield{author}{\bibinfo{person}{Sara Kiesler}, \bibinfo{person}{Robert Kraut}, \bibinfo{person}{Paul Resnick}, {and} \bibinfo{person}{Aniket Kittur}.} \bibinfo{year}{2012}\natexlab{}.
\newblock \showarticletitle{Regulating behavior in online communities}.
\newblock \bibinfo{journal}{\emph{Building successful online communities: Evidence-based social design}}  \bibinfo{volume}{1} (\bibinfo{year}{2012}), \bibinfo{pages}{4--2}.
\newblock


\bibitem[Kou(2021)]%
        {kou2021punishment}
\bibfield{author}{\bibinfo{person}{Yubo Kou}.} \bibinfo{year}{2021}\natexlab{}.
\newblock \showarticletitle{Punishment and Its Discontents: An Analysis of Permanent Ban in an Online Game Community}.
\newblock \bibinfo{journal}{\emph{Proceedings of the ACM on Human-Computer Interaction}} \bibinfo{volume}{5}, \bibinfo{number}{CSCW2} (\bibinfo{year}{2021}), \bibinfo{pages}{1--21}.
\newblock


\bibitem[Kou and Gui(2021)]%
        {kou2021flag}
\bibfield{author}{\bibinfo{person}{Yubo Kou} {and} \bibinfo{person}{Xinning Gui}.} \bibinfo{year}{2021}\natexlab{}.
\newblock \showarticletitle{Flag and flaggability in automated moderation: The case of reporting toxic behavior in an online game community}. In \bibinfo{booktitle}{\emph{Proceedings of the 2021 CHI Conference on Human Factors in Computing Systems}}. \bibinfo{pages}{1--12}.
\newblock


\bibitem[Kraut and Resnick(2012)]%
        {kraut2012building}
\bibfield{author}{\bibinfo{person}{Robert~E Kraut} {and} \bibinfo{person}{Paul Resnick}.} \bibinfo{year}{2012}\natexlab{}.
\newblock \bibinfo{booktitle}{\emph{Building successful online communities: Evidence-based social design}}.
\newblock \bibinfo{publisher}{Mit Press}.
\newblock


\bibitem[Lambert et~al\mbox{.}(2024)]%
        {lambert2024positive}
\bibfield{author}{\bibinfo{person}{Charlotte Lambert}, \bibinfo{person}{Frederick Choi}, {and} \bibinfo{person}{Eshwar Chandrasekharan}.} \bibinfo{year}{2024}\natexlab{}.
\newblock \showarticletitle{“Positive reinforcement helps breed positive behavior”: Moderator Perspectives on Encouraging Desirable Behavior}.
\newblock  (\bibinfo{year}{2024}).
\newblock


\bibitem[Langvardt(2017)]%
        {langvardt2017regulating}
\bibfield{author}{\bibinfo{person}{Kyle Langvardt}.} \bibinfo{year}{2017}\natexlab{}.
\newblock \showarticletitle{Regulating online content moderation}.
\newblock \bibinfo{journal}{\emph{Geo. LJ}}  \bibinfo{volume}{106} (\bibinfo{year}{2017}), \bibinfo{pages}{1353}.
\newblock


\bibitem[Lee et~al\mbox{.}(2019)]%
        {lee2019procedural}
\bibfield{author}{\bibinfo{person}{Min~Kyung Lee}, \bibinfo{person}{Anuraag Jain}, \bibinfo{person}{Hea~Jin Cha}, \bibinfo{person}{Shashank Ojha}, {and} \bibinfo{person}{Daniel Kusbit}.} \bibinfo{year}{2019}\natexlab{}.
\newblock \showarticletitle{Procedural justice in algorithmic fairness: Leveraging transparency and outcome control for fair algorithmic mediation}.
\newblock \bibinfo{journal}{\emph{Proceedings of the ACM on Human-Computer Interaction}} \bibinfo{volume}{3}, \bibinfo{number}{CSCW} (\bibinfo{year}{2019}), \bibinfo{pages}{1--26}.
\newblock


\bibitem[Li et~al\mbox{.}(2022)]%
        {li2022monetary}
\bibfield{author}{\bibinfo{person}{Hanlin Li}, \bibinfo{person}{Brent Hecht}, {and} \bibinfo{person}{Stevie Chancellor}.} \bibinfo{year}{2022}\natexlab{}.
\newblock \showarticletitle{Measuring the Monetary Value of Online Volunteer Work}.
\newblock \bibinfo{journal}{\emph{Proceedings of the International AAAI Conference on Web and Social Media}} \bibinfo{volume}{16}, \bibinfo{number}{1} (\bibinfo{date}{May} \bibinfo{year}{2022}), \bibinfo{pages}{596--606}.
\newblock
\urldef\tempurl%
\url{https://doi.org/10.1609/icwsm.v16i1.19318}
\showDOI{\tempurl}


\bibitem[Lyons et~al\mbox{.}(2022)]%
        {lyons2022s}
\bibfield{author}{\bibinfo{person}{Henrietta Lyons}, \bibinfo{person}{Senuri Wijenayake}, \bibinfo{person}{Tim Miller}, {and} \bibinfo{person}{Eduardo Velloso}.} \bibinfo{year}{2022}\natexlab{}.
\newblock \showarticletitle{What’s the appeal? Perceptions of review processes for algorithmic decisions}. In \bibinfo{booktitle}{\emph{Proceedings of the 2022 CHI Conference on Human Factors in Computing Systems}}. \bibinfo{pages}{1--15}.
\newblock


\bibitem[Lyu et~al\mbox{.}(2024)]%
        {lyu2024blindtokers}
\bibfield{author}{\bibinfo{person}{Yao Lyu}, \bibinfo{person}{Jie Cai}, \bibinfo{person}{Anisa Callis}, \bibinfo{person}{Kelley Cotter}, {and} \bibinfo{person}{John~M. Carroll}.} \bibinfo{year}{2024}\natexlab{}.
\newblock \showarticletitle{"I Got Flagged for Supposed Bullying, Even Though It Was in Response to Someone Harassing Me About My Disability.": A Study of Blind TikTokers’ Content Moderation Experiences}. In \bibinfo{booktitle}{\emph{Proceedings of the CHI Conference on Human Factors in Computing Systems}} (Honolulu, HI, USA) \emph{(\bibinfo{series}{CHI '24})}. \bibinfo{publisher}{Association for Computing Machinery}, \bibinfo{address}{New York, NY, USA}, Article \bibinfo{articleno}{741}, \bibinfo{numpages}{15}~pages.
\newblock
\showISBNx{9798400703300}
\urldef\tempurl%
\url{https://doi.org/10.1145/3613904.3642148}
\showDOI{\tempurl}


\bibitem[Ma and Kou(2022)]%
        {ma2022m}
\bibfield{author}{\bibinfo{person}{Renkai Ma} {and} \bibinfo{person}{Yubo Kou}.} \bibinfo{year}{2022}\natexlab{}.
\newblock \showarticletitle{" I'm not sure what difference is between their content and mine, other than the person itself" A Study of Fairness Perception of Content Moderation on YouTube}.
\newblock \bibinfo{journal}{\emph{Proceedings of the ACM on Human-Computer Interaction}} \bibinfo{volume}{6}, \bibinfo{number}{CSCW2} (\bibinfo{year}{2022}), \bibinfo{pages}{1--28}.
\newblock


\bibitem[Ma et~al\mbox{.}(2023a)]%
        {ma2023transparency}
\bibfield{author}{\bibinfo{person}{Renkai Ma}, \bibinfo{person}{Yao Li}, {and} \bibinfo{person}{Yubo Kou}.} \bibinfo{year}{2023}\natexlab{a}.
\newblock \showarticletitle{Transparency, Fairness, and Coping: How Players Experience Moderation in Multiplayer Online Games}. In \bibinfo{booktitle}{\emph{Proceedings of the 2023 CHI Conference on Human Factors in Computing Systems}}. \bibinfo{pages}{1--21}.
\newblock


\bibitem[Ma et~al\mbox{.}(2023b)]%
        {ma2023litreview}
\bibfield{author}{\bibinfo{person}{Renkai Ma}, \bibinfo{person}{Yue You}, \bibinfo{person}{Xinning Gui}, {and} \bibinfo{person}{Yubo Kou}.} \bibinfo{year}{2023}\natexlab{b}.
\newblock \showarticletitle{How Do Users Experience Moderation?: A Systematic Literature Review}.
\newblock \bibinfo{journal}{\emph{Proc. ACM Hum.-Comput. Interact.}} \bibinfo{volume}{7}, \bibinfo{number}{CSCW2}, Article \bibinfo{articleno}{278} (\bibinfo{date}{oct} \bibinfo{year}{2023}), \bibinfo{numpages}{30}~pages.
\newblock
\urldef\tempurl%
\url{https://doi.org/10.1145/3610069}
\showDOI{\tempurl}


\bibitem[Maddox and Malson(2020)]%
        {maddox2020guidelines}
\bibfield{author}{\bibinfo{person}{Jessica Maddox} {and} \bibinfo{person}{Jennifer Malson}.} \bibinfo{year}{2020}\natexlab{}.
\newblock \showarticletitle{Guidelines without lines, communities without borders: The marketplace of ideas and digital manifest destiny in social media platform policies}.
\newblock \bibinfo{journal}{\emph{Social Media+ Society}} \bibinfo{volume}{6}, \bibinfo{number}{2} (\bibinfo{year}{2020}), \bibinfo{pages}{2056305120926622}.
\newblock


\bibitem[Matias(2019)]%
        {matias2019preventing}
\bibfield{author}{\bibinfo{person}{J~Nathan Matias}.} \bibinfo{year}{2019}\natexlab{}.
\newblock \showarticletitle{Preventing harassment and increasing group participation through social norms in 2,190 online science discussions}.
\newblock \bibinfo{journal}{\emph{Proceedings of the National Academy of Sciences}} \bibinfo{volume}{116}, \bibinfo{number}{20} (\bibinfo{year}{2019}), \bibinfo{pages}{9785--9789}.
\newblock


\bibitem[Matthew~Katsaros and Tyler(2024)]%
        {Katsaros2024justice}
\bibfield{author}{\bibinfo{person}{Jisu~Kim Matthew~Katsaros} {and} \bibinfo{person}{Tom Tyler}.} \bibinfo{year}{2024}\natexlab{}.
\newblock \showarticletitle{Online Content Moderation: Does Justice Need a Human Face?}
\newblock \bibinfo{journal}{\emph{International Journal of Human--Computer Interaction}} \bibinfo{volume}{40}, \bibinfo{number}{1} (\bibinfo{year}{2024}), \bibinfo{pages}{66--77}.
\newblock
\urldef\tempurl%
\url{https://doi.org/10.1080/10447318.2023.2210879}
\showDOI{\tempurl}
\showeprint{https://doi.org/10.1080/10447318.2023.2210879}


\bibitem[Molina and Sundar(2022)]%
        {molina2022ai}
\bibfield{author}{\bibinfo{person}{Maria~D Molina} {and} \bibinfo{person}{S~Shyam Sundar}.} \bibinfo{year}{2022}\natexlab{}.
\newblock \showarticletitle{When AI moderates online content: effects of human collaboration and interactive transparency on user trust}.
\newblock \bibinfo{journal}{\emph{Journal of Computer-Mediated Communication}} \bibinfo{volume}{27}, \bibinfo{number}{4} (\bibinfo{year}{2022}), \bibinfo{pages}{zmac010}.
\newblock


\bibitem[Myers~West(2018)]%
        {myers2018censored}
\bibfield{author}{\bibinfo{person}{Sarah Myers~West}.} \bibinfo{year}{2018}\natexlab{}.
\newblock \showarticletitle{Censored, suspended, shadowbanned: User interpretations of content moderation on social media platforms}.
\newblock \bibinfo{journal}{\emph{New Media \& Society}} \bibinfo{volume}{20}, \bibinfo{number}{11} (\bibinfo{year}{2018}), \bibinfo{pages}{4366--4383}.
\newblock


\bibitem[Naab et~al\mbox{.}(2018)]%
        {naab2018flagging}
\bibfield{author}{\bibinfo{person}{Teresa~K Naab}, \bibinfo{person}{Anja Kalch}, {and} \bibinfo{person}{Tino~GK Meitz}.} \bibinfo{year}{2018}\natexlab{}.
\newblock \showarticletitle{Flagging uncivil user comments: Effects of intervention information, type of victim, and response comments on bystander behavior}.
\newblock \bibinfo{journal}{\emph{New Media \& Society}} \bibinfo{volume}{20}, \bibinfo{number}{2} (\bibinfo{year}{2018}), \bibinfo{pages}{777--795}.
\newblock


\bibitem[Nycyk(2016)]%
        {nycyk2016enforcing}
\bibfield{author}{\bibinfo{person}{Michael Nycyk}.} \bibinfo{year}{2016}\natexlab{}.
\newblock \showarticletitle{Enforcing community guidelines in web-based communities: the case of flame comments on YouTube}.
\newblock \bibinfo{journal}{\emph{International Journal of Web Based Communities}} \bibinfo{volume}{12}, \bibinfo{number}{2} (\bibinfo{year}{2016}), \bibinfo{pages}{131--146}.
\newblock


\bibitem[Ozanne et~al\mbox{.}(2022)]%
        {ozanne2022shall}
\bibfield{author}{\bibinfo{person}{Marie Ozanne}, \bibinfo{person}{Aparajita Bhandari}, \bibinfo{person}{Natalya~N Bazarova}, {and} \bibinfo{person}{Dominic DiFranzo}.} \bibinfo{year}{2022}\natexlab{}.
\newblock \showarticletitle{Shall AI moderators be made visible? Perception of accountability and trust in moderation systems on social media platforms}.
\newblock \bibinfo{journal}{\emph{Big Data \& Society}} \bibinfo{volume}{9}, \bibinfo{number}{2} (\bibinfo{year}{2022}), \bibinfo{pages}{20539517221115666}.
\newblock


\bibitem[Pan et~al\mbox{.}(2022)]%
        {pan2022comparing}
\bibfield{author}{\bibinfo{person}{Christina~A Pan}, \bibinfo{person}{Sahil Yakhmi}, \bibinfo{person}{Tara~P Iyer}, \bibinfo{person}{Evan Strasnick}, \bibinfo{person}{Amy~X Zhang}, {and} \bibinfo{person}{Michael~S Bernstein}.} \bibinfo{year}{2022}\natexlab{}.
\newblock \showarticletitle{Comparing the perceived legitimacy of content moderation processes: Contractors, algorithms, expert panels, and digital juries}.
\newblock \bibinfo{journal}{\emph{Proceedings of the ACM on Human-Computer Interaction}} \bibinfo{volume}{6}, \bibinfo{number}{CSCW1} (\bibinfo{year}{2022}), \bibinfo{pages}{1--31}.
\newblock


\bibitem[Rader et~al\mbox{.}(2018)]%
        {rader2018explanations}
\bibfield{author}{\bibinfo{person}{Emilee Rader}, \bibinfo{person}{Kelley Cotter}, {and} \bibinfo{person}{Janghee Cho}.} \bibinfo{year}{2018}\natexlab{}.
\newblock \showarticletitle{Explanations as mechanisms for supporting algorithmic transparency}. In \bibinfo{booktitle}{\emph{Proceedings of the 2018 CHI conference on human factors in computing systems}}. \bibinfo{pages}{1--13}.
\newblock


\bibitem[Roberts(2016)]%
        {roberts2016commercial}
\bibfield{author}{\bibinfo{person}{Sarah~T Roberts}.} \bibinfo{year}{2016}\natexlab{}.
\newblock \showarticletitle{Commercial content moderation: Digital laborers' dirty work}.
\newblock  (\bibinfo{year}{2016}).
\newblock


\bibitem[Scheuerman et~al\mbox{.}(2021)]%
        {scheuerman2021framework}
\bibfield{author}{\bibinfo{person}{Morgan~Klaus Scheuerman}, \bibinfo{person}{Jialun~Aaron Jiang}, \bibinfo{person}{Casey Fiesler}, {and} \bibinfo{person}{Jed~R. Brubaker}.} \bibinfo{year}{2021}\natexlab{}.
\newblock \showarticletitle{A Framework of Severity for Harmful Content Online}.
\newblock \bibinfo{journal}{\emph{Proc. ACM Hum.-Comput. Interact.}} \bibinfo{volume}{5}, \bibinfo{number}{CSCW2}, Article \bibinfo{articleno}{368} (\bibinfo{date}{oct} \bibinfo{year}{2021}), \bibinfo{numpages}{33}~pages.
\newblock
\urldef\tempurl%
\url{https://doi.org/10.1145/3479512}
\showDOI{\tempurl}


\bibitem[Schoenebeck et~al\mbox{.}(2021)]%
        {schoenebeck2021drawing}
\bibfield{author}{\bibinfo{person}{Sarita Schoenebeck}, \bibinfo{person}{Oliver~L Haimson}, {and} \bibinfo{person}{Lisa Nakamura}.} \bibinfo{year}{2021}\natexlab{}.
\newblock \showarticletitle{Drawing from justice theories to support targets of online harassment}.
\newblock \bibinfo{journal}{\emph{new media \& society}} \bibinfo{volume}{23}, \bibinfo{number}{5} (\bibinfo{year}{2021}), \bibinfo{pages}{1278--1300}.
\newblock


\bibitem[Seering(2020)]%
        {seering2020reconsidering}
\bibfield{author}{\bibinfo{person}{Joseph Seering}.} \bibinfo{year}{2020}\natexlab{}.
\newblock \showarticletitle{Reconsidering self-moderation: the role of research in supporting community-based models for online content moderation}.
\newblock \bibinfo{journal}{\emph{Proceedings of the ACM on Human-Computer Interaction}} \bibinfo{volume}{4}, \bibinfo{number}{CSCW2} (\bibinfo{year}{2020}), \bibinfo{pages}{1--28}.
\newblock


\bibitem[Seering et~al\mbox{.}(2017)]%
        {seering2017shaping}
\bibfield{author}{\bibinfo{person}{Joseph Seering}, \bibinfo{person}{Robert Kraut}, {and} \bibinfo{person}{Laura Dabbish}.} \bibinfo{year}{2017}\natexlab{}.
\newblock \showarticletitle{Shaping pro and anti-social behavior on twitch through moderation and example-setting}. In \bibinfo{booktitle}{\emph{Proceedings of the 2017 ACM conference on computer supported cooperative work and social computing}}. \bibinfo{pages}{111--125}.
\newblock


\bibitem[Shahid and Vashistha(2023)]%
        {shahid2023decolonizing}
\bibfield{author}{\bibinfo{person}{Farhana Shahid} {and} \bibinfo{person}{Aditya Vashistha}.} \bibinfo{year}{2023}\natexlab{}.
\newblock \showarticletitle{Decolonizing Content Moderation: Does Uniform Global Community Standard Resemble Utopian Equality or Western Power Hegemony?}. In \bibinfo{booktitle}{\emph{Proceedings of the 2023 CHI Conference on Human Factors in Computing Systems}}. \bibinfo{pages}{1--18}.
\newblock


\bibitem[Shin and Park(2019)]%
        {shin2019role}
\bibfield{author}{\bibinfo{person}{Donghee Shin} {and} \bibinfo{person}{Yong~Jin Park}.} \bibinfo{year}{2019}\natexlab{}.
\newblock \showarticletitle{Role of fairness, accountability, and transparency in algorithmic affordance}.
\newblock \bibinfo{journal}{\emph{Computers in Human Behavior}}  \bibinfo{volume}{98} (\bibinfo{year}{2019}), \bibinfo{pages}{277--284}.
\newblock


\bibitem[Sunshine and Tyler(2003)]%
        {sunshine2003role}
\bibfield{author}{\bibinfo{person}{Jason Sunshine} {and} \bibinfo{person}{Tom~R Tyler}.} \bibinfo{year}{2003}\natexlab{}.
\newblock \showarticletitle{The role of procedural justice and legitimacy in shaping public support for policing}.
\newblock \bibinfo{journal}{\emph{Law \& society review}} \bibinfo{volume}{37}, \bibinfo{number}{3} (\bibinfo{year}{2003}), \bibinfo{pages}{513--548}.
\newblock


\bibitem[Suzor(2019)]%
        {suzor2019lawless}
\bibfield{author}{\bibinfo{person}{Nicolas~P Suzor}.} \bibinfo{year}{2019}\natexlab{}.
\newblock \bibinfo{booktitle}{\emph{Lawless: The secret rules that govern our digital lives}}.
\newblock \bibinfo{publisher}{Cambridge University Press}.
\newblock


\bibitem[Suzor et~al\mbox{.}(2019)]%
        {suzor2019we}
\bibfield{author}{\bibinfo{person}{Nicolas~P Suzor}, \bibinfo{person}{Sarah~Myers West}, \bibinfo{person}{Andrew Quodling}, {and} \bibinfo{person}{Jillian York}.} \bibinfo{year}{2019}\natexlab{}.
\newblock \showarticletitle{What do we mean when we talk about transparency? Toward meaningful transparency in commercial content moderation}.
\newblock \bibinfo{journal}{\emph{International Journal of Communication}}  \bibinfo{volume}{13} (\bibinfo{year}{2019}), \bibinfo{pages}{18}.
\newblock


\bibitem[Tait(2008)]%
        {tait2008pornographies}
\bibfield{author}{\bibinfo{person}{Sue Tait}.} \bibinfo{year}{2008}\natexlab{}.
\newblock \showarticletitle{Pornographies of Violence? Internet Spectatorship on Body Horror}.
\newblock \bibinfo{journal}{\emph{Critical Studies in Media Communication}} \bibinfo{volume}{25}, \bibinfo{number}{1} (\bibinfo{date}{March} \bibinfo{year}{2008}), \bibinfo{pages}{91–111}.
\newblock
\showISSN{1529-5036, 1479-5809}
\urldef\tempurl%
\url{https://doi.org/10.1080/15295030701851148}
\showDOI{\tempurl}


\bibitem[Thach et~al\mbox{.}(2022)]%
        {thach2022visible}
\bibfield{author}{\bibinfo{person}{Hibby Thach}, \bibinfo{person}{Samuel Mayworm}, \bibinfo{person}{Daniel Delmonaco}, {and} \bibinfo{person}{Oliver Haimson}.} \bibinfo{year}{2022}\natexlab{}.
\newblock \showarticletitle{(In) visible moderation: A digital ethnography of marginalized users and content moderation on Twitch and Reddit}.
\newblock \bibinfo{journal}{\emph{new media \& society}} (\bibinfo{year}{2022}), \bibinfo{pages}{14614448221109804}.
\newblock


\bibitem[Thomas et~al\mbox{.}(2022)]%
        {Thomas2022it's}
\bibfield{author}{\bibinfo{person}{Kurt Thomas}, \bibinfo{person}{Patrick~Gage Kelley}, \bibinfo{person}{Sunny Consolvo}, \bibinfo{person}{Patrawat Samermit}, {and} \bibinfo{person}{Elie Bursztein}.} \bibinfo{year}{2022}\natexlab{}.
\newblock \showarticletitle{“It’s Common and a Part of Being a Content Creator”: Understanding How Creators Experience and Cope with Hate and Harassment Online}. In \bibinfo{booktitle}{\emph{Proceedings of the 2022 CHI Conference on Human Factors in Computing Systems}} (New Orleans, LA, USA) \emph{(\bibinfo{series}{CHI '22})}. \bibinfo{publisher}{Association for Computing Machinery}, \bibinfo{address}{New York, NY, USA}, Article \bibinfo{articleno}{121}, \bibinfo{numpages}{15}~pages.
\newblock
\showISBNx{9781450391573}
\urldef\tempurl%
\url{https://doi.org/10.1145/3491102.3501879}
\showDOI{\tempurl}


\bibitem[Tyler(2006)]%
        {tyler2006psychological}
\bibfield{author}{\bibinfo{person}{Tom~R Tyler}.} \bibinfo{year}{2006}\natexlab{}.
\newblock \showarticletitle{Psychological perspectives on legitimacy and legitimation}.
\newblock \bibinfo{journal}{\emph{Annu. Rev. Psychol.}}  \bibinfo{volume}{57} (\bibinfo{year}{2006}), \bibinfo{pages}{375--400}.
\newblock


\bibitem[Vaccaro et~al\mbox{.}(2020)]%
        {vaccaro2020end}
\bibfield{author}{\bibinfo{person}{Kristen Vaccaro}, \bibinfo{person}{Christian Sandvig}, {and} \bibinfo{person}{Karrie Karahalios}.} \bibinfo{year}{2020}\natexlab{}.
\newblock \showarticletitle{" At the End of the Day Facebook Does What ItWants" How Users Experience Contesting Algorithmic Content Moderation}.
\newblock \bibinfo{journal}{\emph{Proceedings of the ACM on human-computer interaction}} \bibinfo{volume}{4}, \bibinfo{number}{CSCW2} (\bibinfo{year}{2020}), \bibinfo{pages}{1--22}.
\newblock


\bibitem[Vaccaro et~al\mbox{.}(2021)]%
        {vaccaro2021contestability}
\bibfield{author}{\bibinfo{person}{Kristen Vaccaro}, \bibinfo{person}{Ziang Xiao}, \bibinfo{person}{Kevin Hamilton}, {and} \bibinfo{person}{Karrie Karahalios}.} \bibinfo{year}{2021}\natexlab{}.
\newblock \showarticletitle{Contestability for content moderation}.
\newblock \bibinfo{journal}{\emph{Proceedings of the ACM on human-computer interaction}} \bibinfo{volume}{5}, \bibinfo{number}{CSCW2} (\bibinfo{year}{2021}), \bibinfo{pages}{1--28}.
\newblock


\bibitem[Westermann and Coscia(2022)]%
        {westermann2022potential}
\bibfield{author}{\bibinfo{person}{Camilla~Jung Westermann} {and} \bibinfo{person}{Michele Coscia}.} \bibinfo{year}{2022}\natexlab{}.
\newblock \showarticletitle{A potential mechanism for low tolerance feedback loops in social media flagging systems}.
\newblock \bibinfo{journal}{\emph{Plos one}} \bibinfo{volume}{17}, \bibinfo{number}{5} (\bibinfo{year}{2022}), \bibinfo{pages}{e0268270}.
\newblock


\bibitem[Wilson and Land(2020)]%
        {wilson2020hate}
\bibfield{author}{\bibinfo{person}{Richard~Ashby Wilson} {and} \bibinfo{person}{Molly~K Land}.} \bibinfo{year}{2020}\natexlab{}.
\newblock \showarticletitle{Hate speech on social media: Content moderation in context}.
\newblock \bibinfo{journal}{\emph{Conn. L. Rev.}}  \bibinfo{volume}{52} (\bibinfo{year}{2020}), \bibinfo{pages}{1029}.
\newblock


\bibitem[Wojcieszak et~al\mbox{.}(2021)]%
        {wojcieszak2021can}
\bibfield{author}{\bibinfo{person}{Magdalena Wojcieszak}, \bibinfo{person}{Arti Thakur}, \bibinfo{person}{Jo{\~a}o~Fernando Ferreira~Gon{\c{c}}alves}, \bibinfo{person}{Andreu Casas}, \bibinfo{person}{Ericka Menchen-Trevino}, {and} \bibinfo{person}{\&~Miriam Boon}.} \bibinfo{year}{2021}\natexlab{}.
\newblock \showarticletitle{Can AI enhance people’s support for online moderation and their openness to dissimilar political views?}
\newblock \bibinfo{journal}{\emph{Journal of Computer-Mediated Communication}} \bibinfo{volume}{26}, \bibinfo{number}{4} (\bibinfo{year}{2021}), \bibinfo{pages}{223--243}.
\newblock


\bibitem[Yurrita et~al\mbox{.}(2023)]%
        {yurrita2023disentangling}
\bibfield{author}{\bibinfo{person}{Mireia Yurrita}, \bibinfo{person}{Tim Draws}, \bibinfo{person}{Agathe Balayn}, \bibinfo{person}{Dave Murray-Rust}, \bibinfo{person}{Nava Tintarev}, {and} \bibinfo{person}{Alessandro Bozzon}.} \bibinfo{year}{2023}\natexlab{}.
\newblock \showarticletitle{Disentangling Fairness Perceptions in Algorithmic Decision-Making: the Effects of Explanations, Human Oversight, and Contestability}. In \bibinfo{booktitle}{\emph{Proceedings of the 2023 CHI Conference on Human Factors in Computing Systems}}. \bibinfo{pages}{1--21}.
\newblock


\bibitem[Zeng and Kaye(2022)]%
        {zeng2022content}
\bibfield{author}{\bibinfo{person}{Jing Zeng} {and} \bibinfo{person}{D~Bondy~Valdovinos Kaye}.} \bibinfo{year}{2022}\natexlab{}.
\newblock \showarticletitle{From content moderation to visibility moderation: A case study of platform governance on TikTok}.
\newblock \bibinfo{journal}{\emph{Policy \& Internet}} \bibinfo{volume}{14}, \bibinfo{number}{1} (\bibinfo{year}{2022}), \bibinfo{pages}{79--95}.
\newblock


\bibitem[Zhang et~al\mbox{.}(2023)]%
        {zhang2023cleaning}
\bibfield{author}{\bibinfo{person}{Alice~Qian Zhang}, \bibinfo{person}{Kaitlin Montague}, {and} \bibinfo{person}{Shagun Jhaver}.} \bibinfo{year}{2023}\natexlab{}.
\newblock \showarticletitle{Cleaning Up the Streets: Understanding Motivations, Mental Models, and Concerns of Users Flagging Social Media Posts}.
\newblock \bibinfo{journal}{\emph{arXiv preprint arXiv:2309.06688}} (\bibinfo{year}{2023}).
\newblock


\bibitem[Zolides(2021)]%
        {zolides2021gender}
\bibfield{author}{\bibinfo{person}{Andrew Zolides}.} \bibinfo{year}{2021}\natexlab{}.
\newblock \showarticletitle{Gender moderation and moderating gender: Sexual content policies in Twitch’s community guidelines}.
\newblock \bibinfo{journal}{\emph{New Media \& Society}} \bibinfo{volume}{23}, \bibinfo{number}{10} (\bibinfo{year}{2021}), \bibinfo{pages}{2999--3015}.
\newblock


\end{thebibliography}

\appendix

\section{Survey Sample Description}
\label{sec:appendix_survey_sample}
\begin{table}[ht]
\centering
\caption{Demographics of Survey Respondents.}
\resizebox{0.7\textwidth}{!}{%
\label{table:Demo_stat}
\small
\begin{tabular}{lll}
\toprule
 \textbf{Demographic Factor} & \textbf{Category} & \textbf{Number (\%)} \\
\midrule
Gender
 & \quad Male & 1441 (48.6\%) \\
 & \quad Female & 1495 (50.4\%) \\
\midrule
Age
 & \quad Range: 18-89 (Mean = 45) & - \\
\midrule
Ethnicity
 & \quad White & 2123 (72.3\%) \\
 & \quad Black or African American & 359 (12.2\%) \\
 & \quad Asian & 152 (5.2\%) \\
 & \quad Pacific Islander & 254 (8.7\%) \\
 & \quad American Indian or Alaska Native & 48 (1.6\%) \\
\midrule
Hispanic, Latino, or Spanish Origin
 & \quad Yes & 377 (12.8\%) \\
 & \quad No & 2559 (87.2\%) \\
 
\midrule
Income
& \quad Less than \$25,000 & 736 (25\%) \\
& \quad \$25,000 to \$49,999 & 738 (25.1\%) \\
& \quad \$50,000 to \$74,999 & 543 (18.5\%) \\
& \quad \$75,000 to \$124,999 & 555 (19\%) \\
& \quad \$125,000 and above & 336 (11.4\%) \\
& \quad Prefer not to answer & 28 (1\%) \\
\midrule
Political affiliation
 & \quad Democrat & 1217 (41.5\%) \\
 & \quad Republican & 1090 (37.1\%) \\
 & \quad Neutral & 659 (21.4\%) \\
\midrule
Geographic region
 & \quad South & 1104 (37.6\%) \\
 & \quad West & 699 (23.8\%) \\
 & \quad Northeast & 587 (20\%) \\
 & \quad Midwest & 546 (18.6\%) \\
\midrule
Social media use frequency
 & \quad Never & 168 (5.7\%) \\
 & \quad Once a week & 193 (6.6\%) \\
 & \quad 2-3 times a week & 291 (9.9\%) \\
 & \quad 4-6 times a week & 308 (10.5\%) \\
 & \quad Daily & 1976 (67.3\%) \\
\midrule
 Educational Attainment
 & \quad Some high school or less & 996 (33.9\%) \\
 & \quad Some college including AD, BA & 1565 (53.3\%) \\
 & \quad Master's degree or equivalent & 281 (9.6\%) \\
 & \quad Doctorate degree & 72 (2.5\%) \\
\bottomrule
\end{tabular}
}
\label{table:Demo_stat}
\end{table}

Table \ref{table:Demo_stat} shows the demographic characteristics of the survey respondents. Our participants included 1,441 males and 1,495 females, and they had a mean age of 45. 
The majority of income brackets of our sample were between \$25,000 and \$49,999 (25.0\%) and less than \$25,000 (24.9\%).
%\sandy{Missing content after "participants."}
%, while smaller proportions fall into higher income brackets.
Study participants were predominantly White (72.3\%), followed by Black or African American (12.2\%), and other ethnic groups (15.5\%). 
% Political affiliations and geographic distribution were evenly distributed among participants. 
Geographic distribution showed that respondents were predominantly from the South (37.6\%), followed by the West (23.8\%), the Northeast (20\%), and the Midwest (18.6\%).
Daily social media usage was the most prevalent (67.3\%), with only 12.3\% of participants reporting less frequent than weekly use. Educational attainment varied widely: 33.9\% attended some high school or less; 53.3\% attended some college, including AD and BA; and 12.8\% had graduate degrees.

\section{Fairness Perceptions for Different Flagged Posts}
\label{sec:appendix_posts}

Table \ref{table:rule_violation_example_means} shows variations in fairness perceptions for different selections of inappropriate posts to flag by survey respondents.

\begin{table}[h!]
    \centering
    \caption{The mean values of perceived consistency, transparency, and voice for each of the rule violation examples that participants choose to flag in the survey. Standard deviation (SD) values are shown in brackets.}
    \label{table:rule_violation_example_means}
    \resizebox{0.9\textwidth}{!}{
    \begin{tabular}{lccccc}
    \toprule
        \multirow{2}{*}{\textbf{Category}} 
        & \multirow{2}{*}{N}
        & \multicolumn{3}{c}{\textbf{{Mean (SD)}}} \\
                \cline{3-5}
        & 
        & \textbf{Consistency} 
        & \textbf{Transparency} 
        & \textbf{Voice}\\
    \hline
        Mexicans come from an uncivilized ...
        & 621 
        & 5.57 (1.45) & 5.33 (1.31) & 5.31 (1.49) \\

        @Sean11 I hate all you F*** ...
        & 272
        & 5.50 (1.52) & 5.39 (1.29) & 5.33 (1.43) \\

        Drinking bleach has been scientifically ...
        & 694 & 5.68 (1.37) & 5.42 (1.36) & 5.35 (1.56) \\

        Today is a great day! I ate 723 calories ...
        & 165
        & 5.09 (1.62) & 4.98 (1.64) & 5.01 (1.74) \\

        Download the software for credit card ...
        & 178
        & 5.34 (1.57) & 5.01 (1.49) & 5.00 (1.60) \\

        Omg just got tons of Bucks from here! ...
        & 134
        & 5.24 (1.56) & 5.00 (1.41) & 4.86 (1.68) \\

        I have a masturbation video of @Janny12 ...
        & 872
        & 5.60 (1.39) & 5.34 (1.37) & 5.43 (1.51) \\
    \hline
        \textbf{Total}
        & \textbf{2,936} 
        & \textbf{5.54 (1.45)} 
        & \textbf{5.30 (1.38)} 
        & \textbf{5.30 (1.55)} \\
    \bottomrule
    \end{tabular}
    }
\end{table}

\section{Additional Analyses}
\label{sec:appendix_additional_analyses}
Table \ref{table:interaction_effect} below presents the GLM analysis results, which indicate how different combinations of flagging components interact, affecting perceived consistency, transparency, and voice. 
Next, we present Table \ref{table:Interaction_voice}, which details our results on interaction effects of the classification scheme and a text box on perceived voice.
Tables \ref{table:cognitive_burden} and \ref{table:future_use} summarize our results on how different flagging components influence usability.
We comment on these results in sec. \ref{sec:Additional Analyses}.

\begin{table}[ht]
\captionsetup{justification=centering}
\centering
\caption{GLM Results, Indicating the Interaction Effects on Perceived Consistency, Transparency, and Voice.}
\resizebox{0.9\textwidth}{!}{%
\begin{threeparttable}
\begin{tabular}{lllllll}
\toprule
\textbf{Fairness aspect} & \textbf{Interaction between variables} & \textbf{SS} & \textbf{df} & \textbf{MS} & \textbf{\textit{F}} & \textbf{\textit{p}} \\
\midrule
\multirow{11}{*}{Consistency}
 & Classification * Guidelines & 10.68 & 4 & 2.67 & 1.27 & .28 \\
 & Classification * Text box & 8.53 & 2 & 4.27 & 2.02 & .13 \\
 & Classification * Moderator & 4.72 & 4 & 1.18 & .56 & .69 \\
 & Guidelines * Text box & 4.89 & 2 & 2.44 & 1.16 & .32 \\
 & Guidelines * Moderator & 3.95 & 4 & .99 & .47 & .76 \\
 & Text box * Moderator & 6.71 & 2 & 3.35 & 1.59 & .20 \\
 & Classification * Guidelines * Text box & 5.17 & 4 & 1.29 & .61 & .65 \\
 & Classification * Guidelines * Moderator & 27.54 & 8 & 3.44 & 1.63 & .11 \\
 & Classification * Text box * Moderator & 1.63 & 4 & .41 & .19 & .94 \\
 & Guidelines * Text box * Moderator & 11.79 & 4 & 2.95 & 1.40 & .23 \\
 & Classification * Guidelines * Text box * Moderator & 17.37 & 8 & 2.17 & 1.03 & .41 \\
\midrule
\multirow{11}{*}{Transparency}
  & Classification * Guidelines & 1.78 & 4 & .44 & .24 & .92 \\
  & Classification * Text box & 4.17 & 2 & 2.09 & 1.11 & .33 \\
  & Classification * Moderator & 8.27 & 4 & 2.07 & 1.10 & .36 \\
  & Guidelines * Text box & 15.89 & 2 & 7.95 & 4.22 & .02 \\
  & Guidelines * Moderator & 6.88 & 4 & 1.72 & .92 & .45 \\
  & Text box * Moderator & 9.13 & 2 & 4.57 & 2.43 & .09 \\
  & Classification * Guidelines * Text box & 9.28 & 4 & 2.32 & 1.23 & .29 \\
  & Classification * Guidelines * Moderator & 13.11 & 8 & 1.64 & .87 & .54 \\
  & Classification * Text box * Moderator & 13.71 & 4 & 3.43 & 1.82 & .12 \\
  & Guidelines * Text box * Moderator & 7.91 & 4 & 1.98 & 1.05 & .38 \\
  & Classification * Guidelines * Text box * Moderator & 12.69 & 8 & 1.59 & .84 & .56 \\
\midrule
\multirow{11}{*}{Voice}
 & Classification * Guidelines & 10.11 & 4 & 2.53 & 1.12 & .34 \\
 & Classification * Text box & 37.33 & 2 & 18.67 & 8.29 & <.001 \\
 & Classification * Moderator & 14.30 & 4 & 3.58 & 1.59 & .18 \\
 & Guidelines * Text box & 4.66 & 2 & 2.33 & 1.04 & .36 \\
 & Guidelines * Moderator & 1.41 & 4 & .35 & .16 & .96 \\
 & Text box * Moderator & 1.63 & 2 & .82 & .36 & .70 \\
 & Classification * Guidelines * Text box & 2.65 & 4 & .66 & .29 & .88 \\
 & Classification * Guidelines * Moderator & 12.85 & 8 & 1.61 & .71 & .68 \\
 & Classification * Text box * Moderator & 3.41 & 4 & .85 & .38 & .82 \\
 & Guidelines * Text box * Moderator & 5.58 & 4 & 1.39 & .62 & .65 \\
 & Classification * Guidelines * Text box * Moderator & 9.55 & 8 & 1.19 & .53 & .83 \\
\bottomrule
\end{tabular}
\end{threeparttable}
}
\label{table:interaction_effect}
\end{table}

%Interaction_voice
\begin{table}[ht]
\captionsetup{justification=centering}  % Centering the table title
\centering
\caption{Interaction Effects of Classification Scheme and Text Box on Perceived Voice.}
\resizebox{0.7\textwidth}{!}{%
\begin{threeparttable}
\begin{tabular}{lllccc}
\toprule
\textbf{Fairness aspect} & \multicolumn{2}{c}{\textbf{Interaction between the two variables}} & \textbf{MD} & \textbf{SE} & \textbf{\textit{p}} \\
\midrule
\multirow{6}{*}{Voice} 
    & \multirow{3}{*}{No text box} 
       & Simple - No classification  & .34  & .10 & <.001 \\
    &  & Detailed - No classification & .35 & .10 & <.001 \\
    &  & Detailed - Simple classification & .01 & .10 & 1.00   \\
\cmidrule{2-6}
    & \multirow{3}{*}{Text box is provided} 
       & Simple - No classification  & -.05  & .10 & 1.00 \\
    &  & Detailed - No classification & -.20 & .10 & .13 \\
    &  & Detailed - Simple classification & -.15 & .10 & .38 \\
\bottomrule
\end{tabular}
\end{threeparttable}
}
\label{table:Interaction_voice}
\end{table}

%Table is adjusted
\begin{table}[ht]
\centering
\begin{minipage}{0.45\textwidth}
\centering
\resizebox{1.08\textwidth}{!}{%
\begin{threeparttable}
\caption{\fontsize{8}{9}\selectfont Results Summarizing Whether Different Choices of Flagging Components Impact Participants' Cognitive Burden.}
\begin{tabular}{lccccc}
\toprule
\textbf{Components} & \textbf{SS} & \textbf{df} & \textbf{MS} & \textbf{f/t} & \textbf{\textit{p}} \\
\midrule
Classification & 8.14 & 2 & 4.07 & 1.31 & .27 \\ 
Guidelines & 17.63 & 2 & 8.81 & 2.83 & .06 \\ 
Text box & - & 2933.87 & - & -3.03 & .01 \\
Moderator & 4.01 & 2 & 2.00 & 0.64 & .53 \\ 
\bottomrule
\end{tabular}
\label{table:cognitive_burden}
\end{threeparttable}
}
\end{minipage}
\hspace{20pt}
\begin{minipage}{0.45\textwidth}
\centering
\resizebox{1.06\textwidth}{!}{%
\begin{threeparttable}
\caption{Results Summarizing Whether Different Choices of Flagging Components Impact Participants' Future Use.}
\begin{tabular}{lccccc}
\toprule
\textbf{Components} & \textbf{SS} & \textbf{df} & \textbf{MS} & \textbf{f/t} & \textbf{\textit{p}} \\
\midrule
Classification & .91 & 2 & .46 & .19 & .83 \\ 
Guidelines & .12 & 2 & .06 & .03 & .98 \\ 
Text box & - & 2930.17 & - & -1.07 & .14 \\
Moderator & 8.40 & 2 & 4.20 & 1.77 & .17 \\ 
\bottomrule
\end{tabular}
\label{table:future_use}
\end{threeparttable}
}
\end{minipage}
\vspace{5pt}
\hspace{0.02\textwidth}
\parbox{\textwidth}{%
\raggedright
\small In the above tables, ANOVA results are presented for `Classification,' `Guidelines,' and `Moderator' components and \textit{t}-test results are presented for the `Text Box' component.
}
\label{table:combined}
\end{table}

\end{document}